\pdfoutput=1

\documentclass[%
reprint,
 superscriptaddress,
 amsmath,amssymb,
 aps,
]{revtex4-1}

\RequirePackage[l2tabu, orthodox]{nag}
\usepackage[all, warning]{onlyamsmath}

\usepackage[utf8]{inputenc}

\usepackage{graphicx}

\usepackage[svgnames]{xcolor} % better color for hyperref
\usepackage[colorlinks,citecolor=DarkGreen,linkcolor=FireBrick,linktocpage,unicode]{hyperref} % add hypertext capabilities

\usepackage{amsmath,amssymb,mathtools}
\usepackage{url}
\usepackage[version=3]{mhchem}
\usepackage{color}
\usepackage{bm}
\usepackage{multirow}

\usepackage{tikz}
\usetikzlibrary{positioning,arrows.meta,shapes}
\usepackage{algorithm, algpseudocode}
\usepackage{float}

\newcommand{\argmin}{\mathop{\mathrm{argmin}}}

\newcommand{\code}[1]{\texttt{#1}}

\begin{document}

\preprint{APS/123-QED}

\title{Efficient Crystal Structure Prediction Using Universal Neural Network Potential with Diversity Preservation in Genetic Algorithms}

\author{Takuya Shibayama}
  \affiliation{Preferred Networks, Inc., Tokyo 100-0004, Japan.}
\author{Hideaki Imamura}
  \affiliation{Preferred Networks, Inc., Tokyo 100-0004, Japan.}
\author{Katsuhiko Nishimra}
  \affiliation{Preferred Networks, Inc., Tokyo 100-0004, Japan.}
\author{Kohei Shinohara}
  \email{kshinohara@preferred.jp}
  \affiliation{Preferred Networks, Inc., Tokyo 100-0004, Japan.}
\author{Chikashi Shinagawa}
  \affiliation{Preferred Networks, Inc., Tokyo 100-0004, Japan.}
\author{So Takamoto}
  \affiliation{Preferred Networks, Inc., Tokyo 100-0004, Japan.}
\author{Ju Li}
  \affiliation{Department of Nuclear Science and Engineering, and Department of Materials Science and Engineering, Massachusetts Institute of Technology, Cambridge, MA 02139, USA.}

\date{\today}

\begin{abstract}
Crystal structure prediction (CSP) is crucial for identifying stable crystal structures in given systems and is a prerequisite for computational atomistic simulations.
Recent advances in neural network potentials (NNPs) have reduced the computational cost of CSP.
However, searching for stable crystal structures across the entire composition space in multicomponent systems remains a significant challenge.
Here, we propose an improvement of genetic algorithm (GA) -based CSP method using a universal NNP.
Our GA-based methods are designed to efficiently expand convex hull volumes while preserving the diversity of crystal structures.
Our hull-informed filtering and elitist-selection procedures incorporate an aging mechanism that prioritizes recently improved compositions.
We also employ niching to prevent convergence to a small set of stoichiometries, thereby preserving a diverse, high-quality population.
Our evaluation shows that the present method outperforms the symmetry-aware random structure generation and existing CSP methods, achieving a larger convex hull with fewer trials.
We demonstrated that our approach, combined with the developed universal NNP (PFP), can accurately reproduce and explore phase diagrams obtained through DFT calculations; this indicates the validity of PFP across a wide range of crystal structures and element combinations.
This study, which integrates a universal NNP with a GA-based CSP method, highlights the promise of these methods in materials discovery.
\end{abstract}

\maketitle

\section{\label{sec:introduction}Introduction}

Computational atomistic simulations based on a quantum-mechanical description enhances our understanding of materials and even contribute to modern materials design \cite{https://doi.org/10.1002/jcc.21057}.
Crystal structure prediction (CSP), a process to predict stable crystal structures in given systems, is a crucial prerequisite to harness the computational atomic simulations \cite{oganov2011modern,Oganov2019,wales_2004,Jain2016}.
Despite the well-established methodology for analyzing crystal structures from experimental data, CSP is essential for accelerating materials discovery through the ab initio approach.

While CSP plays an integral role in predicting stable crystal structures, it represents a daunting global optimization task owning to the vast energy landscape.
Substituting elements from known crystal structure prototypes is a prevailing method for generating candidate structures \cite{doi:10.1021/ic102031h}.
Although the prototype substitution method often gives reasonable results, its coverage falls short, especially in multicomponent systems.
Metaheuristic algorithms offer another approach to creating novel crystal structures.
These include methods such as random structure search \cite{Pickard_2011}, basin hopping \cite{doi:10.1021/jp970984n}, minima hopping \cite{doi:10.1063/1.3512900,KRUMMENACHER2024101632}, genetic algorithm (GA) \cite{PhysRevLett.75.288,doi:10.1063/1.2210932,GLASS2006713,doi:10.1021/ar1001318,LYAKHOV20131172,LONIE2011372,doi:10.1021/acs.jpcc.0c09531}, particle swarm optimization \cite{PhysRevB.82.094116}, and Bayesian optimization \cite{PhysRevMaterials.2.013803,PhysRevLett.124.086102}.
Typically, these CSP methods are combined with density functional theory (DFT) calculations to evaluate the formation energy of the candidate structures.
However, the time-consuming DFT calculations, which limit the exploration of structures, significantly hinder the efficiency of CSP.

Recent advancements in machine learning potential, interatomic potential fitted by DFT calculations, enable an efficient approach to CSP thanks to their fast and accurate energy evaluation \cite{PhysRevLett.98.146401,PhysRevLett.104.136403,doi:10.1137/15M1054183,PhysRevB.99.014104,PhysRevB.99.064114}.
Because we are often interested in multicomponent systems like ternary or quaternary systems in CSP, it is preferable for the machine learning potential to exhibit scalability with the number of elements and transferability across various systems.
Also, high accuracy is required to capture subtle energy differences among distinct structures in CSP.
A universal neural network potential (NNP) trained with extensive datasets has been gaining significant attention owing to its scalability and transferability, which includes
M3GNet \cite{doi:10.1021/acs.chemmater.9b01294,Chen2022},
ALIGNN-FF \cite{D2DD00096B},
CHGNet \cite{Deng2023},
MACE-MP-0 \cite{NEURIPS2022_4a36c3c5,batatia2023foundation},
GNoME \cite{Merchant2023},
Orb \cite{neumann2024orbfastscalableneural},
EquiformerV2-OMat24 \cite{equiformer_v2,barroso_omat24},
MatterSim \cite{yang2024mattersim},
and PFP \cite{TAKAMOTO2022111280,10.1038/s41467-022-30687-9,JACOBS2025101214}.

Applying a machine learning potential to CSP is relatively straightforward and it still achieves a significant improvement in efficiency \cite{PhysRevLett.120.026102,PhysRevB.99.064114,C8FD00055G,10.3389/fchem.2020.589795,PhysRevMaterials.4.063801,Kang2022,Cheng2022,https://doi.org/10.1002/adma.202210788,Merchant2023,Chang2024}.
Nevertheless, further development of CSP algorithms employing universal NNPs have room for enhancement.
While the existing ab initio CSP methods often implicitly assume the number of energy evaluations to be limited due to the computational cost of DFT, the universal NNPs can allow us the vast number of energy evaluations and consider an entire convex hull concurrently, thereby appreciating the diversity of crystal structures all at once.

\begin{figure}[!tb]
  \centering
  \includegraphics[width=0.5\textwidth]{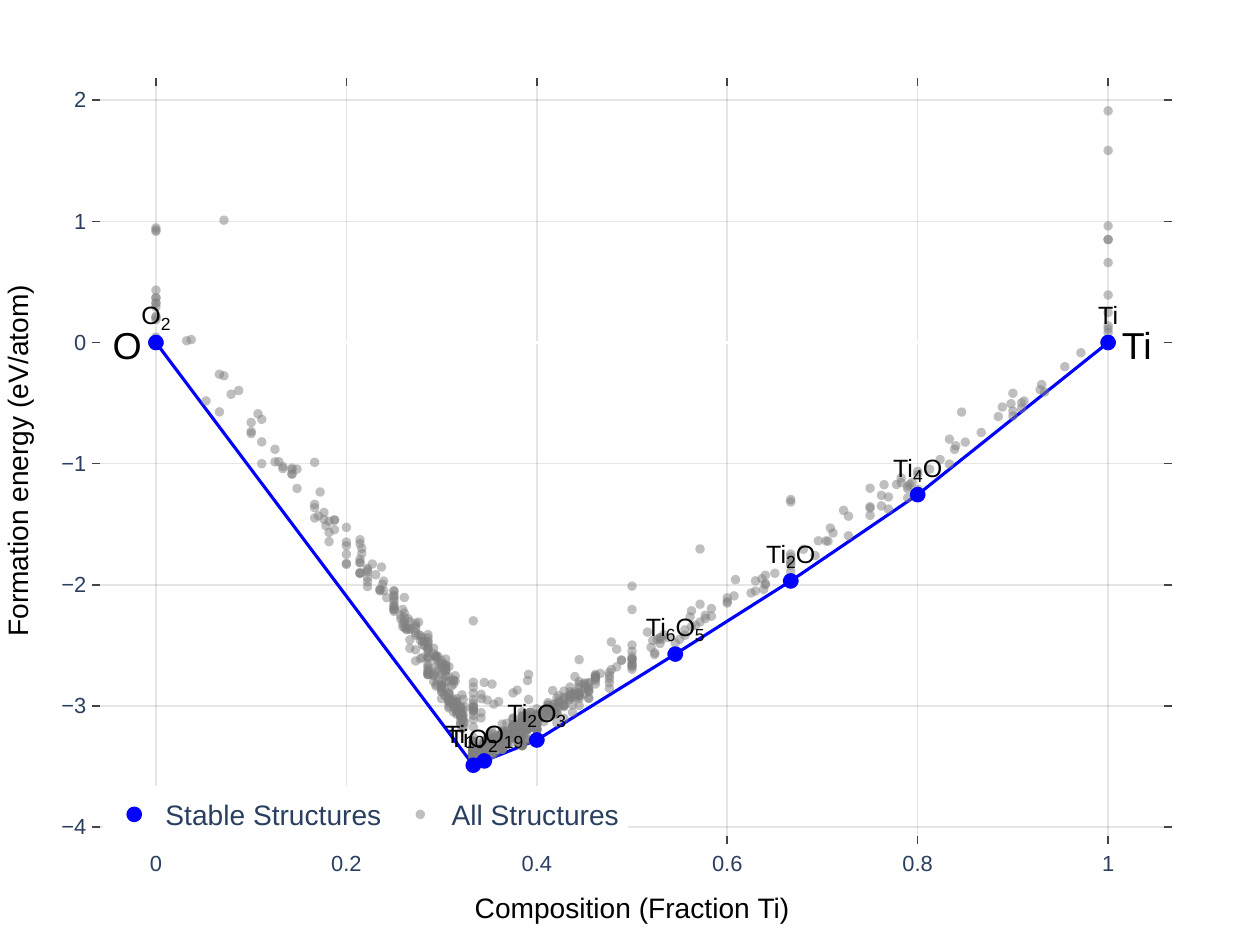}
  \caption{Biased sampling of the convex hull in the Ti--O system by a standard GA-based CSP method.}
  \label{fig:2-element-biased}
\end{figure}

GA-based CSP methods exemplified by \textsc{USPEX} provide variable-composition searches and have enabled numerous discoveries~\cite{doi:10.1063/1.2210932,oganov2011modern,GLASS2006713,LYAKHOV20131172}.
However, they are not explicitly designed to optimize the entire convex hull.
As a consequence, selection pressure and operator design tend to bias the search toward a few low-energy stoichiometries.
Figure~\ref{fig:2-element-biased} illustrates this effect in the Ti--O system: when the composition is allowed to vary, a traditional GA-based CSP concentrates trials near \ce{TiO2}, leaving large regions of the hull insufficiently explored.

Recently, convex hull genetic algorithm (CHGA) reframes variable-composition CSP as convex hull optimization, leveraging an analogy to Pareto-front optimization to act on the hull as a whole~\cite{Donaldson2024AGA}.
While promising, published validations have focused on two-element systems with on the order of $10^4$ trials.
In practice, exploring the entire convex hull requires much more massive simulations than composition-fixed searches.
This becomes especially pronounced for ternary and higher-order systems, where $10^4$ trials are insufficient to cover the hull.
Moreover, as shown in Section~\ref{sec:results-and-discussion}, existing approaches often saturate early, slipping into local optima and thus failing to benefit from prolonged runs.

In this work, we propose a GA-based method that remains effective over long optimization by dynamically shifting the search region through controlled forgetting (aging) of outdated structures, while preserving diversity across populations (niching).
Conceptually, our method stands on both traditional GA-based CSP and CHGA:
(i) we adopt the GA pipeline with standard variation operators and the use of energy/hull information for ranking, and
(ii) we explicitly shape the selection pressure at the level of the elitist selection so that the search does not collapse to a few stoichiometries during long optimization.
Concretely, we (a) filter out unpromising trials using an age-weighted and hull-aware score, and (b) form the elite population by a hull-informed non-dominated sorting coupled with niching tie-breakers.

The remainder of this paper is organized as follows.
Section~\ref{sec:method-pfp} describes PFP and a procedure for calculating formation energy with PFP in this study.
Section~\ref{sec:method-ga} provides the problem setting and basic terminology, the background of GA-based CSP and CHGA, and the proposed method.
Section~\ref{sec:results-and-discussion} shows the application of the present GA to a chemically diverse range of elements, from binary to octonary systems.

\section{\label{sec:method-pfp}Formation energy calculation with PFP}

\subsection{\label{sec:method-pfp-pfp}PFP}

In this study, we used our developed universal NNP called the PFP version 6.0.0 \cite{10.1038/s41467-022-30687-9}, which was trained on around 42 million structures.
The PFP can be applied to arbitrary combinations of 72 supported elements (all elements from H to Bi except Tc, Pm, Eu, Tb, Dy, Ho, Er, Tm, Yb, Lu, and Tl).
Its training dataset was collected with DFT calculations using plane-wave basis sets and the projector augmented wave (PAW) \cite{PhysRevB.50.17953,PhysRevB.59.1758} with the Perdew--Burke--Ernzerhof (PBE) exchange-correlation functional \cite{PhysRevLett.77.3865} implemented in the Vienna ab-initio simulation package (VASP)~\cite{PhysRevB.47.558,PhysRevB.54.11169,KRESSE199615}.
Details of the DFT calculations are described in Ref.~\onlinecite{10.1038/s41467-022-30687-9}.

For systems containing V, Cr, Mn, Fe, Co, Ni, Cu, Mo, or W, we prepared the DFT calculations with the generalized gradient approximation (GGA) with PBE functional and GGA with Hubbard $U$ corrections (GGA+U) introduced by Dudarev et al. \cite{PhysRevB.57.1505}.
The $U-J$ parameters in GGA+U were adopted from the Materials Project (MP) \cite{10.1063/1.4812323} except for Cu and from Wang et al. \cite{PhysRevB.73.195107} for Cu.
We trained PFP simultaneously for the DFT calculations with and without Hubbard $U$ corrections.
The PFP has two modes to predict energies with and without Hubbard $U$ corrections.
We used PFP trained with Hubbard $U$ corrections for oxides and fluorides containing V, Cr, Mn, Fe, Co, Ni, Cu, Mo, or W, and otherwise without Hubbard $U$ corrections.
This choice is consistent with MP except for Cu.

\subsection{Reference simple systems}

For formation energy calculations, we selected reference crystal structures for the 72 supported elements and 2 partially supported elements (Tc and Tl), as listed in Table~\ref{tab:simple-systems}.
For each element, we collected crystal structures from MP and searched for the lowest-energy structure after structural relaxation with PFP.
The structural relaxation of atomic positions and lattice constants was performed until residual forces were less than $5 \times 10^{-3}$ eV/\AA.
We matched the reference structures with the AFLOW standard encyclopedia of crystallographic prototypes \cite{MEHL2017S1} using \code{pymatgen}~\cite{ONG2013314}.
The third column of Table~\ref{tab:simple-systems} displays the matched prototype structures within 0.1 meV/atom from the lowest-energy structure, which captures the diversity of stable prototype structures across various elements.
The fourth column shows the PFP energy differences in meV/atom between two structures: one is the energy-lowest structure in PFP and the other is the energy-lowest structure in DFT calculation in MP.
While the energy differences are particularly noticeable for layered or molecular crystal structures of B, O, As, and Br, possibly due to convergence issues in structural relaxations, PFP accurately evaluates stable crystal structures up to approximately 10 meV/atom.

\begin{table*}[htb]
  \centering
  \caption{
    Reference simple systems for each element in this study.
    The second column shows initial structures for structural relaxation taken from MP, recorded with \code{material\_id}.
    The third column shows prototype structures within 0.1 meV/atom from the lowest-energy structure.
    When a prototype is not found, \code{material\_id} is denoted instead.
    The fourth column shows the PFP energy differences (meV/atom) between the present reference structures and the on-the-hull structures in MP.
  }
  \label{tab:simple-systems}
  \scalebox{0.9}{
  \begin{tabular}{cccc|cccc}
    \hline \hline
    H & mp-973783 & $\beta$-O, mp-731827 & 4.3  &  Sr & mp-867202 & A1 (fcc), A3 (hcp), A3' ($\alpha$-La), C19 ($\alpha$-Sm) & 0.0 \\
    He & mp-614456 & A1 (fcc), A3 (hcp) & 0.8  &  Y & mp-112 & A3 (hcp) & 0.0 \\
    Li & mp-135 & A2 (bcc) & 0.7  &  Zr & mp-8635 & A1 (fcc) & 2.2 \\
    Be & mp-87 & A3 (hcp) & 0.0  &  Nb & mp-75 & A2 (bcc) & 0.0 \\
    B & mp-161 & $\beta$-B & 40.5  &  Mo & mp-129 & A2 (bcc) & 0.0 \\
    C & mp-568286 & mp-568286, mp-990424 & 0.6  &  Tc & mp-113 & A3 (hcp) & 0.0 \\
    N & mp-570747 & $\gamma$-N & 5.7  &  Ru & mp-33 & A3 (hcp) & 0.0 \\
    O & mp-1180036 & mp-1180036 & 41.3  &  Rh & mp-74 & A1 (fcc) & 0.0 \\
    F & mp-561203 & A14 (molecular iodine) & 0.0  &  Pd & mp-2 & A1 (fcc) & 0.0 \\
    Ne & mp-111 & A1 (fcc) & 0.0  &  Ag & mp-10597 & A3 (hcp) & 0.6 \\
    Na & mp-974920 & A1 (fcc), A3 (hcp), A3' ($\alpha$-La), C19 ($\alpha$-Sm) & 0.0  &  Cd & mp-94 & A3 (hcp) & 0.0 \\
    Mg & mp-1056702 & A1 (fcc) & 1.1  &  In & mp-973111 & A3 (hcp) & 1.9 \\
    Al & mp-134 & A1 (fcc) & 0.0  &  Sn & mp-117 & A4 (diamond) & 0.0 \\
    Si & mp-165 & Lonsdaleite & 5.9  &  Sb & mp-104 & A7 ($\alpha$-As) & 0.0 \\
    P & mp-1198724 & mp-1198724 & 9.6  &  Te & mp-567313 & A8 ($\gamma$-Se) & 0.0 \\
    S & mp-666931 & mp-666931 & 0.7  &  I & mp-23153 & A14 (molecular iodine) & 0.0 \\
    Cl & mp-1008394 & A14 (molecular iodine) & 0.0  &  Xe & mp-570510 & A1 (fcc), A3 (hcp), A3' ($\alpha$-La), C19 ($\alpha$-Sm) & 0.0 \\
    Ar & mp-568145 & A1 (fcc), A3 (hcp) & 0.0  &  Cs & mp-639727 & A1 (fcc), A3 (hcp), A3' ($\alpha$-La) & 0.0 \\
    K & mp-1184764 & A1 (fcc), A3 (hcp), A3' ($\alpha$-La), C19 ($\alpha$-Sm) & 0.0  &  Ba & mp-1096840 & A2 (bcc) & 0.0 \\
    Ca & mp-1183455 & A1 (fcc), A3 (hcp), A3' ($\alpha$-La) & 0.0  &  La & mp-156 & A1 (fcc), A3' ($\alpha$-La) & 0.0 \\
    Sc & mp-67 & A3 (hcp) & 0.0  &  Ce & mp-64 & A20 ($\alpha$-U) & 5.4 \\
    Ti & mp-46 & A3 (hcp) & 16.5  &  Pr & mp-38 & A1 (fcc), A3 (hcp), A3' ($\alpha$-La), C19 ($\alpha$-Sm) & 0.0 \\
    V & mp-146 & A2 (bcc) & 0.0  &  Nd & mp-123 & A1 (fcc), A3' ($\alpha$-La), C19 ($\alpha$-Sm) & 0.0 \\
    Cr & mp-90 & A2 (bcc) & 0.0  &  Sm & mp-21377 & A1 (fcc), A3 (hcp), A3' ($\alpha$-La), C19 ($\alpha$-Sm) & 0.0 \\
    Mn & mp-35 & A12 ($\alpha$-Mn) & 0.0  &  Gd & mp-155 & A3 (hcp), C19 ($\alpha$-Sm) & 0.0 \\
    Fe & mp-1271068 & A2 (bcc) & 0.0  &  Hf & mp-103 & A3 (hcp) & 0.0 \\
    Co & mp-54 & A3 (hcp) & 6.0  &  Ta & mp-50 & A2 (bcc) & 29.6 \\
    Ni & mp-23 & A1 (fcc) & 0.0  &  W & mp-91 & A2 (bcc) & 0.0 \\
    Cu & mp-30 & A1 (fcc) & 0.0  &  Re & mp-8 & A3 (hcp) & 23.6 \\
    Zn & mp-1187812 & A3 (hcp), C19 ($\alpha$-Sm) & 0.1  &  Os & mp-49 & A3 (hcp) & 0.0 \\
    Ga & mp-142 & A11 ($\alpha$-Ga) & 0.0  &  Ir & mp-101 & A1 (fcc) & 0.0 \\
    Ge & mp-1007760 & Lonsdaleite & 4.1  &  Pt & mp-126 & A1 (fcc) & 0.0 \\
    As & mp-11 & A7 ($\alpha$-As) & 23.0  &  Au & mp-1008634 & A3 (hcp) & 2.0 \\
    Se & mp-542605 & mp-542605 & 1.1  &  Hg & mp-1184554 & A3 (hcp), A3' ($\alpha$-La), C19 ($\alpha$-Sm) & 2.4 \\
    Br & mp-1120813 & mp-1120813 & 10.3  &  Tl & mp-151 & A1 (fcc) & 0.3 \\
    Kr & mp-976347 & A1 (fcc), A3 (hcp), A3' ($\alpha$-La), C19 ($\alpha$-Sm) & 0.0  &  Pb & mp-20745 & A1 (fcc), A3 (hcp), A3' ($\alpha$-La) & 0.1 \\
    Rb & mp-12628 & A1 (fcc), A3 (hcp), A3' ($\alpha$-La), C19 ($\alpha$-Sm) & 12.8  &  Bi & mp-23152 & A7 ($\alpha$-As) & 0.0 \\
    \hline \hline
  \end{tabular}
  }
\end{table*}

\subsection{\label{sec:method-pfp-correction}MP compatibility and energy corrections}

Because we used almost the same DFT settings as MP, we applied anion and GGA/GGA+U mixing scheme corrections \cite{Wang2021} into PFP total energies for better predictions across diverse chemical systems.
The policy regarding which modes of PFP to use with and without Hubbard $U$ corrections in Section~\ref{sec:method-pfp-pfp} is compatible with the GGA/GGA+U mixing scheme in MP.
The exception is an oxide with Cu, in which case we use a correction for Cu in Ref.~\onlinecite{PhysRevB.84.045115}.
Consequently, we can directly compare a convex hull with PFP and MP for all systems except oxides with Cu.

\section{\label{sec:method-ga}GA-based CSP method}

\subsection{\label{sec:method-setting}Problem setting and terminology}

We consider CSP under variable compositions for a given list of elements~\cite{oganov2011modern,Oganov2019}.
Let $M$ denote the number of elements, and let $r\in[0,1]^M$ be a reduced composition with $\sum_{a=1}^{M} r_a = 1$.
For a crystal structure $i$, let $r(i)$ be its reduced composition and $E(i)$ be its formation energy per atom after structural relaxation (Section~\ref{sec:method-pfp}).
Thermodynamic stability at $0$~K is evaluated by the distance (``energy above hull'') to the lower convex envelope of the energy--composition space; structures on the convex hull are thermodynamically stable with respect to decomposition into competing phases~\cite{PhysRevB.80.092101,Bartel2022}.
Our goal is to efficiently discover low-energy structures near or on the convex hull by exploring the joint space of compositions and periodic atomic configurations with up to a user-specified maximum number of atoms per unit cell.
Throughout the paper, we assume that a list of elements (e.g., O--Sr--Ti) is given in advance, and the search spans all stoichiometries composed of those elements.

In this work we adopt a \emph{genetic algorithm (GA)} approach, which has achieved substantial success in CSP over the last two decades~\cite{PhysRevLett.75.288,doi:10.1063/1.2210932,GLASS2006713,doi:10.1021/ar1001318,LYAKHOV20131172,LONIE2011372,doi:10.1021/acs.jpcc.0c09531,oganov2011modern,Oganov2019,omee2023crystalstructurepredictionusing,Donaldson2024AGA,omee2025polymorphismcrystalstructureprediction}.
Our implementation specializes this GA paradigm to the variable-composition setting with convex hull aware optimization~\cite{Donaldson2024AGA} for multicomponent systems, particularly for improving performance in the case where $M > 2$.

We use standard GA terminology~\cite{oganov2011modern}.
A \emph{generation} is one cycle of variation and selection, and a \emph{population} is the set of candidates in a generation.
\emph{Parents} are selected structures that produce \emph{offspring} via variation operators.
The variation operators which produce offspring from one or more parents are called \emph{mutation} and \emph{crossover}, respectively.
\emph{Selection} determines which candidates survive to the next generation; we use \emph{elitist} selection to guarantee carryover of top performers and \emph{niching} to maintain diversity across compositions and structures (e.g., fitness sharing~\cite{goldberg1987genetic}, crowding-based schemes as in NSGA-II~\cite{996017}, or hyperplane-based methods in NSGA-III~\cite{6600851}).
While not a term commonly used in genetic algorithm contexts, the term \emph{trial} as employed in general iterative black-box optimization algorithms~\cite{akiba2019optuna} refers to a pipeline of variation, structural relaxation, and formation energy evaluation for a single structure within a population.

The remainder of this section is organized as follows.
In Section~\ref{sec:method-background}, we summarize the standard GA-based CSP pipeline exemplified by \textsc{USPEX}~\cite{oganov2011modern} (initialization, crossover/mutation, and fitness-based selection), and we briefly review a recent \emph{convex-hull genetic algorithm (CHGA)} that explicitly casts variable-composition search as an optimization of the convex hull itself~\cite{Donaldson2024AGA}.
In Section~\ref{sec:method-proposed}, motivated by multicomponent systems ($M>2$), we describe our hull-informed filtering and elitist selection procedures that integrate \emph{aging} (to prefer recently improved compositions) and \emph{niching} (to avoid collapse to a few stoichiometries), thereby maintaining a high-quality and diverse population across populations.

\subsection{\label{sec:method-background}Background: standard GA and convex-hull GA for CSP}

\begin{figure}[!tb]
  \centering
  \begin{tikzpicture}
    \tikzset{process/.style={rectangle, draw, text centered, rounded corners, minimum height = 1.0cm}};  % minimum height = 1.5cm
    \tikzset{branch/.style={diamond, draw, text centered, aspect=5.5}};  % minimum height = 1.5cm
    \tikzset{rightarrow/.style={arrows={-{Latex[length=2mm]}}}};

    \node[process,align=center] (init) {Generate initial population};
    \node[process,align=center] (retrieve) [below = of init.center] {Retrieve structures from current population};
    \node[process,align=center] (select) [below = of retrieve.center] {Select population to produce next generation};
    \node[process,align=center] (generate) [below = of select.center] {Produce next generation \\ by variation operators};
    \node[process,align=center] (opt) [below = of generate.center] {Calculate formation energy \\ after structural relaxation};
    \node[branch] (check) [below = of opt.center] {Enough trials?};
    \node[align=center] (end) [below = of check.center] {End};

    \draw[rightarrow] (init) -- (retrieve);
    \draw[rightarrow] (retrieve) -- (select);
    \draw[rightarrow] (select) -- (generate);
    \draw[rightarrow] (generate) -- (opt);
    \draw[rightarrow] (opt) -- (check);
    \draw[rightarrow] (check) -- (end) node[midway,right] {Yes};
    \draw[rightarrow] (check.east) -- node[midway,below] {No} ++ (2cm,0) |- (retrieve.east);
  \end{tikzpicture}
  \caption{Schematic diagram of the standard GA-based CSP method.}
  \label{fig:csp-workflow}
\end{figure}

\subsubsection{Standard GA-based CSP}

Figure~\ref{fig:csp-workflow} outlines a standard GA-based CSP method.
After an initial population is produced by sampling, the algorithm iterates until a sufficient number of trials have been completed: from each population, structures that should survive to the next generation and those that should serve as parents are selected; offspring are then created by applying variation operators to the selected parents, locally relaxed, and evaluated before the next iteration~\cite{oganov2011modern,GLASS2006713,LYAKHOV20131172,LONIE2011372,ase2017}.

The initial population is commonly generated via random structure generation, which is also reused as a variation operator to maintain diversity~\cite{GLASS2006713,LYAKHOV20131172,ase2017}.
Purely unconstrained random sampling tends to produce chemically similar, as the number of atoms per cell grows~\cite{LYAKHOV20101623}.
To mitigate this, existing methods split the simulation cell into multiple subcells, perform random sampling in a single subcell, and then copy it to the remaining subcells~\cite{LYAKHOV20101623}.

Because thermodynamically stable crystals often exhibit (partial) symmetry~\cite{Pickard_2011}, symmetry-constrained generators are also employed at the subcell stage, in which atoms are assigned to compatible Wyckoff positions of a sampled space group.
Popular software that implements these strategies includes \code{PyXtal}~\cite{FREDERICKS2021107810} and \code{AIRSS}~\cite{Pickard_2011}.

Selection of survivors and parents to produce next generation typically relies on a per-structure fitness value after local relaxation~\cite{oganov2011modern,Oganov2019}.
In GA-based CSP this is commonly an energy-based criterion (e.g., formation energy or enthalpy per atom; lower is better), optionally augmented with similarity-aware penalties to avoid duplicates at the same stoichiometry~\cite{LYAKHOV20131172,valle2008crystal,oganov2009quantify}.
To preserve diversity, deterministic truncation is often combined with stochastic schemes such as fitness-proportionate (roulette-wheel) selection~\cite{oganov2011modern}.

Variation operators that produce offspring from one or more parents are called \emph{mutation} and \emph{crossover}, respectively; standard operators for periodic crystals include heredity (cut-and-splice), lattice strain, atomic ``rattle'', and species permutation/transmutation~\cite{10.2138/rmg.2010.71.13,doi:10.1021/ar1001318,GLASS2006713,LYAKHOV20131172}.

\subsubsection{Convex hull GA (CHGA)}

\begin{figure}[!tb]
  \centering
  \includegraphics[width=0.5\textwidth]{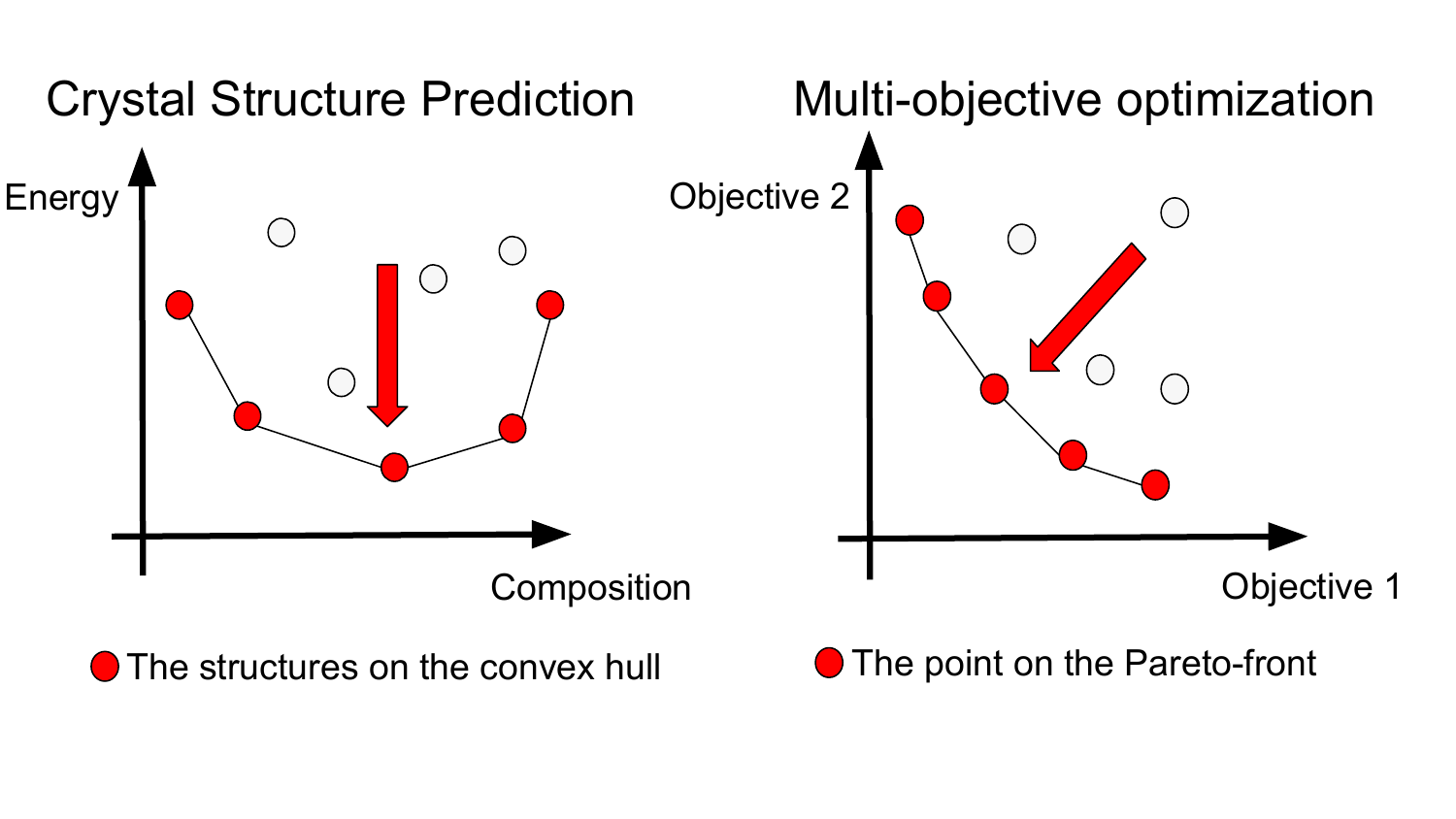}
  \caption{The similarity between the convex hull in the energy--composition space and the Pareto front in the multi-objective optimization problem.}
  \label{fig:similarity-of-convex-hull-and-pareto-front}
\end{figure}

CHGA reformulates variable-composition CSP as a direct \emph{convex hull} optimization problem: instead of optimizing individual structures in isolation, the algorithm aims to optimize the hull itself, leveraging an analogy to Pareto-front optimization in multi-objective genetic algorithms~\cite{Donaldson2024AGA}.
Figure~\ref{fig:similarity-of-convex-hull-and-pareto-front} illustrates the similarity between the convex hull in the energy--composition space and the Pareto front in multi-objective optimization.

In multi-objective optimization, NSGA-II/III~\cite{996017,6600851} are widely used ``gold standards'':
they apply non-dominated sorting to partition a population into fronts and adopt elitist survival from lower-rank (better) fronts; niching then maintains diversity—via crowding distance in low dimensions (NSGA-II) or reference points on a hyperplane/simplex in higher dimensions (NSGA-III).

In CHGA, the dominance relation for non-dominated sorting is defined via the convex hull~\cite{Donaldson2024AGA}.
For a given population, one first draws the convex hull in the energy--composition space and assigns \emph{rank~0} to structures on the hull.
Removing these and re-drawing the hull yields \emph{rank~1}, and repeating this ``peeling'' assigns a rank to every structure.
Thus realizing a hull-based non-dominated sort.

Survivor selection follows an elitist strategy analogous to NSGA families.
As for tie-breaking, CHGA explores niching mechanisms tailored to hull optimization and reports, in two-elements system tests (Li--Si), that niching was \emph{not} critical; they instead used the convex hull hypervolume as a convergence indicator and recommended full elitism with a mix of bred and high-symmetry random generation~\cite{Donaldson2024AGA}.

\subsection{\label{sec:method-proposed}Proposed method: aging and niching for convex hull expansion}

We propose a GA-based method that remains effective over long runs by dynamically shifting the search region through controlled forgetting (aging) of outdated structures, while preserving diversity across populations (niching).
The difference from standard GA-based CSP and CHGA (Section~\ref{sec:method-background}) lies in the selection step (Fig.~\ref{fig:csp-workflow}).
Concretely, we (a) filter out unpromising trials using an age-weighted, hull-aware score (Section~\ref{sec:method-ga-population-filtering}) and (b) form the elite population by a hull-informed non-dominated sorting coupled with niching tie-breakers (Section~\ref{sec:method-ga-elite-population-selector}).
The overall workflow remains that of Fig.~\ref{fig:csp-workflow} and supports asynchronous trial evaluation.

\subsubsection{\label{sec:method-ga-population-filtering}Population filtering with aging and niching}

The population filtering is performed to construct a subset of the parent population and the elite population.
This step is just after retrieving the current population and before selecting the elite population, as shown in Fig.~\ref{fig:csp-workflow}.

In this step, we selected structures close to the convex hull with the most recently updated composition.
Our algorithm design is based on two intuitions:
(1) lower-energy structures should exist near the structures close to the convex hull, and (2) there should be some room for improvement in the neighbors of the compositions that observed recent updates on their lowest energy.
Population filtering is similar to the aging evolution strategy~\cite{C2CE06642D,real2019regularized} in that it ignores the structures that have not been updated for a long time.

\begin{figure}[!tb]
  \centering
  \includegraphics[width=0.5\textwidth]{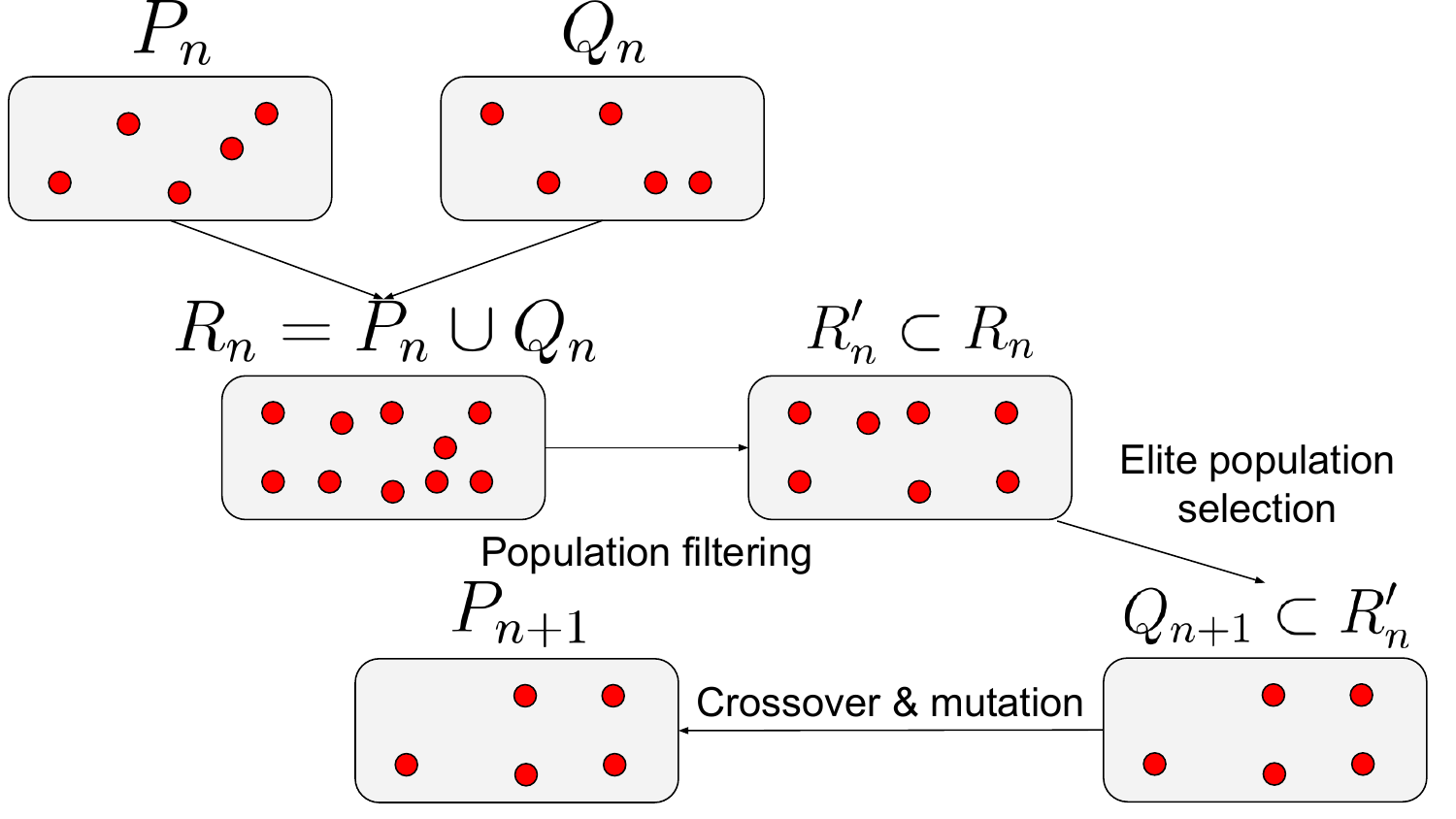}
  \caption{The relationship between the parent population, elite population, filtered population, next elite population, and the next population.}
  \label{fig:pqr-relations}
\end{figure}

Let $P_n$ be the parent generation, $Q_n$ be the elite population used to select the parent generation, and $R_n = P_n \cup Q_n$, as shown in Fig.~\ref{fig:pqr-relations}.
Let $j_*(i)$ be the structure with the smallest energy in the neighborhood of the structure $i$ defined as follows:
\begin{equation}
  \begin{aligned}
    j_*(i) = \argmin_{j \in \bigcup_{k=1}^n P_k} \{ E(j) \mid \| r(i), r(j)\|_1 < \delta \},
  \end{aligned}
\end{equation}
where $\| x \|_1 = \sum_{k=1}^d |x_i|$ is the $1$-norm, and $\delta > 0$ is the threshold of the $1$-norm of the composition.
Here, we use all the structures we have explored so far $\bigcup_{k=1}^n P_k$ as neighbors instead of $R_n$ to maintain the diversity of the elite population.
We define the distance to the convex hull of the structure $i$ as
\begin{align}
  E(i) - E(j_*(i)).
\end{align}
Let $n^*(i)$ be the generation in which the structure $j_*(i)$ was generated.
Let $E_{\max}$ and $E_{\min}$ be the maximum and minimum energies over all the structures explored so far defined as follows:
\begin{equation}
  \begin{aligned}
    E_{\max} = \max \{ E(i) \mid i \in \bigcup_{k=1}^n P_k \}, \\
    E_{\min} = \min \{ E(i) \mid i \in \bigcup_{k=1}^n P_k \}.
  \end{aligned}
\end{equation}
We later normalize the distance to the convex hull $E(i) - E(j_*(i))$ to be between $0$ and $1$ by dividing it by $E_{\max} - E_{\min}$.

The population filtering is a procedure to construct a subset $R'_n \subset R_n$.
For each element $i$ in $R_n$, we calculated the following quantity by considering the distance to the convex hull $E(i) - E(j_*(i))$ (niching) and the generation difference $n - n^*(i)$ (aging):
\begin{align}
    D(i) = \frac{E(i) - E(j_*(i))}{E_{\max} - E_{\min}} \times \alpha^{n - n^*(i)},
\end{align}
where $\alpha > 1$ is a tuning parameter.
The parameter $\alpha$ controls the importance of the generation difference.
The smaller $D(i)$ indicates that the structure is closer to the convex hull, and composition has been updated more recently.
We adopt only the structures $i$ that satisfy $D(i) < \epsilon$, where $\epsilon$ is a tuning parameter.
The parameter $\epsilon$ controls the threshold of the distance to the convex hull and the generation difference.
Finally, we define the following set $R'_n$ as the output of this step,
\begin{align}
    R'_n = \left\{ i \in R_n \mid D(i) < \epsilon \right\}.
\end{align}

\subsubsection{\label{sec:method-ga-elite-population-selector}Niching to maintain diversity across population}
Niching is performed to construct the elite population $Q_{n+1}$ from the filtered population $R'_n$.
The elite population is used to select the parent population $P_{n+1}$ for the next generation, as shown in Fig.~\ref{fig:pqr-relations}.
We can reduce computational costs and search for stable structures more efficiently by applying crossover and mutation only to promising structures.

While CHGA reported from binary-system benchmarks that explicit niching was unnecessary~\cite{Donaldson2024AGA}, our experiments on ternary and higher systems demonstrate that niching materially improves hull coverage and prevents collapse to a few stoichiometries (see Section~\ref{sec:results-and-discussion}).
Accordingly, we retain convex hull-based non-dominated sorting for elitist selection, but for tie-breaking within the same hull layer we adopt niching mechanisms inspired by NSGA-II/III~\cite{996017,6600851} and adapt them to the energy--composition geometry of CSP, thereby maintaining diversity across the population.
We observed that the NSGA-III-like niching procedure outperformed the NSGA-II-like one in our preliminary experiments, consistent with the general trend that NSGA-III is more effective in higher systems.

First, we define a procedure using the crowding distance similar to NSGA-II.
The definition of crowding distance is different because we consider a different problem from the usual multi-objective optimization problem.

Let $H^{(k,a)}$ be the set of $H^{k}$, the output of the population filtering with rank $k$, sorted by $r(i)_a$.
For each $a \in \{1, \ldots, M\}$ and structure $i$, let $r(i)_{a, \mathrm{left}}$ and $r(i)_{a, \mathrm{right}}$ be the ratio of the $a$-th element in the structures on both sides of $i$ in $H^{(k,a)}$, and let $\Delta_r(i, a)$ be the difference between them,
\begin{align}
  \Delta_r(i, a) = r(i)_{a, \mathrm{right}} - r(i)_{a, \mathrm{left}}.
\end{align}
Similarly, for the energy $E(i)$, let $H^{(k,E)}$ be the set of $H^{k}$ sorted by $E(i)$.
For each structure $i$, let $E(i)_{\mathrm{right}}$ and $E(i)_{\mathrm{left}}$ be energies of the structures on both sides of $i$ in $H^{(k,E)}$ and $\Delta_E(i)$ be the difference of them as follows:
\begin{align}
  \Delta_E(i) = E(i)_{\mathrm{right}} - E(i)_{\mathrm{left}}.
\end{align}
Note that the energy of each structure is min-max normalized for stability.

Using the $\Delta_r(i, a)$ and $\Delta_E(i)$ defined above, we define the crowding distance $c(i)$ for the structures $i$ in $H^{k}$ as follows:
\begin{align}
  c(i) = \Delta_E(i) \times \sum_{a=1}^M \Delta_r(i, a).
\end{align}
We can order structures with the same rank by selecting structures with small crowding distance $c(i)$.
The algorithm is summarized in Algorithm~\ref{alg:select_low_crowding}.

\begin{algorithm}[H]
  \caption{NSGA-II inspired niching}
  \label{alg:select_low_crowding}
  \begin{algorithmic}[1]
    \Require Population set $H^k$ with rank $k$, number of selected structures $\ell$
    \Ensure Set $\hat{H}^k$ containing $\ell$ structures with smallest $c(i)$

    \For{each $a \in \{1, \ldots, M\}$}
        \State Sort $H^k$ by $r(i)_a$ to create $H^{(k,a)}$
        \For{each structure $i \in H^k$}
            \State Identify neighboring structures in $H^{(k,a)}$
            \State Compute $r(i)_{a, \mathrm{left}}$ and $r(i)_{a, \mathrm{right}}$
            \State Compute $\Delta_r(i, a) = r(i)_{a, \mathrm{right}} - r(i)_{a, \mathrm{left}}$
        \EndFor
    \EndFor

    \State Sort $H^k$ by energy $E(i)$ to create $H^{(k,E)}$
    \For{each structure $i \in H^k$}
        \State Identify neighboring structures in $H^{(k,E)}$
        \State Compute $E(i)_{\mathrm{left}}$ and $E(i)_{\mathrm{right}}$
        \State Compute $\Delta_E(i) = E(i)_{\mathrm{right}} - E(i)_{\mathrm{left}}$
    \EndFor

    \For{each structure $i \in H^k$}
        \State Compute crowding distance:
        \[
          c(i) = \Delta_E(i) \times \sum_{a=1}^{M} \Delta_r(i, a)
        \]
    \EndFor

    \State Sort structures in $H^k$ by $c(i)$ in ascending order
    \State Select the first $\ell$ structures with the smallest $c(i)$ to form $\hat{H}^k$
    \State \Return $\hat{H}^k$
  \end{algorithmic}
\end{algorithm}

Next, we define the hyperplane-based method similar to NSGA-III.
To clarify the method, we need to define reference points and how to measure distances.
We adopt the following distance:
\begin{align}
  d(i,j) = |E(i) - E(j)| \times \sum_{a=1}^M |r(i)_a - r(j)_a|.
\end{align}
We constructed the reference points as follows.
For each composition axis $a \in \{1, \ldots, M\}$, select the structure with the maximum $r(i)_a$ in $H^{k}$ and call them $i_1, \ldots, i_M$.
Then, consider an $(M-1)$-dimensional simplex with $i_1, \ldots, i_M$ as vertices, and place points $r_1, \ldots, r_K$ uniformly at regular intervals on it, and use these as reference points.
Here, $K$ is determined by the number of points to be placed on the edge of the $(M-1)$-dimensional simplex, and if $k+1$ points are placed on the edge, $K = \binom{M+k-1}{k}$.
The algorithm is summarized in Algorithm~\ref{alg:reference_line_selection}.

\begin{algorithm}[H]
  \begin{algorithmic}[1]
    \Require Population set $H^k$ with rank $k$, number of selected structures $\ell$
    \Ensure Set $\hat{H}^k$ containing $\ell$ structures distributed along the reference lines

    \State \textbf{Normalize} each objective variable to be non-negative

    \State \textbf{Construct reference lines}:
    \State Define $K$ reference points by selecting structures $i_1, \ldots, i_M$ with maximum $r(i)_a$ for each axis $a \in \{1, \ldots, M\}$
    \State Place $K$ points $r_1, \ldots, r_K$ uniformly on an $(M-1)$-dimensional simplex

    \For{each structure $i \in H^k$}
        \For{each reference line $j$}
            \State Project $i$ onto reference line $j$
            \State Compute distance $d(i,j) = |E(i) - E(j)| \times \sum_{a=1}^{M} |r(i)_a - r(j)_a|$
        \EndFor
        \State Assign $i$ to the reference line $j^*$ with the smallest $d(i,j)$
        \State Add $i$ to the neighborhood set of $j^*$
    \EndFor

    \While{size of $\hat{H}^k$ is less than $\ell$}
        \State Select reference line $j^*$ with the smallest neighborhood set
        \State Select structure $i^*$ in the neighborhood set of $j^*$ closest to the reference line
        \State Add $i^*$ to $\hat{H}^k$ and remove it from the neighborhood set of $j^*$
    \EndWhile

    \State \Return $\hat{H}^k$
  \end{algorithmic}
  \caption{NSGA-III inspired niching}
  \label{alg:reference_line_selection}
\end{algorithm}

\subsubsection{Other implementation details}

\paragraph{Local relaxation}
For the next population, the formation energy is calculated after structural relaxation with PFP, as described in Section~\ref{sec:method-pfp}.

\paragraph{Asynchronous parallelization}
Note that we present the workflow in Fig.~\ref{fig:csp-workflow} with clarity, assuming a synchronous search per generation.
In practice, each trial can be evaluated asynchronously for high parallelism; once a structure returns, its parent generation is retrieved (by its generation index) and selection data structures are updated.
Such trial-level asynchrony is naturally supported by modern optimization frameworks such as \code{Optuna}~\cite{akiba2019optuna}.

\paragraph{\code{PyXtal}-based random structure generation}
We use symmetry-aware random structure generation by \code{PyXtal}~\cite{FREDERICKS2021107810} for the random structure generation in the sub-cells, thereby improving the search efficiency because stable crystal structures tend to have high symmetry~\cite{Pickard_2011}.
The random structure generation by \code{PyXtal} requires a composition and a space-group type as input.
We sampled the number of atoms in the unit cell uniformly up to the given maximum number of atoms in the unit cell and selected a composition from a multinomial distribution where each element has an equal probability.
We sampled a space-group type uniformly from all space-group types other than $P1$.
As some combinations of a composition and a space-group type are infeasible to generate a crystal structure, we retried the entire sampling process until a crystal structure is successfully generated.

\begin{table*}[htb]
  \centering
  \caption{Crossover and mutation methods used in the proposed GA-based CSP method.}
  \label{tab:crossover-and-mutation}
  \scalebox{0.8}{
  \begin{tabular}{lll}
    \hline \hline
    Method & Description & \begin{tabular}{@{}l}Selection \\probability\end{tabular} \\
    \hline
    Cut-and-splice crossover
      & \begin{tabular}{@{}l}
          Generate a child structure by cutting two parent structures \\
          at their unit cells along a crystal plane and joining them \cite{10.2138/rmg.2010.71.13,doi:10.1021/ar1001318}.
        \end{tabular}
      & $1/2$ \\
    Random composition mutation
      & Generate a child structure by randomly changing its composition \cite{GLASS2006713,10.2138/rmg.2010.71.13,LYAKHOV20131172}.
      & $1/12$ \\
    Strain mutation
      & \begin{tabular}{@{}l}
          Distort the unit cell by applying a random strain matrix \\
          to the basis vectors \cite{GLASS2006713,10.2138/rmg.2010.71.13,LYAKHOV20131172}.
        \end{tabular}
      & $1/12$ \\
    Rattle mutation
      & Perturb the atomic positions by a random displacement \cite{GLASS2006713,10.2138/rmg.2010.71.13,LYAKHOV20131172}.
      & $1/12$ \\
    Permutation mutation
      & Swap the atomic positions of two selected atoms \cite{GLASS2006713,10.2138/rmg.2010.71.13,LYAKHOV20131172}.
      & $1/12$ \\
    Random atom deletion mutation
      & \begin{tabular}{@{}l}
          Generate a child structure by randomly removing one atom from \\
          the parent structure.
        \end{tabular}
      & $1/12$ \\
    Random structure generation mutation
      & Generate a structure by the random structure generation \cite{Donaldson2024AGA}.
      & $1/12$ \\
    \hline \hline
  \end{tabular}
  }
\end{table*}

\paragraph{Variation operators}
Crossover and mutation are the processes of generating new structures $P_{n+1}$ from the elite population $Q_{n+1}$.
Crossover involves combining two parent structures to generate a child structure, and mutation involves changing the structure of a parent structure.
The crossover and mutations used in our proposed method are listed in Table~\ref{tab:crossover-and-mutation}.
Several methods are implemented in \code{ASE}~\cite{ase2017} and we modify some of the implementation \footnote{
  The modified crossover and mutation methods are released and available at \url{https://pypi.org/project/pfn-ase-extras/}.
}.

The cut-and-splice crossover is a method to generate a child structure by cutting two parent structures at their unit cells along a crystal plane and joining them \cite{PhysRevLett.75.288,PhysRevLett.108.126101}.
Because its implementation in ASE only allowed crossover between parent structures with the same composition, we modify it to allow crossover between parent structures with different compositions, which is already known in the field of CSP \cite{10.2138/rmg.2010.71.13,doi:10.1021/ar1001318}.

The random atom deletion mutation is a method to generate a child structure by randomly removing one atom from the parent structure.
This mutation is effective to change the composition of the structure and explore the composition space.

USPEX and CHGA uses random structure generation in half of the trials to maintain the diversity of the composition in the population~\cite{oganov2011modern,Donaldson2024AGA},
but our proposed method incorporates it as a standard mutation operation.
The random structure generation mutation is a method to generate a child structure by the random structure generation by \code{PyXtal}.
This mutation can effectively maintain the diversity of the composition in the population.

\section{\label{sec:results-and-discussion}Results and discussion}

In this section, we present the results of our proposed method for CSP using PFP and GA.
First, we conducted an ablation study of the population filtering in Section~\ref{sec:results-ablation-study} to demonstrate the effectiveness of the half of proposed method.
Secondly, we conducted an ablation study of the niching methods in Section~\ref{sec:results-ablation-study-niching} to demonstrate the effectiveness of the other half of proposed method.
Thirdly, we evaluated the performance of the proposed method by comparing it with existing methods in Section~\ref{sec:results-quantitative}.
We then demonstrated the effectiveness of the proposed method by performing CSP searches for binary and ternary systems with MP in Section~\ref{sec:results-binary-ternary}.
The results demonstrate that the proposed method can efficiently explore the search space and discover new crystal structure candidates that update the convex hull of MP.
For the energy evaluation, we used the energy corrections described in Section~\ref{sec:method-pfp-correction}.

\subsection{\label{sec:results-ablation-study}Ablation study of population filtering}

\begin{figure*}[!tb]
  \includegraphics[width=\textwidth]{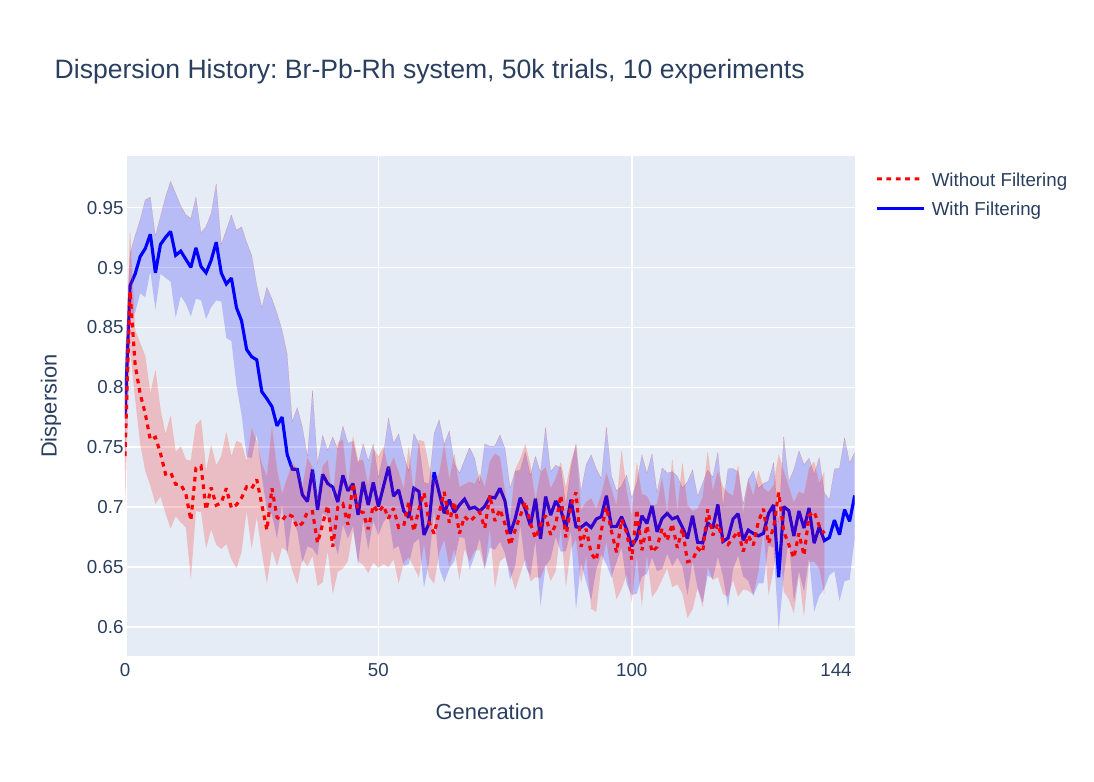}
  \caption{
    Compositional dispersion history of the Br--Pb--Rh system.
    The Average Nearest Neighbor Index (ANNI) is plotted against the generation number.
    The blue line indicates the case with population filtering, while the red line indicates the case without population filtering.
    The shaded area represents the standard deviation of ANNI values across 10 experiments.
    Lower ANNI values indicate greater compositional skewness.
  }
  \label{fig:dispersion-history}
\end{figure*}

\begin{figure*}[!tb]
  \includegraphics[width=\textwidth]{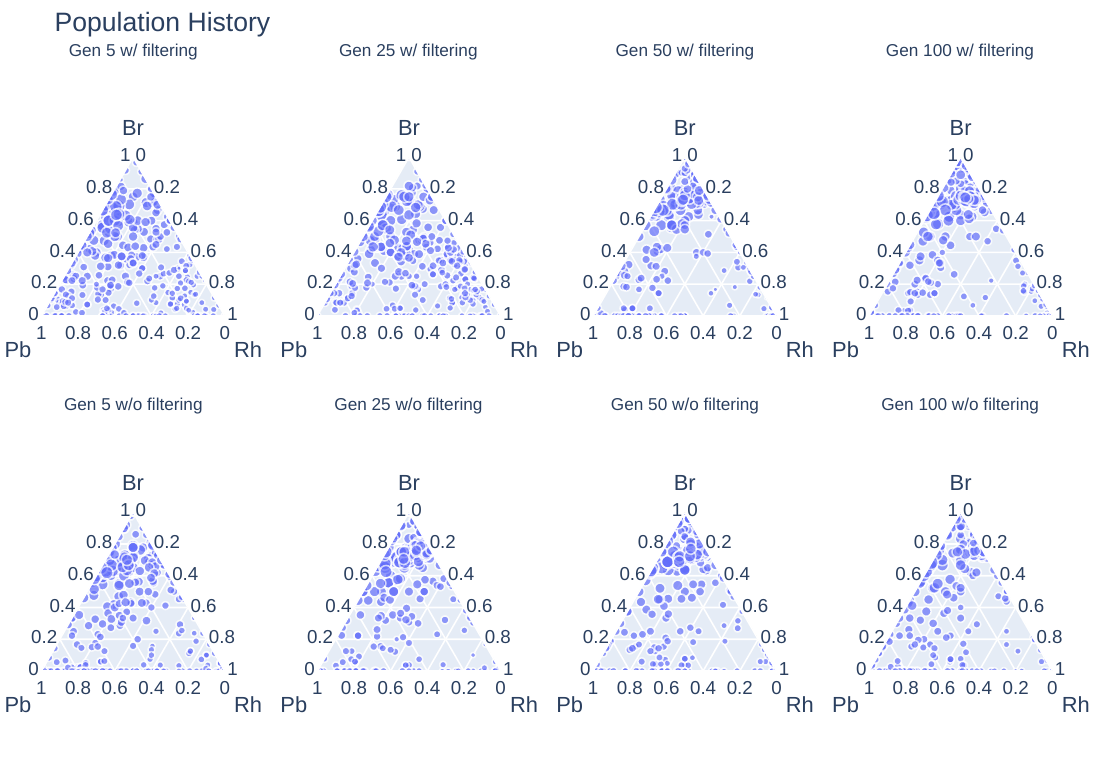}
  \caption{
    Compositional distribution of structures explored at each generation in the Br--Pb--Rh system.
    The upper panel shows the case with population filtering, while the lower panel shows the case without population filtering.
    From left to right, the generations are 5, 25, 50, and 100.
  }
  \label{fig:population-history}
\end{figure*}

In this section, we conducted an ablation study of the population filtering.
Population filtering is performed immediately prior to elitist selection in each generation to preserve the diversity of candidate compositions.
The ablation study compared two cases: one with population filtering and one without.
We demonstrated that implementing population filtering enables exploration of a wider range of structures across various compositions compared to its absence.

In this experiment, we examined the Br--Pb--Rh system with the following parameters: maximum number of atoms in the unit cell was 128, the population size was 250, the total number of trials was 50,000, $\alpha = 2.4$, and $\epsilon = 0.25$.
We repeated the search 10 times for each case to account for the stochastic nature of the CSP search.
The genetic operations and their selection probabilities are specified in Table~\ref{tab:crossover-and-mutation}.
We conducted comparisons under two conditions: with and without enabling population filtering as described in Section~\ref{sec:method-ga-population-filtering}.

To quantitatively assess how compositional dispersion evolves across generations, we employed the Average Nearest Neighbor Index (ANNI) \cite{clark1954distance} to evaluate the compositional homogeneity of structures explored in each generation.
Lower ANNI values indicate greater compositional skewness, which implies that the explored structures are concentrated in a narrower compositional range.
On the other hand, higher ANNI values indicate a more diverse distribution of compositions.
As shown in Fig.~\ref{fig:dispersion-history}, when population filtering is enabled, the ANNI values are significantly higher during early generations compared to when filtering is disabled, indicating greater compositional dispersion.
However, beginning approximately from the 40th generation onward, ANNI values become comparable regardless of whether population filtering is active or not, suggesting that compositional distribution becomes similarly skewed throughout both search phases.

Figure~\ref{fig:population-history} illustrates the distribution of compositions of structures explored at each generation for one search.
At generations 5 and 25, we observe that enabling population filtering results in a more broadly distributed range of structural compositions.
From generation 50 onward, even when population filtering is active, the distribution of compositions becomes narrower, exhibiting patterns similar to when population filtering is disabled.
Conversely, when population filtering is disabled, the explored structures show concentration within a narrow compositional range, and this range remains relatively stable across generations.
This demonstrates that population filtering enables exploration of structures across a wider portion of the search space.

\subsection{\label{sec:results-ablation-study-niching}Ablation study of niching methods}

\begin{figure*}[tb]
  \centering
  \includegraphics[width=\textwidth]{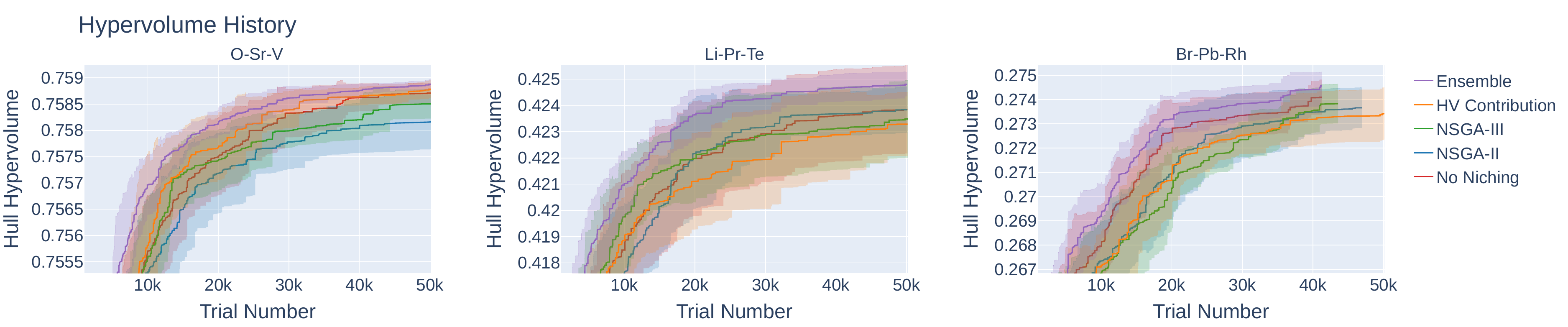}
  \caption{
    Comparison of hull volume history for different niching methods in (left) O--Sr--V, (middle) Br--Pb--Rh, and (right) Li--Pr--Te systems. The no-niching method (red), the proposed NSGA-II based method (blue), the proposed NSGA-III based method (green), the existing hypervolume contribution based method (orange), and those ensembled method (purple) are compared.
  }
  \label{fig:niching-ablation}
\end{figure*}

In this section, we present the results of an ablation study conducted to evaluate the impact of different niching methods on search performance.
Here, niching refers to the techniques employed during elite population selection in each generation, as described in Section~\ref{sec:method-ga-elite-population-selector}, to maintain diversity across population.

We compared the proposed NSGA-II and NSGA-III based niching methods with a no-niching method and an existing hypervolume contribution based niching method~\cite{Donaldson2024AGA}.
Additionally, considering that previous studies \cite{Donaldson2024AGA} suggest that niching has a minor impact on search performance, we also included an ensembled method that randomly selects one of the niching methods for each trial to enhance search performance.
To evaluate the influence of different element systems, we conducted experiments on three ternary systems: O--Sr--V, Br--Pb--Rh, and Li--Pr--Te.
The search conditions were the same as those in Section~\ref{sec:results-ablation-study}.

Figure~\ref{fig:niching-ablation} shows the result of search performance for each niching method.
For average performance, the best methods were HV Contribution, No Niching, and NSGA-II for O--Sr--V, Br--Pb--Rh, and Li--Pr--Te respectively. 
However, most methods fell within ±1 standard deviation, indicating that the performance differences were not particularly significant.
Although performance differences vary by element system, the ensembled method generally demonstrates the highest search performance.
This suggests that different niching methods explore varying diversities of structures, and by ensembling them, a broader range of structures can be explored, leading to improved search performance.

\subsection{\label{sec:results-quantitative}Quantitative comparison with existing methods and MP}

\begin{figure*}[!tb]
  \includegraphics[width=\textwidth]{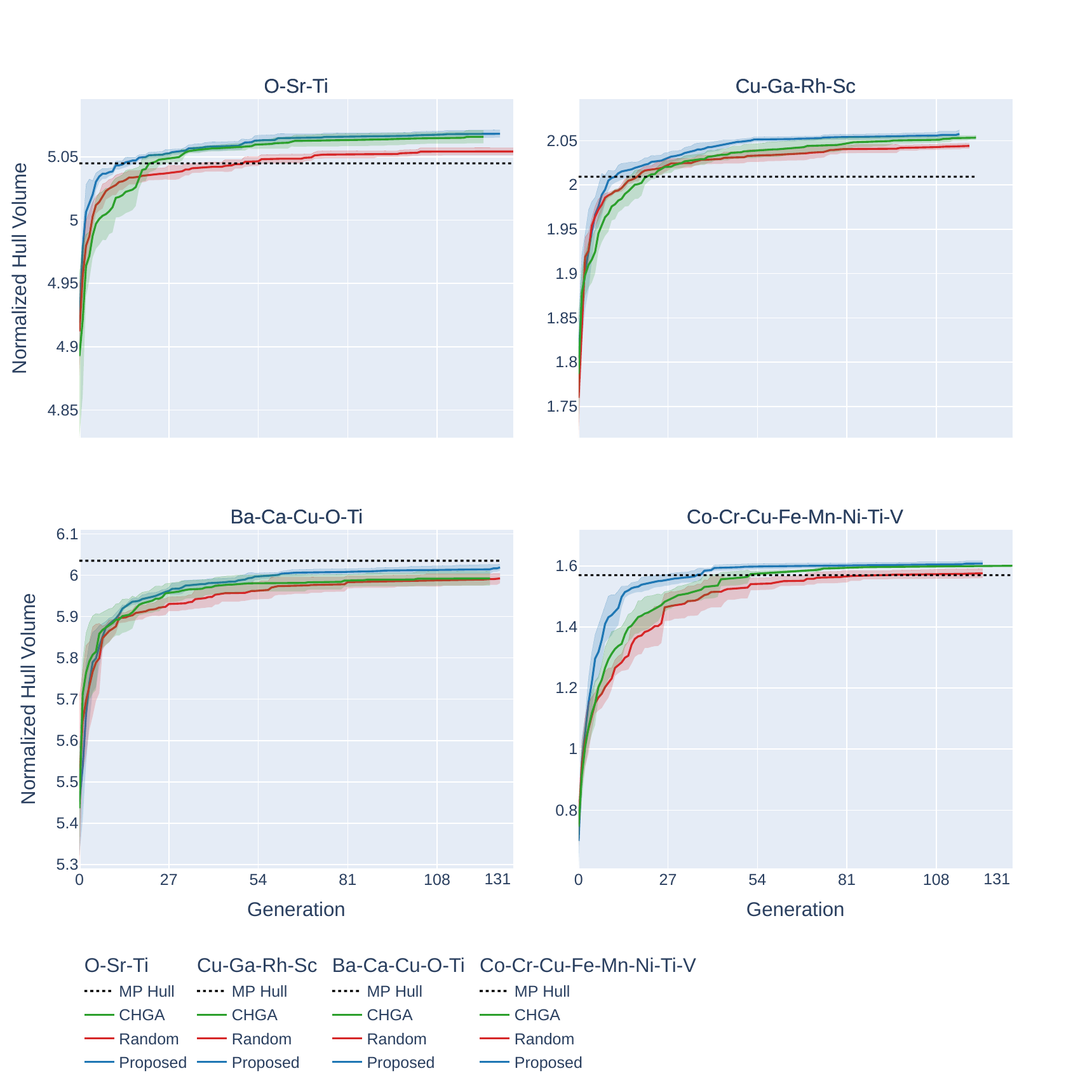}
  \caption{
    Comparison between random search (red), CHGA (green) and the proposed method (blue) by the hull volume history.
    The shaded area represents the standard deviation of hull volumes across five experiments.
    The black dashed line indicates the normalized volumes for the convex hull constructed from MP structures.
  }
  \label{fig:hull-volume}
\end{figure*}

We quantitatively evaluated the proposed method compared with the symmetry-aware random search implemented in \code{PyXtal}~\cite{FREDERICKS2021107810}, the traditional GA-based method implemented in \code{ASE}~\cite{ase2017}, and CHGA~\cite{Donaldson2024AGA}.
In this experiment, we fixed the following inputs and parameters: the maximum number of atoms in the unit cell was 128, the population size was 250, the total number of trials was 50,000, and $\alpha = 2.4$, $\epsilon = 0.25$.
The genetic operations and their selection probabilities are listed in Table~\ref{tab:crossover-and-mutation}.
We chose four systems ranging from ternary to octonary systems: O--Sr--Ti and Ba--Ca--Cu--O--Tl as well known oxide systems~\cite{MatterGen2025}, and Cu--Ga--Rh--Sc and Co--Cr--Cu--Fe--Mn--Ni--Ti--V as random alloy systems.
The Hubbard $U$ correction for Cu is applied for our method, while it is not applied for MP, which may affect the comparison in the Ba--Ca--Cu--O--Tl system, as mentioned in Section~\ref{sec:method-pfp-pfp}.

We evaluated each search by the volume of the convex hull, which can assess both the depth of the convex hull and the diversity of the included structures~\cite{Donaldson2024AGA}.
The hull volume value of the $M$-element system is normalized by the unit hull volume $V_{{\rm unit}, M} = 1\,\mathrm{eV} / M!$, which corresponds to the volume of the convex hull with only one crystal structure on the hull with formation energy of $\mathord{-}1$~eV/atom.
We performed the search by five times for each system because the CSP search involves randomness.
Additionally, we compared the hull volume with that of MP.
We performed structure optimization by PFP for MP structures within 0.2 eV/atom in MP convex hull to construct the MP convex hull.
The absolute value of the hull volume depends on the element system and cannot be compared across different systems.

Figure~\ref{fig:hull-volume} illustrates how the hull volume changed as the search progressed for each system, indicating that the proposed method outperformed random search for all the element systems.
The final mean hull volumes for each system are listed in Table~\ref{tab:hull-reproducibility}.
Here, we denote the mean hull volume of the proposed method as $V_{\mathrm{proposed}}$, and that of the random search as $V_{\mathrm{random}}$.
The improvements of final mean hull volumes of the proposed method compared with random search are 0.014, 0.014, 0.027, and 0.024 for O--Sr--Ti, Cu--Ga--Rh--Sc, Ba--Ca--Cu--O--Tl, and Co--Cr--Cu--Fe--Mn--Ni--Ti--V respectively.
These improvements correspond to updates of the convex hull by 14 to 27 meV/atom if only one entry exists on the convex hull.
We mention that the conversion of a hull volume to a formation energy update depends on the shape of the convex hull.

The proposed method also outperformed CHGA for all the element systems, as shown in Fig.~\ref{fig:hull-volume}.
The difference in performance varies qualitatively depending on the element system.
For the ternary system of O--Sr--Ti, both methods achieved similar hull volumes, but the proposed method outperformed CHGA in the early generations.
For the quaternary system of Cu--Ga--Rh--Sc, the proposed method consistently outperformed CHGA throughout the search, especially in the early generations.
For the quinary system of Ba--Ca--Cu--O--Tl, both methods achieved similar hull volumes at the early generations, but the proposed method outperformed CHGA in the later generations.
For the octonary system of Co--Cr--Cu--Fe--Mn--Ni--Ti--V, the proposed method significantly outperformed CHGA at early generations, while both methods achieved similar hull volumes in the later generations.
These results indicate that the proposed method can efficiently explore the search space fastter than CHGA, especially in multicomponent systems.
The final mean hull volumes for each system are listed in Table~\ref{tab:hull-reproducibility}.
We denote the mean hull volume of CHGA as $V_{\mathrm{chga}}$.
The improvements of final mean hull volumes of the proposed method compared with CHGA are 0.002, 0.005, 0.014, and 0.0 for O--Sr--Ti, Cu--Ga--Rh--Sc, Ba--Ca--Cu--O--Tl, and Co--Cr--Cu--Fe--Mn--Ni--Ti--V respectively.

The updated hull volume by the proposed method compared with that of MP is 0.023, 0.049, -0.016, and 0.031 for O--Sr--Ti, Cu--Ga--Rh--Sc, Ba--Ca--Cu--O--Tl, and Co--Cr--Cu--Fe--Mn--Ni--Ti--V, respectively; this corresponds to 0.5, 2.4, -0.3, and 2.0\% of MP hull volume, respectively.
The proposed method achieved larger convex hull volumes than MP for systems except for the Ba--Ca--Cu--O--Tl.

Figure~\ref{fig:hull-volume} also depicts the efficiency of the CSP search with the proposed method.
The number of generations required to reach 99\% of each final hull volume for O--Sr--Ti, Cu--Ga--Rh--Sc, Ba--Ca--Cu--O--Tl, and Co--Cr--Cu--Fe--Mn--Ni--Ti--V were 4, 34, 26, and 641, respectively, for the proposed method, 5, 28, 33, and 72, respectively, for a random search, and 11, 41, 20, 71, respectively, for CHGA; this indicates that the proposed method can efficiently explore the search space, especially in multicomponent systems.

\begin{table*}[!tb]
  \centering
  \caption{
    Mean normalized hull volume and average number of MP hull structures compared with our CSP hull.
    The second columns show the mean normalized hull volume of the proposed method, while the last three columns show that for the random search, CHGA, and the normalized hull volume of MP, respectively.
    The third to fifth columns show the average number of MP structures below, around, and above the CSP hull by the proposed method, respectively (See the main text for the detailed definition).
    The sixth column shows the average ratio of the number of MP structures around or above the CSP hull.
  }
  \label{tab:hull-reproducibility}
  \begin{tabular}{cccccccccc}
    \hline \hline
    System & $V_{\mathrm{proposed}}$  & Below hull & Around hull & Above hull & Reproducibility & $V_{\mathrm{random}}$ & $V_{\mathrm{chga}}$ &$V_{\mathrm{MP}}$ \\
    \hline
      O--Sr--Ti & 5.068 & 0 & 12.0 & 6.0 & 1.00 & 5.054 & 5.066 & 5.045\\
      Cu--Ga--Rh--Sc & 2.058 & 3.0 & 17.2 & 4.8 & 0.88 & 2.044 & 2.053 & 2.009\\
      Ba--Ca--Cu--O--Tl & 6.019 & 10.0 & 29.8 & 11.2 & 0.80 & 5.992 & 5.992 & 6.035\\
      \begin{tabular}{c}
        Co--Cr--Cu--Fe--\\
        Mn--Ni--Ti--V
      \end{tabular}
    & 1.601 & 8.4 & 38.4 & 10.2 & 0.85 & 1.577 & 1.601 & 1.570\\
    \hline \hline
  \end{tabular}
\end{table*}

\begin{figure*}[!tb]
  \includegraphics[width=\textwidth]{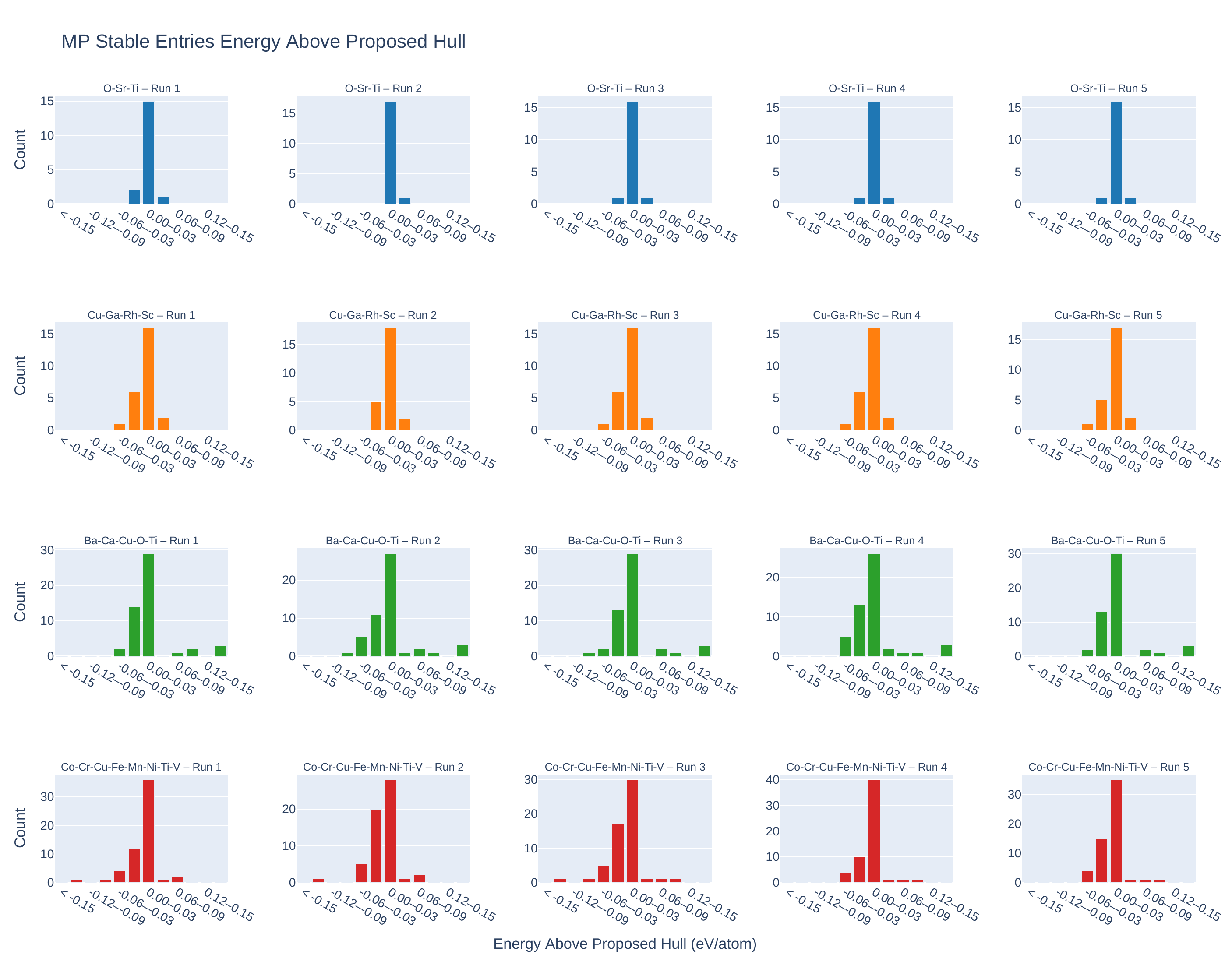}
  \caption{
    Distribution of the energy above the hull of the crystal structures of MP.
    The convex hull is reconstructed using the CSP results obtained using the method proposed in Section~\ref{sec:results-quantitative}.
    The results of the five searches are shown for each column.
  }
  \label{fig:e_above_hull_distribution}
\end{figure*}

We validated the effectiveness of our proposed method for actual applications by comparing the convex hulls of CSP and that of MP.
Table~\ref{tab:hull-reproducibility} shows the average number of MP structures below, around, and above the convex hull of CSP searches.
The structures around the hull are defined as those with energy above hull within 0.01 eV/atom.
The structures below and above the hull are those with energy above hull less than -0.01 eV/atom and greater than 0.01 eV/atom, respectively.
The average ratio of the number of MP structures around and above the convex hull to the total number of structures is 1.00, 0.88, 0.80, and 0.85 for O--Sr--Ti, Cu--Ga--Rh--Sc, Ba--Ca--Cu--O--Tl, and Co--Cr--Cu--Fe--Mn--Ni--Ti--V, respectively.
Although the trends vary by element systems, around 80\% of MP structures are reproduced by CSP.
In particular, all MP structures are reproduced in the O--Sr--Ti system.

Figure~\ref{fig:e_above_hull_distribution} illustrates an energy above hull distribution of MP structures with the CSP hull by the proposed method.
The results of the five searches are shown in each column for each row of element system.
As demonstrated, many MP structures are concentrated around the CSP hull.
In this figure, MP structures with negative energy above hull indicate that the CSP search by the proposed method missed more stable structures in MP.
For the octonary system of Co--Cr--Cu--Fe--Mn--Ni--Ti--V, two crystal structures shows energy above hull less than -0.1 eV/atom, Cr$_2$Ti (mp-1425) and Mn$_2$Ti (mp-1949).
These structures are not found in all the five CSP searches, while random search was able to find Cr$_2$Ti twice and Mn$_2$Ti once in five searches in the potential energy landscape of PFP.

\subsection{\label{sec:results-binary-ternary}CSP for binary and ternary systems}

\begin{figure*}[!tb]
  \centering
  \includegraphics[width=\textwidth]{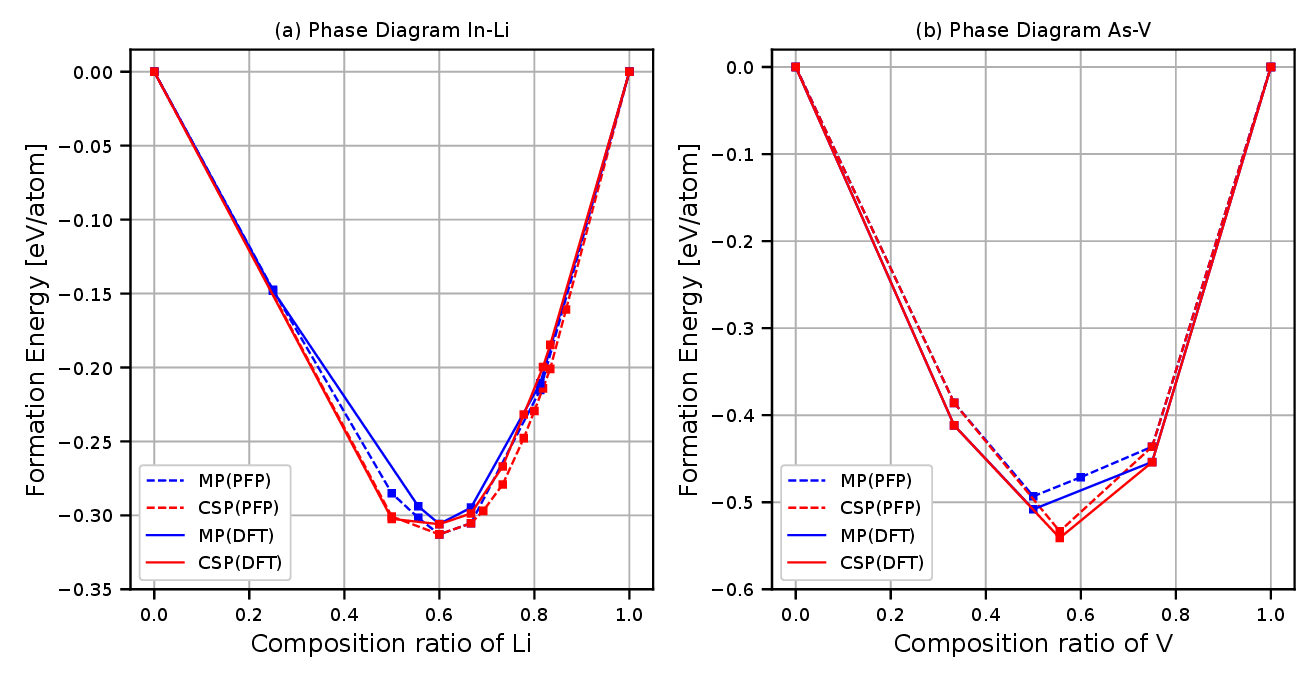}
  \caption{
    Binary phase diagrams of (a) In--Li and (b) As--V systems.
    Red and blue lines show the convex hulls of the formation energy of our CSP results and MP, respectively.
    The energy is evaluated by VASP for solid lines, while PFP is used for dashed lines.
  }
  \label{fig:phasediagram-binary}
\end{figure*}

\begin{figure*}[!tb]
  \centering
  \includegraphics[width=\textwidth]{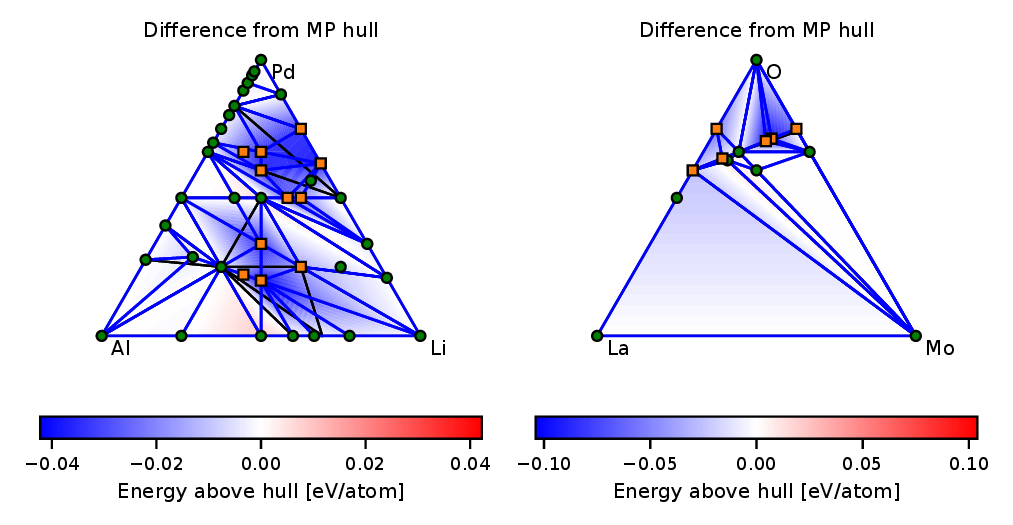}
  \caption{
    Phase diagrams of ternary systems.
    Black and blue lines show the convex hulls of MP and our CSP results, respectively.
    Here, formation energies were evaluated with DFT calculations.
    Blue and red hatchings indicate the differences between the MP and our CSP convex hulls.
    The blue region shows the CSP results updating the MP convex hull.
    Markers display the simplices of the CSP convex hull, while orange squares indicate those with energy updates greater than 10 meV/atom.
  }
  \label{fig:phasediagram-ternary}
\end{figure*}

We demonstrated the capability of our CSP method using PFP to discover new crystal structure candidates for binary and ternary systems.
The search conditions were the same as those in Section~\ref{sec:results-quantitative}; however, the structures from MP are added to the initial population of the search.
The crystal structures obtained from the search were re-evaluated using DFT calculations to verify their validity.
We report the results for systems In--Li, As--V, Al--Li--Pd, and La--Mo--O shown in Figs.~\ref{fig:phasediagram-binary} and \ref{fig:phasediagram-ternary}.

Figure~\ref{fig:phasediagram-binary} depicts the convex hulls by our CSP methods and MP, respectively.
DFT calculations with VASP were performed for these structures using the same settings as in Section~\ref{sec:method-pfp-pfp}.
By comparing convex hulls, the structures identified through our CSP method are confirmed to be also stable in DFT calculations.
These findings suggest that the CSP approach using PFP is highly effective in discovering new crystal structure candidates.

Figure~\ref{fig:phasediagram-ternary} shows comparisons of MP and CSP phase diagrams for ternary systems.
Numerous promising new crystal candidates have also been discovered in the ternary systems.
Our CSP methodology demonstrates the capability to explore and update the entire phase diagram comprehensively.

\begin{figure*}[!tb]
  \centering
    \includegraphics[width=\textwidth]{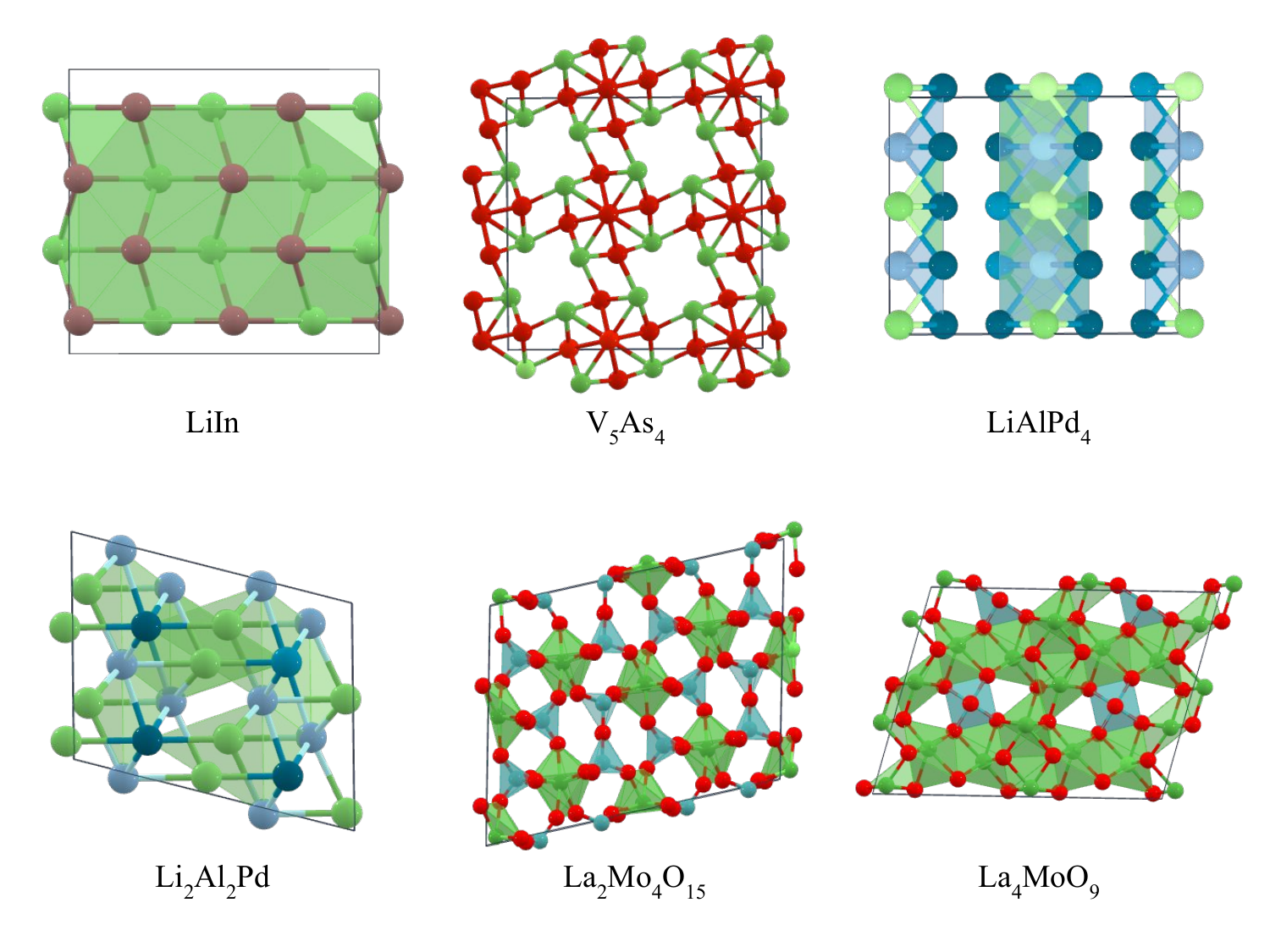}
  \caption{Crystal structures that update the convex hull.}
  \label{fig:hull-crystal-structures}
\end{figure*}

Crystal structures that update the convex hull of MP in the systems shown in Figs.~\ref{fig:phasediagram-binary} and \ref{fig:phasediagram-ternary} are listed in Table~\ref{tab:hull-crystal-structures} \footnote{
  These crystal structures in xyz formats are available at Supplemental Materials.
}.
The space-group type was identified using \code{spglib} with \code{symprec = 0.01} and \code{angle\_tolerance = 5}~\cite{spglib}.
Additionally, we matched these crystal structures with AFLOW prototypes~\cite{MEHL2017S1,HICKS2019S1,HICKS2021110450,ECKERT2024112988} by assigning the AFLOW label using \code{aviary.wren}~\cite{goodall_2022_rapid}.
Several crystal structures were not found as prototype structures in the AFLOW database, indicating that our CSP method can discover novel crystal structures.
Some of the listed crystal structures are visualized in Fig.~\ref{fig:hull-crystal-structures}.

\begin{table*}[!tb]
  \centering
  \caption{
    List of crystal structures that update the convex hull.
    The fourth column shows the matched AFLOW prototype if it exists.
  }
  \label{tab:hull-crystal-structures}
  \begin{tabular}{ccccc}
    \hline \hline
    Formula & Space-group type & \begin{tabular}{@{}c} Hull energy update \\ $[\mathrm{meV/atom}]$ \end{tabular} & AFLOW prototype \\
    \hline
    In$_{2}$Li$_{2}$ & $Pmma$ & $\mathord{-}35$ & AB\_oP4\_51\_e\_f-001 \\
    In$_{2}$Li$_{7}$ & $P\overline{1}$ & $\mathord{-}1$ &  \\
    In$_{4}$Li$_{11}$ & $P\overline{1}$ & $\mathord{-}10$ &  \\
    In$_{4}$Li$_{9}$ & $C2/m$ & $\mathord{-}5$ &  \\
    As$_{4}$V$_{5}$ & $I4/m$ & $\mathord{-}45$ & A4B5\_tI18\_87\_h\_ah-001 \\
    Al$_{2}$Li$_{2}$Pd & $C2/m$ & $\mathord{-}42$ &  \\
    Al$_{2}$Li$_{2}$Pd$_{2}$ & $P\overline{3}m1$ & $\mathord{-}34$ &  \\
    Al$_{2}$Li$_{2}$Pd$_{6}$ & $P2_{1}/m$ & $\mathord{-}33$ &  \\
    Al$_{2}$Li$_{4}$Pd$_{2}$ & $P4_{2}/mmc$ & $\mathord{-}14$ &  \\
    Al$_{2}$Li$_{4}$Pd$_{6}$ & $P2_{1}/m$ & $\mathord{-}25$ &  \\
    Al$_{2}$Li$_{6}$Pd$_{8}$ & $Im\overline{3}m$ & $\mathord{-}19$ &  \\
    Al$_{2}$Li$_{7}$ & $P\overline{1}$ & $\mathord{-}4$ &  \\
    Al$_{2}$LiPd$_{6}$ & $P\overline{1}$ & $\mathord{-}18$ &  \\
    Al$_{3}$Pd$_{7}$ & $C2/m$ & $\mathord{-}4$ &  \\
    Al$_{4}$Li$_{3}$Pd$_{2}$ & $P\overline{3}m1$ & $\mathord{-}22$ &  \\
    Al$_{4}$LiPd$_{2}$ & $P\overline{1}$ & $\mathord{-}8$ &  \\
    AlLi$_{5}$Pd$_{2}$ & $P\overline{4}m2$ & $\mathord{-}4$ &  \\
    AlLiPd$_{4}$ & $Cmmm$ & $\mathord{-}33$ &  \\
    AlPd$_{11}$ & $Cmmm$ & $\mathord{-}3$ &  \\
    AlPd$_{17}$ & $P4/mmm$ & $\mathord{-}1$ &  \\
    Li$_{33}$Pd$_{55}$ & $P\overline{1}$ & $\mathord{-}36$ &  \\
    LiPd$_{3}$ & $Cmmm$ & $\mathord{-}33$ & AB3\_oC8\_65\_a\_bf-001 \\
    La$_{2}$O$_{6}$ & $C2/m$ & $\mathord{-}58$ &  \\
    La$_{2}$Mo$_{2}$O$_{6}$ & $R3c$ & $\mathord{-}1$ &  \\
    La$_{2}$Mo$_{4}$O$_{15}$ & $P\overline{1}$ & $\mathord{-}100$ &  \\
    La$_{4}$Mo$_{6}$O$_{24}$ & $P\overline{1}$ & $\mathord{-}105$ &  \\
    La$_{4}$MoO$_{9}$ & $C2$ & $\mathord{-}12$ &  \\
    Mo$_{3}$O$_{9}$ & $Cm$ & $\mathord{-}60$ &  \\
    \hline \hline
  \end{tabular}
\end{table*}

\section{\label{sec:conclusion}Conclusion}

We have presented the efficient GA-based CSP method using our developed universal NNP, PFP.
Our method comprehensively explores the entire composition space and successfully discovers numerous stable crystal structures.
The proposed method even identified unregistered stable crystal structures in MP, and subsequent DFT calculations validated their stability.
Given that CSP requires precise formation energy accuracy to distinguish various polymorphs, these results also demonstrate the validity of PFP for a wide range of crystal structures and element combinations.
To efficiently search the entire composition space, we have proposed the novel elitist selection method inspired from multi-objective optimization techniques, which considers both formation energies and compositions.
Since the proposed GA-based CSP method is agnostic to how to generate initial populations, it would be fruitful in future works to combine it with other structure generation methods to tackle the vast search spaces in CSP, such as generative models for crystal structures.
Our present method and results show significant promise for accelerating materials discovery.

\begin{acknowledgments}
We would like to thank our colleagues from Preferred Networks, Inc. for helpful discussions and support, including Shuhei Watanabe.
PFP v6 was developed using the National Institute of Advanced Industrial Science and Technology's AI Bridging Cloud Infrastructure (ABCI) in addition to in-house supercomputers in Preferred Networks, Inc.
\end{acknowledgments}

\section*{Competing interests}

T.S, H.I, K.N, K.S, C.S, and S.T are employees of Preferred Networks, Inc.
Matlantis Corp., a joint venture among Preferred Networks Inc., ENEOS Corporation, and Mitsubishi Corporation, offers Matlantis, a software-as-a-service platform that incorporates a product based on this research.

\bibliography{references}

%merlin.mbs apsrev4-1.bst 2010-07-25 4.21a (PWD, AO, DPC) hacked
%Control: key (0)
%Control: author (8) initials jnrlst
%Control: editor formatted (1) identically to author
%Control: production of article title (-1) disabled
%Control: page (0) single
%Control: year (1) truncated
%Control: production of eprint (0) enabled
\begin{thebibliography}{87}%
\makeatletter
\providecommand \@ifxundefined [1]{%
 \@ifx{#1\undefined}
}%
\providecommand \@ifnum [1]{%
 \ifnum #1\expandafter \@firstoftwo
 \else \expandafter \@secondoftwo
 \fi
}%
\providecommand \@ifx [1]{%
 \ifx #1\expandafter \@firstoftwo
 \else \expandafter \@secondoftwo
 \fi
}%
\providecommand \natexlab [1]{#1}%
\providecommand \enquote  [1]{``#1''}%
\providecommand \bibnamefont  [1]{#1}%
\providecommand \bibfnamefont [1]{#1}%
\providecommand \citenamefont [1]{#1}%
\providecommand \href@noop [0]{\@secondoftwo}%
\providecommand \href [0]{\begingroup \@sanitize@url \@href}%
\providecommand \@href[1]{\@@startlink{#1}\@@href}%
\providecommand \@@href[1]{\endgroup#1\@@endlink}%
\providecommand \@sanitize@url [0]{\catcode `\\12\catcode `\$12\catcode
  `\&12\catcode `\#12\catcode `\^12\catcode `\_12\catcode `\%12\relax}%
\providecommand \@@startlink[1]{}%
\providecommand \@@endlink[0]{}%
\providecommand \url  [0]{\begingroup\@sanitize@url \@url }%
\providecommand \@url [1]{\endgroup\@href {#1}{\urlprefix }}%
\providecommand \urlprefix  [0]{URL }%
\providecommand \Eprint [0]{\href }%
\providecommand \doibase [0]{http://dx.doi.org/}%
\providecommand \selectlanguage [0]{\@gobble}%
\providecommand \bibinfo  [0]{\@secondoftwo}%
\providecommand \bibfield  [0]{\@secondoftwo}%
\providecommand \translation [1]{[#1]}%
\providecommand \BibitemOpen [0]{}%
\providecommand \bibitemStop [0]{}%
\providecommand \bibitemNoStop [0]{.\EOS\space}%
\providecommand \EOS [0]{\spacefactor3000\relax}%
\providecommand \BibitemShut  [1]{\csname bibitem#1\endcsname}%
\let\auto@bib@innerbib\@empty
%</preamble>
\bibitem [{\citenamefont {Hafner}(2008)}]{https://doi.org/10.1002/jcc.21057}%
  \BibitemOpen
  \bibfield  {author} {\bibinfo {author} {\bibfnamefont {J.}~\bibnamefont
  {Hafner}},\ }\href {\doibase https://doi.org/10.1002/jcc.21057} {\bibfield
  {journal} {\bibinfo  {journal} {J. Comput. Chem.}\ }\textbf {\bibinfo
  {volume} {29}},\ \bibinfo {pages} {2044} (\bibinfo {year}
  {2008})}\BibitemShut {NoStop}%
\bibitem [{\citenamefont {Oganov}(2011)}]{oganov2011modern}%
  \BibitemOpen
  \bibfield  {author} {\bibinfo {author} {\bibfnamefont {A.~R.}\ \bibnamefont
  {Oganov}},\ }\href {\doibase 10.1002/9783527632831} {\emph {\bibinfo {title}
  {Modern methods of crystal structure prediction}}}\ (\bibinfo  {publisher}
  {John Wiley \& Sons},\ \bibinfo {year} {2011})\BibitemShut {NoStop}%
\bibitem [{\citenamefont {Oganov}\ \emph {et~al.}(2019)\citenamefont {Oganov},
  \citenamefont {Pickard}, \citenamefont {Zhu},\ and\ \citenamefont
  {Needs}}]{Oganov2019}%
  \BibitemOpen
  \bibfield  {author} {\bibinfo {author} {\bibfnamefont {A.~R.}\ \bibnamefont
  {Oganov}}, \bibinfo {author} {\bibfnamefont {C.~J.}\ \bibnamefont {Pickard}},
  \bibinfo {author} {\bibfnamefont {Q.}~\bibnamefont {Zhu}}, \ and\ \bibinfo
  {author} {\bibfnamefont {R.~J.}\ \bibnamefont {Needs}},\ }\href {\doibase
  10.1038/s41578-019-0101-8} {\bibfield  {journal} {\bibinfo  {journal} {Nat.
  Rev. Mater.}\ }\textbf {\bibinfo {volume} {4}},\ \bibinfo {pages} {331}
  (\bibinfo {year} {2019})}\BibitemShut {NoStop}%
\bibitem [{\citenamefont {Wales}(2004)}]{wales_2004}%
  \BibitemOpen
  \bibfield  {author} {\bibinfo {author} {\bibfnamefont {D.}~\bibnamefont
  {Wales}},\ }\href {\doibase 10.1017/CBO9780511721724} {\emph {\bibinfo
  {title} {Energy Landscapes: Applications to Clusters, Biomolecules and
  Glasses}}}\ (\bibinfo  {publisher} {Cambridge University Press},\ \bibinfo
  {year} {2004})\BibitemShut {NoStop}%
\bibitem [{\citenamefont {Jain}\ \emph {et~al.}(2016)\citenamefont {Jain},
  \citenamefont {Shin},\ and\ \citenamefont {Persson}}]{Jain2016}%
  \BibitemOpen
  \bibfield  {author} {\bibinfo {author} {\bibfnamefont {A.}~\bibnamefont
  {Jain}}, \bibinfo {author} {\bibfnamefont {Y.}~\bibnamefont {Shin}}, \ and\
  \bibinfo {author} {\bibfnamefont {K.~A.}\ \bibnamefont {Persson}},\ }\href
  {\doibase 10.1038/natrevmats.2015.4} {\bibfield  {journal} {\bibinfo
  {journal} {Nat. Rev. Mater.}\ }\textbf {\bibinfo {volume} {1}},\ \bibinfo
  {pages} {15004} (\bibinfo {year} {2016})}\BibitemShut {NoStop}%
\bibitem [{\citenamefont {Hautier}\ \emph {et~al.}(2011)\citenamefont
  {Hautier}, \citenamefont {Fischer}, \citenamefont {Ehrlacher}, \citenamefont
  {Jain},\ and\ \citenamefont {Ceder}}]{doi:10.1021/ic102031h}%
  \BibitemOpen
  \bibfield  {author} {\bibinfo {author} {\bibfnamefont {G.}~\bibnamefont
  {Hautier}}, \bibinfo {author} {\bibfnamefont {C.}~\bibnamefont {Fischer}},
  \bibinfo {author} {\bibfnamefont {V.}~\bibnamefont {Ehrlacher}}, \bibinfo
  {author} {\bibfnamefont {A.}~\bibnamefont {Jain}}, \ and\ \bibinfo {author}
  {\bibfnamefont {G.}~\bibnamefont {Ceder}},\ }\href {\doibase
  10.1021/ic102031h} {\bibfield  {journal} {\bibinfo  {journal} {Inorg. Chem.}\
  }\textbf {\bibinfo {volume} {50}},\ \bibinfo {pages} {656} (\bibinfo {year}
  {2011})}\BibitemShut {NoStop}%
\bibitem [{\citenamefont {Pickard}\ and\ \citenamefont
  {Needs}(2011)}]{Pickard_2011}%
  \BibitemOpen
  \bibfield  {author} {\bibinfo {author} {\bibfnamefont {C.~J.}\ \bibnamefont
  {Pickard}}\ and\ \bibinfo {author} {\bibfnamefont {R.~J.}\ \bibnamefont
  {Needs}},\ }\href {\doibase 10.1088/0953-8984/23/5/053201} {\bibfield
  {journal} {\bibinfo  {journal} {J. Phys.: Condens. Matter.}\ }\textbf
  {\bibinfo {volume} {23}},\ \bibinfo {pages} {053201} (\bibinfo {year}
  {2011})}\BibitemShut {NoStop}%
\bibitem [{\citenamefont {Wales}\ and\ \citenamefont
  {Doye}(1997)}]{doi:10.1021/jp970984n}%
  \BibitemOpen
  \bibfield  {author} {\bibinfo {author} {\bibfnamefont {D.~J.}\ \bibnamefont
  {Wales}}\ and\ \bibinfo {author} {\bibfnamefont {J.~P.~K.}\ \bibnamefont
  {Doye}},\ }\href {\doibase 10.1021/jp970984n} {\bibfield  {journal} {\bibinfo
   {journal} {J. Phys. Chem. A}\ }\textbf {\bibinfo {volume} {101}},\ \bibinfo
  {pages} {5111} (\bibinfo {year} {1997})}\BibitemShut {NoStop}%
\bibitem [{\citenamefont {Amsler}\ and\ \citenamefont
  {Goedecker}(2010)}]{doi:10.1063/1.3512900}%
  \BibitemOpen
  \bibfield  {author} {\bibinfo {author} {\bibfnamefont {M.}~\bibnamefont
  {Amsler}}\ and\ \bibinfo {author} {\bibfnamefont {S.}~\bibnamefont
  {Goedecker}},\ }\href {\doibase 10.1063/1.3512900} {\bibfield  {journal}
  {\bibinfo  {journal} {J. Chem. Phys.}\ }\textbf {\bibinfo {volume} {133}},\
  \bibinfo {pages} {224104} (\bibinfo {year} {2010})}\BibitemShut {NoStop}%
\bibitem [{\citenamefont {Krummenacher}\ \emph {et~al.}(2024)\citenamefont
  {Krummenacher}, \citenamefont {Gubler}, \citenamefont {Finkler},
  \citenamefont {Huber}, \citenamefont {Sommer-J\"{o}rgensen},\ and\
  \citenamefont {Goedecker}}]{KRUMMENACHER2024101632}%
  \BibitemOpen
  \bibfield  {author} {\bibinfo {author} {\bibfnamefont {M.}~\bibnamefont
  {Krummenacher}}, \bibinfo {author} {\bibfnamefont {M.}~\bibnamefont
  {Gubler}}, \bibinfo {author} {\bibfnamefont {J.~A.}\ \bibnamefont {Finkler}},
  \bibinfo {author} {\bibfnamefont {H.}~\bibnamefont {Huber}}, \bibinfo
  {author} {\bibfnamefont {M.}~\bibnamefont {Sommer-J\"{o}rgensen}}, \ and\
  \bibinfo {author} {\bibfnamefont {S.}~\bibnamefont {Goedecker}},\ }\href
  {\doibase https://doi.org/10.1016/j.softx.2024.101632} {\bibfield  {journal}
  {\bibinfo  {journal} {SoftwareX}\ }\textbf {\bibinfo {volume} {25}},\
  \bibinfo {pages} {101632} (\bibinfo {year} {2024})}\BibitemShut {NoStop}%
\bibitem [{\citenamefont {Deaven}\ and\ \citenamefont
  {Ho}(1995)}]{PhysRevLett.75.288}%
  \BibitemOpen
  \bibfield  {author} {\bibinfo {author} {\bibfnamefont {D.~M.}\ \bibnamefont
  {Deaven}}\ and\ \bibinfo {author} {\bibfnamefont {K.~M.}\ \bibnamefont
  {Ho}},\ }\href {\doibase 10.1103/PhysRevLett.75.288} {\bibfield  {journal}
  {\bibinfo  {journal} {Phys. Rev. Lett.}\ }\textbf {\bibinfo {volume} {75}},\
  \bibinfo {pages} {288} (\bibinfo {year} {1995})}\BibitemShut {NoStop}%
\bibitem [{\citenamefont {Oganov}\ and\ \citenamefont
  {Glass}(2006)}]{doi:10.1063/1.2210932}%
  \BibitemOpen
  \bibfield  {author} {\bibinfo {author} {\bibfnamefont {A.~R.}\ \bibnamefont
  {Oganov}}\ and\ \bibinfo {author} {\bibfnamefont {C.~W.}\ \bibnamefont
  {Glass}},\ }\href {\doibase 10.1063/1.2210932} {\bibfield  {journal}
  {\bibinfo  {journal} {J. Chem. Phys.}\ }\textbf {\bibinfo {volume} {124}},\
  \bibinfo {pages} {244704} (\bibinfo {year} {2006})}\BibitemShut {NoStop}%
\bibitem [{\citenamefont {Glass}\ \emph {et~al.}(2006)\citenamefont {Glass},
  \citenamefont {Oganov},\ and\ \citenamefont {Hansen}}]{GLASS2006713}%
  \BibitemOpen
  \bibfield  {author} {\bibinfo {author} {\bibfnamefont {C.~W.}\ \bibnamefont
  {Glass}}, \bibinfo {author} {\bibfnamefont {A.~R.}\ \bibnamefont {Oganov}}, \
  and\ \bibinfo {author} {\bibfnamefont {N.}~\bibnamefont {Hansen}},\ }\href
  {\doibase https://doi.org/10.1016/j.cpc.2006.07.020} {\bibfield  {journal}
  {\bibinfo  {journal} {Comput. Phys. Commun.}\ }\textbf {\bibinfo {volume}
  {175}},\ \bibinfo {pages} {713} (\bibinfo {year} {2006})}\BibitemShut
  {NoStop}%
\bibitem [{\citenamefont {Oganov}\ \emph {et~al.}(2011)\citenamefont {Oganov},
  \citenamefont {Lyakhov},\ and\ \citenamefont
  {Valle}}]{doi:10.1021/ar1001318}%
  \BibitemOpen
  \bibfield  {author} {\bibinfo {author} {\bibfnamefont {A.~R.}\ \bibnamefont
  {Oganov}}, \bibinfo {author} {\bibfnamefont {A.~O.}\ \bibnamefont {Lyakhov}},
  \ and\ \bibinfo {author} {\bibfnamefont {M.}~\bibnamefont {Valle}},\ }\href
  {\doibase 10.1021/ar1001318} {\bibfield  {journal} {\bibinfo  {journal} {Acc.
  Chem. Res.}\ }\textbf {\bibinfo {volume} {44}},\ \bibinfo {pages} {227}
  (\bibinfo {year} {2011})},\ \bibinfo {note} {pMID: 21361336}\BibitemShut
  {NoStop}%
\bibitem [{\citenamefont {Lyakhov}\ \emph {et~al.}(2013)\citenamefont
  {Lyakhov}, \citenamefont {Oganov}, \citenamefont {Stokes},\ and\
  \citenamefont {Zhu}}]{LYAKHOV20131172}%
  \BibitemOpen
  \bibfield  {author} {\bibinfo {author} {\bibfnamefont {A.~O.}\ \bibnamefont
  {Lyakhov}}, \bibinfo {author} {\bibfnamefont {A.~R.}\ \bibnamefont {Oganov}},
  \bibinfo {author} {\bibfnamefont {H.~T.}\ \bibnamefont {Stokes}}, \ and\
  \bibinfo {author} {\bibfnamefont {Q.}~\bibnamefont {Zhu}},\ }\href {\doibase
  https://doi.org/10.1016/j.cpc.2012.12.009} {\bibfield  {journal} {\bibinfo
  {journal} {Comput. Phys. Commun.}\ }\textbf {\bibinfo {volume} {184}},\
  \bibinfo {pages} {1172} (\bibinfo {year} {2013})}\BibitemShut {NoStop}%
\bibitem [{\citenamefont {Lonie}\ and\ \citenamefont
  {Zurek}(2011)}]{LONIE2011372}%
  \BibitemOpen
  \bibfield  {author} {\bibinfo {author} {\bibfnamefont {D.~C.}\ \bibnamefont
  {Lonie}}\ and\ \bibinfo {author} {\bibfnamefont {E.}~\bibnamefont {Zurek}},\
  }\href {\doibase https://doi.org/10.1016/j.cpc.2010.07.048} {\bibfield
  {journal} {\bibinfo  {journal} {Comput. Phys. Commun.}\ }\textbf {\bibinfo
  {volume} {182}},\ \bibinfo {pages} {372} (\bibinfo {year}
  {2011})}\BibitemShut {NoStop}%
\bibitem [{\citenamefont {Falls}\ \emph {et~al.}(2021)\citenamefont {Falls},
  \citenamefont {Avery}, \citenamefont {Wang}, \citenamefont {Hilleke},\ and\
  \citenamefont {Zurek}}]{doi:10.1021/acs.jpcc.0c09531}%
  \BibitemOpen
  \bibfield  {author} {\bibinfo {author} {\bibfnamefont {Z.}~\bibnamefont
  {Falls}}, \bibinfo {author} {\bibfnamefont {P.}~\bibnamefont {Avery}},
  \bibinfo {author} {\bibfnamefont {X.}~\bibnamefont {Wang}}, \bibinfo {author}
  {\bibfnamefont {K.~P.}\ \bibnamefont {Hilleke}}, \ and\ \bibinfo {author}
  {\bibfnamefont {E.}~\bibnamefont {Zurek}},\ }\href {\doibase
  10.1021/acs.jpcc.0c09531} {\bibfield  {journal} {\bibinfo  {journal} {J.
  Phys. Chem. C}\ }\textbf {\bibinfo {volume} {125}},\ \bibinfo {pages} {1601}
  (\bibinfo {year} {2021})}\BibitemShut {NoStop}%
\bibitem [{\citenamefont {Wang}\ \emph {et~al.}(2010)\citenamefont {Wang},
  \citenamefont {Lv}, \citenamefont {Zhu},\ and\ \citenamefont
  {Ma}}]{PhysRevB.82.094116}%
  \BibitemOpen
  \bibfield  {author} {\bibinfo {author} {\bibfnamefont {Y.}~\bibnamefont
  {Wang}}, \bibinfo {author} {\bibfnamefont {J.}~\bibnamefont {Lv}}, \bibinfo
  {author} {\bibfnamefont {L.}~\bibnamefont {Zhu}}, \ and\ \bibinfo {author}
  {\bibfnamefont {Y.}~\bibnamefont {Ma}},\ }\href {\doibase
  10.1103/PhysRevB.82.094116} {\bibfield  {journal} {\bibinfo  {journal} {Phys.
  Rev. B}\ }\textbf {\bibinfo {volume} {82}},\ \bibinfo {pages} {094116}
  (\bibinfo {year} {2010})}\BibitemShut {NoStop}%
\bibitem [{\citenamefont {Yamashita}\ \emph {et~al.}(2018)\citenamefont
  {Yamashita}, \citenamefont {Sato}, \citenamefont {Kino}, \citenamefont
  {Miyake}, \citenamefont {Tsuda},\ and\ \citenamefont
  {Oguchi}}]{PhysRevMaterials.2.013803}%
  \BibitemOpen
  \bibfield  {author} {\bibinfo {author} {\bibfnamefont {T.}~\bibnamefont
  {Yamashita}}, \bibinfo {author} {\bibfnamefont {N.}~\bibnamefont {Sato}},
  \bibinfo {author} {\bibfnamefont {H.}~\bibnamefont {Kino}}, \bibinfo {author}
  {\bibfnamefont {T.}~\bibnamefont {Miyake}}, \bibinfo {author} {\bibfnamefont
  {K.}~\bibnamefont {Tsuda}}, \ and\ \bibinfo {author} {\bibfnamefont
  {T.}~\bibnamefont {Oguchi}},\ }\href {\doibase
  10.1103/PhysRevMaterials.2.013803} {\bibfield  {journal} {\bibinfo  {journal}
  {Phys. Rev. Mater.}\ }\textbf {\bibinfo {volume} {2}},\ \bibinfo {pages}
  {013803} (\bibinfo {year} {2018})}\BibitemShut {NoStop}%
\bibitem [{\citenamefont {Bisbo}\ and\ \citenamefont
  {Hammer}(2020)}]{PhysRevLett.124.086102}%
  \BibitemOpen
  \bibfield  {author} {\bibinfo {author} {\bibfnamefont {M.~K.}\ \bibnamefont
  {Bisbo}}\ and\ \bibinfo {author} {\bibfnamefont {B.}~\bibnamefont {Hammer}},\
  }\href {\doibase 10.1103/PhysRevLett.124.086102} {\bibfield  {journal}
  {\bibinfo  {journal} {Phys. Rev. Lett.}\ }\textbf {\bibinfo {volume} {124}},\
  \bibinfo {pages} {086102} (\bibinfo {year} {2020})}\BibitemShut {NoStop}%
\bibitem [{\citenamefont {Behler}\ and\ \citenamefont
  {Parrinello}(2007)}]{PhysRevLett.98.146401}%
  \BibitemOpen
  \bibfield  {author} {\bibinfo {author} {\bibfnamefont {J.}~\bibnamefont
  {Behler}}\ and\ \bibinfo {author} {\bibfnamefont {M.}~\bibnamefont
  {Parrinello}},\ }\href {\doibase 10.1103/PhysRevLett.98.146401} {\bibfield
  {journal} {\bibinfo  {journal} {Phys. Rev. Lett.}\ }\textbf {\bibinfo
  {volume} {98}},\ \bibinfo {pages} {146401} (\bibinfo {year}
  {2007})}\BibitemShut {NoStop}%
\bibitem [{\citenamefont {Bart\'ok}\ \emph {et~al.}(2010)\citenamefont
  {Bart\'ok}, \citenamefont {Payne}, \citenamefont {Kondor},\ and\
  \citenamefont {Cs\'anyi}}]{PhysRevLett.104.136403}%
  \BibitemOpen
  \bibfield  {author} {\bibinfo {author} {\bibfnamefont {A.~P.}\ \bibnamefont
  {Bart\'ok}}, \bibinfo {author} {\bibfnamefont {M.~C.}\ \bibnamefont {Payne}},
  \bibinfo {author} {\bibfnamefont {R.}~\bibnamefont {Kondor}}, \ and\ \bibinfo
  {author} {\bibfnamefont {G.}~\bibnamefont {Cs\'anyi}},\ }\href {\doibase
  10.1103/PhysRevLett.104.136403} {\bibfield  {journal} {\bibinfo  {journal}
  {Phys. Rev. Lett.}\ }\textbf {\bibinfo {volume} {104}},\ \bibinfo {pages}
  {136403} (\bibinfo {year} {2010})}\BibitemShut {NoStop}%
\bibitem [{\citenamefont {Shapeev}(2016)}]{doi:10.1137/15M1054183}%
  \BibitemOpen
  \bibfield  {author} {\bibinfo {author} {\bibfnamefont {A.~V.}\ \bibnamefont
  {Shapeev}},\ }\href {\doibase 10.1137/15M1054183} {\bibfield  {journal}
  {\bibinfo  {journal} {Multiscale Model. Simul.}\ }\textbf {\bibinfo {volume}
  {14}},\ \bibinfo {pages} {1153} (\bibinfo {year} {2016})}\BibitemShut
  {NoStop}%
\bibitem [{\citenamefont {Drautz}(2019)}]{PhysRevB.99.014104}%
  \BibitemOpen
  \bibfield  {author} {\bibinfo {author} {\bibfnamefont {R.}~\bibnamefont
  {Drautz}},\ }\href {\doibase 10.1103/PhysRevB.99.014104} {\bibfield
  {journal} {\bibinfo  {journal} {Phys. Rev. B}\ }\textbf {\bibinfo {volume}
  {99}},\ \bibinfo {pages} {014104} (\bibinfo {year} {2019})}\BibitemShut
  {NoStop}%
\bibitem [{\citenamefont {Podryabinkin}\ \emph {et~al.}(2019)\citenamefont
  {Podryabinkin}, \citenamefont {Tikhonov}, \citenamefont {Shapeev},\ and\
  \citenamefont {Oganov}}]{PhysRevB.99.064114}%
  \BibitemOpen
  \bibfield  {author} {\bibinfo {author} {\bibfnamefont {E.~V.}\ \bibnamefont
  {Podryabinkin}}, \bibinfo {author} {\bibfnamefont {E.~V.}\ \bibnamefont
  {Tikhonov}}, \bibinfo {author} {\bibfnamefont {A.~V.}\ \bibnamefont
  {Shapeev}}, \ and\ \bibinfo {author} {\bibfnamefont {A.~R.}\ \bibnamefont
  {Oganov}},\ }\href {\doibase 10.1103/PhysRevB.99.064114} {\bibfield
  {journal} {\bibinfo  {journal} {Phys. Rev. B}\ }\textbf {\bibinfo {volume}
  {99}},\ \bibinfo {pages} {064114} (\bibinfo {year} {2019})}\BibitemShut
  {NoStop}%
\bibitem [{\citenamefont {Chen}\ \emph {et~al.}(2019)\citenamefont {Chen},
  \citenamefont {Ye}, \citenamefont {Zuo}, \citenamefont {Zheng},\ and\
  \citenamefont {Ong}}]{doi:10.1021/acs.chemmater.9b01294}%
  \BibitemOpen
  \bibfield  {author} {\bibinfo {author} {\bibfnamefont {C.}~\bibnamefont
  {Chen}}, \bibinfo {author} {\bibfnamefont {W.}~\bibnamefont {Ye}}, \bibinfo
  {author} {\bibfnamefont {Y.}~\bibnamefont {Zuo}}, \bibinfo {author}
  {\bibfnamefont {C.}~\bibnamefont {Zheng}}, \ and\ \bibinfo {author}
  {\bibfnamefont {S.~P.}\ \bibnamefont {Ong}},\ }\href {\doibase
  10.1021/acs.chemmater.9b01294} {\bibfield  {journal} {\bibinfo  {journal}
  {Chem. Mat.}\ }\textbf {\bibinfo {volume} {31}},\ \bibinfo {pages} {3564}
  (\bibinfo {year} {2019})}\BibitemShut {NoStop}%
\bibitem [{\citenamefont {Chen}\ and\ \citenamefont {Ong}(2022)}]{Chen2022}%
  \BibitemOpen
  \bibfield  {author} {\bibinfo {author} {\bibfnamefont {C.}~\bibnamefont
  {Chen}}\ and\ \bibinfo {author} {\bibfnamefont {S.~P.}\ \bibnamefont {Ong}},\
  }\href {\doibase 10.1038/s43588-022-00349-3} {\bibfield  {journal} {\bibinfo
  {journal} {Nat. Comput. Sci.}\ }\textbf {\bibinfo {volume} {2}},\ \bibinfo
  {pages} {718} (\bibinfo {year} {2022})}\BibitemShut {NoStop}%
\bibitem [{\citenamefont {Choudhary}\ \emph {et~al.}(2023)\citenamefont
  {Choudhary}, \citenamefont {DeCost}, \citenamefont {Major}, \citenamefont
  {Butler}, \citenamefont {Thiyagalingam},\ and\ \citenamefont
  {Tavazza}}]{D2DD00096B}%
  \BibitemOpen
  \bibfield  {author} {\bibinfo {author} {\bibfnamefont {K.}~\bibnamefont
  {Choudhary}}, \bibinfo {author} {\bibfnamefont {B.}~\bibnamefont {DeCost}},
  \bibinfo {author} {\bibfnamefont {L.}~\bibnamefont {Major}}, \bibinfo
  {author} {\bibfnamefont {K.}~\bibnamefont {Butler}}, \bibinfo {author}
  {\bibfnamefont {J.}~\bibnamefont {Thiyagalingam}}, \ and\ \bibinfo {author}
  {\bibfnamefont {F.}~\bibnamefont {Tavazza}},\ }\href {\doibase
  10.1039/D2DD00096B} {\bibfield  {journal} {\bibinfo  {journal} {Digit.
  Discov.}\ }\textbf {\bibinfo {volume} {2}},\ \bibinfo {pages} {346} (\bibinfo
  {year} {2023})}\BibitemShut {NoStop}%
\bibitem [{\citenamefont {Deng}\ \emph {et~al.}(2023)\citenamefont {Deng},
  \citenamefont {Zhong}, \citenamefont {Jun}, \citenamefont {Riebesell},
  \citenamefont {Han}, \citenamefont {Bartel},\ and\ \citenamefont
  {Ceder}}]{Deng2023}%
  \BibitemOpen
  \bibfield  {author} {\bibinfo {author} {\bibfnamefont {B.}~\bibnamefont
  {Deng}}, \bibinfo {author} {\bibfnamefont {P.}~\bibnamefont {Zhong}},
  \bibinfo {author} {\bibfnamefont {K.}~\bibnamefont {Jun}}, \bibinfo {author}
  {\bibfnamefont {J.}~\bibnamefont {Riebesell}}, \bibinfo {author}
  {\bibfnamefont {K.}~\bibnamefont {Han}}, \bibinfo {author} {\bibfnamefont
  {C.~J.}\ \bibnamefont {Bartel}}, \ and\ \bibinfo {author} {\bibfnamefont
  {G.}~\bibnamefont {Ceder}},\ }\href {\doibase 10.1038/s42256-023-00716-3}
  {\bibfield  {journal} {\bibinfo  {journal} {Nat. Mach. Intell.}\ }\textbf
  {\bibinfo {volume} {5}},\ \bibinfo {pages} {1031} (\bibinfo {year}
  {2023})}\BibitemShut {NoStop}%
\bibitem [{\citenamefont {Batatia}\ \emph {et~al.}(2022)\citenamefont
  {Batatia}, \citenamefont {Kovacs}, \citenamefont {Simm}, \citenamefont
  {Ortner},\ and\ \citenamefont {Csanyi}}]{NEURIPS2022_4a36c3c5}%
  \BibitemOpen
  \bibfield  {author} {\bibinfo {author} {\bibfnamefont {I.}~\bibnamefont
  {Batatia}}, \bibinfo {author} {\bibfnamefont {D.~P.}\ \bibnamefont {Kovacs}},
  \bibinfo {author} {\bibfnamefont {G.}~\bibnamefont {Simm}}, \bibinfo {author}
  {\bibfnamefont {C.}~\bibnamefont {Ortner}}, \ and\ \bibinfo {author}
  {\bibfnamefont {G.}~\bibnamefont {Csanyi}},\ }in\ \href
  {https://proceedings.neurips.cc/paper_files/paper/2022/file/4a36c3c51af11ed9f34615b81edb5bbc-Paper-Conference.pdf}
  {\emph {\bibinfo {booktitle} {Advances in Neural Information Processing
  Systems}}},\ Vol.~\bibinfo {volume} {35},\ \bibinfo {editor} {edited by\
  \bibinfo {editor} {\bibfnamefont {S.}~\bibnamefont {Koyejo}}, \bibinfo
  {editor} {\bibfnamefont {S.}~\bibnamefont {Mohamed}}, \bibinfo {editor}
  {\bibfnamefont {A.}~\bibnamefont {Agarwal}}, \bibinfo {editor} {\bibfnamefont
  {D.}~\bibnamefont {Belgrave}}, \bibinfo {editor} {\bibfnamefont
  {K.}~\bibnamefont {Cho}}, \ and\ \bibinfo {editor} {\bibfnamefont
  {A.}~\bibnamefont {Oh}}}\ (\bibinfo  {publisher} {Curran Associates, Inc.},\
  \bibinfo {year} {2022})\ pp.\ \bibinfo {pages} {11423--11436}\BibitemShut
  {NoStop}%
\bibitem [{\citenamefont {Batatia}\ \emph {et~al.}(2023)\citenamefont
  {Batatia}, \citenamefont {Benner}, \citenamefont {Chiang}, \citenamefont
  {Elena}, \citenamefont {Kov\'{a}cs}, \citenamefont {Riebesell}, \citenamefont
  {Advincula}, \citenamefont {Asta}, \citenamefont {Baldwin}, \citenamefont
  {Bernstein}, \citenamefont {Bhowmik}, \citenamefont {Blau}, \citenamefont
  {C\u{a}rare}, \citenamefont {Darby}, \citenamefont {De}, \citenamefont {Pia},
  \citenamefont {Deringer}, \citenamefont {Elijo\v{s}ius}, \citenamefont
  {El-Machachi}, \citenamefont {Fako}, \citenamefont {Ferrari}, \citenamefont
  {Genreith-Schriever}, \citenamefont {George}, \citenamefont {Goodall},
  \citenamefont {Grey}, \citenamefont {Han}, \citenamefont {Handley},
  \citenamefont {Heenen}, \citenamefont {Hermansson}, \citenamefont {Holm},
  \citenamefont {Jaafar}, \citenamefont {Hofmann}, \citenamefont {Jakob},
  \citenamefont {Jung}, \citenamefont {Kapil}, \citenamefont {Kaplan},
  \citenamefont {Karimitari}, \citenamefont {Kroupa}, \citenamefont {Kullgren},
  \citenamefont {Kuner}, \citenamefont {Kuryla}, \citenamefont {Liepuoniute},
  \citenamefont {Margraf}, \citenamefont {Magd\u{a}u}, \citenamefont
  {Michaelides}, \citenamefont {Moore}, \citenamefont {Naik}, \citenamefont
  {Niblett}, \citenamefont {Norwood}, \citenamefont {O'Neill}, \citenamefont
  {Ortner}, \citenamefont {Persson}, \citenamefont {Reuter}, \citenamefont
  {Rosen}, \citenamefont {Schaaf}, \citenamefont {Schran}, \citenamefont
  {Sivonxay}, \citenamefont {Stenczel}, \citenamefont {Svahn}, \citenamefont
  {Sutton}, \citenamefont {van~der Oord}, \citenamefont {Varga-Umbrich},
  \citenamefont {Vegge}, \citenamefont {Vondr\'{a}k}, \citenamefont {Wang},
  \citenamefont {Witt}, \citenamefont {Zills},\ and\ \citenamefont
  {Cs\'{a}nyi}}]{batatia2023foundation}%
  \BibitemOpen
  \bibfield  {author} {\bibinfo {author} {\bibfnamefont {I.}~\bibnamefont
  {Batatia}}, \bibinfo {author} {\bibfnamefont {P.}~\bibnamefont {Benner}},
  \bibinfo {author} {\bibfnamefont {Y.}~\bibnamefont {Chiang}}, \bibinfo
  {author} {\bibfnamefont {A.~M.}\ \bibnamefont {Elena}}, \bibinfo {author}
  {\bibfnamefont {D.~P.}\ \bibnamefont {Kov\'{a}cs}}, \bibinfo {author}
  {\bibfnamefont {J.}~\bibnamefont {Riebesell}}, \bibinfo {author}
  {\bibfnamefont {X.~R.}\ \bibnamefont {Advincula}}, \bibinfo {author}
  {\bibfnamefont {M.}~\bibnamefont {Asta}}, \bibinfo {author} {\bibfnamefont
  {W.~J.}\ \bibnamefont {Baldwin}}, \bibinfo {author} {\bibfnamefont
  {N.}~\bibnamefont {Bernstein}}, \bibinfo {author} {\bibfnamefont
  {A.}~\bibnamefont {Bhowmik}}, \bibinfo {author} {\bibfnamefont {S.~M.}\
  \bibnamefont {Blau}}, \bibinfo {author} {\bibfnamefont {V.}~\bibnamefont
  {C\u{a}rare}}, \bibinfo {author} {\bibfnamefont {J.~P.}\ \bibnamefont
  {Darby}}, \bibinfo {author} {\bibfnamefont {S.}~\bibnamefont {De}}, \bibinfo
  {author} {\bibfnamefont {F.~D.}\ \bibnamefont {Pia}}, \bibinfo {author}
  {\bibfnamefont {V.~L.}\ \bibnamefont {Deringer}}, \bibinfo {author}
  {\bibfnamefont {R.}~\bibnamefont {Elijo\v{s}ius}}, \bibinfo {author}
  {\bibfnamefont {Z.}~\bibnamefont {El-Machachi}}, \bibinfo {author}
  {\bibfnamefont {E.}~\bibnamefont {Fako}}, \bibinfo {author} {\bibfnamefont
  {A.~C.}\ \bibnamefont {Ferrari}}, \bibinfo {author} {\bibfnamefont
  {A.}~\bibnamefont {Genreith-Schriever}}, \bibinfo {author} {\bibfnamefont
  {J.}~\bibnamefont {George}}, \bibinfo {author} {\bibfnamefont {R.~E.~A.}\
  \bibnamefont {Goodall}}, \bibinfo {author} {\bibfnamefont {C.~P.}\
  \bibnamefont {Grey}}, \bibinfo {author} {\bibfnamefont {S.}~\bibnamefont
  {Han}}, \bibinfo {author} {\bibfnamefont {W.}~\bibnamefont {Handley}},
  \bibinfo {author} {\bibfnamefont {H.~H.}\ \bibnamefont {Heenen}}, \bibinfo
  {author} {\bibfnamefont {K.}~\bibnamefont {Hermansson}}, \bibinfo {author}
  {\bibfnamefont {C.}~\bibnamefont {Holm}}, \bibinfo {author} {\bibfnamefont
  {J.}~\bibnamefont {Jaafar}}, \bibinfo {author} {\bibfnamefont
  {S.}~\bibnamefont {Hofmann}}, \bibinfo {author} {\bibfnamefont {K.~S.}\
  \bibnamefont {Jakob}}, \bibinfo {author} {\bibfnamefont {H.}~\bibnamefont
  {Jung}}, \bibinfo {author} {\bibfnamefont {V.}~\bibnamefont {Kapil}},
  \bibinfo {author} {\bibfnamefont {A.~D.}\ \bibnamefont {Kaplan}}, \bibinfo
  {author} {\bibfnamefont {N.}~\bibnamefont {Karimitari}}, \bibinfo {author}
  {\bibfnamefont {N.}~\bibnamefont {Kroupa}}, \bibinfo {author} {\bibfnamefont
  {J.}~\bibnamefont {Kullgren}}, \bibinfo {author} {\bibfnamefont {M.~C.}\
  \bibnamefont {Kuner}}, \bibinfo {author} {\bibfnamefont {D.}~\bibnamefont
  {Kuryla}}, \bibinfo {author} {\bibfnamefont {G.}~\bibnamefont {Liepuoniute}},
  \bibinfo {author} {\bibfnamefont {J.~T.}\ \bibnamefont {Margraf}}, \bibinfo
  {author} {\bibfnamefont {I.-B.}\ \bibnamefont {Magd\u{a}u}}, \bibinfo
  {author} {\bibfnamefont {A.}~\bibnamefont {Michaelides}}, \bibinfo {author}
  {\bibfnamefont {J.~H.}\ \bibnamefont {Moore}}, \bibinfo {author}
  {\bibfnamefont {A.~A.}\ \bibnamefont {Naik}}, \bibinfo {author}
  {\bibfnamefont {S.~P.}\ \bibnamefont {Niblett}}, \bibinfo {author}
  {\bibfnamefont {S.~W.}\ \bibnamefont {Norwood}}, \bibinfo {author}
  {\bibfnamefont {N.}~\bibnamefont {O'Neill}}, \bibinfo {author} {\bibfnamefont
  {C.}~\bibnamefont {Ortner}}, \bibinfo {author} {\bibfnamefont {K.~A.}\
  \bibnamefont {Persson}}, \bibinfo {author} {\bibfnamefont {K.}~\bibnamefont
  {Reuter}}, \bibinfo {author} {\bibfnamefont {A.~S.}\ \bibnamefont {Rosen}},
  \bibinfo {author} {\bibfnamefont {L.~L.}\ \bibnamefont {Schaaf}}, \bibinfo
  {author} {\bibfnamefont {C.}~\bibnamefont {Schran}}, \bibinfo {author}
  {\bibfnamefont {E.}~\bibnamefont {Sivonxay}}, \bibinfo {author}
  {\bibfnamefont {T.~K.}\ \bibnamefont {Stenczel}}, \bibinfo {author}
  {\bibfnamefont {V.}~\bibnamefont {Svahn}}, \bibinfo {author} {\bibfnamefont
  {C.}~\bibnamefont {Sutton}}, \bibinfo {author} {\bibfnamefont
  {C.}~\bibnamefont {van~der Oord}}, \bibinfo {author} {\bibfnamefont
  {E.}~\bibnamefont {Varga-Umbrich}}, \bibinfo {author} {\bibfnamefont
  {T.}~\bibnamefont {Vegge}}, \bibinfo {author} {\bibfnamefont
  {M.}~\bibnamefont {Vondr\'{a}k}}, \bibinfo {author} {\bibfnamefont
  {Y.}~\bibnamefont {Wang}}, \bibinfo {author} {\bibfnamefont {W.~C.}\
  \bibnamefont {Witt}}, \bibinfo {author} {\bibfnamefont {F.}~\bibnamefont
  {Zills}}, \ and\ \bibinfo {author} {\bibfnamefont {G.}~\bibnamefont
  {Cs\'{a}nyi}},\ }\href@noop {} {\enquote {\bibinfo {title} {A foundation
  model for atomistic materials chemistry},}\ } (\bibinfo {year} {2023}),\
  \Eprint {http://arxiv.org/abs/arXiv:2401.00096} {arXiv:2401.00096}
  \BibitemShut {NoStop}%
\bibitem [{\citenamefont {Merchant}\ \emph {et~al.}(2023)\citenamefont
  {Merchant}, \citenamefont {Batzner}, \citenamefont {Schoenholz},
  \citenamefont {Aykol}, \citenamefont {Cheon},\ and\ \citenamefont
  {Cubuk}}]{Merchant2023}%
  \BibitemOpen
  \bibfield  {author} {\bibinfo {author} {\bibfnamefont {A.}~\bibnamefont
  {Merchant}}, \bibinfo {author} {\bibfnamefont {S.}~\bibnamefont {Batzner}},
  \bibinfo {author} {\bibfnamefont {S.~S.}\ \bibnamefont {Schoenholz}},
  \bibinfo {author} {\bibfnamefont {M.}~\bibnamefont {Aykol}}, \bibinfo
  {author} {\bibfnamefont {G.}~\bibnamefont {Cheon}}, \ and\ \bibinfo {author}
  {\bibfnamefont {E.~D.}\ \bibnamefont {Cubuk}},\ }\href {\doibase
  10.1038/s41586-023-06735-9} {\bibfield  {journal} {\bibinfo  {journal}
  {Nature}\ }\textbf {\bibinfo {volume} {624}},\ \bibinfo {pages} {80}
  (\bibinfo {year} {2023})}\BibitemShut {NoStop}%
\bibitem [{\citenamefont {Neumann}\ \emph {et~al.}(2024)\citenamefont
  {Neumann}, \citenamefont {Gin}, \citenamefont {Rhodes}, \citenamefont
  {Bennett}, \citenamefont {Li}, \citenamefont {Choubisa}, \citenamefont
  {Hussey},\ and\ \citenamefont {Godwin}}]{neumann2024orbfastscalableneural}%
  \BibitemOpen
  \bibfield  {author} {\bibinfo {author} {\bibfnamefont {M.}~\bibnamefont
  {Neumann}}, \bibinfo {author} {\bibfnamefont {J.}~\bibnamefont {Gin}},
  \bibinfo {author} {\bibfnamefont {B.}~\bibnamefont {Rhodes}}, \bibinfo
  {author} {\bibfnamefont {S.}~\bibnamefont {Bennett}}, \bibinfo {author}
  {\bibfnamefont {Z.}~\bibnamefont {Li}}, \bibinfo {author} {\bibfnamefont
  {H.}~\bibnamefont {Choubisa}}, \bibinfo {author} {\bibfnamefont
  {A.}~\bibnamefont {Hussey}}, \ and\ \bibinfo {author} {\bibfnamefont
  {J.}~\bibnamefont {Godwin}},\ }\href@noop {} {\enquote {\bibinfo {title}
  {Orb: A fast, scalable neural network potential},}\ } (\bibinfo {year}
  {2024}),\ \Eprint {http://arxiv.org/abs/arXiv:2410.22570} {arXiv:2410.22570}
  \BibitemShut {NoStop}%
\bibitem [{\citenamefont {Liao}\ \emph {et~al.}(2024)\citenamefont {Liao},
  \citenamefont {Wood}, \citenamefont {Das},\ and\ \citenamefont
  {Smidt}}]{equiformer_v2}%
  \BibitemOpen
  \bibfield  {author} {\bibinfo {author} {\bibfnamefont {Y.-L.}\ \bibnamefont
  {Liao}}, \bibinfo {author} {\bibfnamefont {B.}~\bibnamefont {Wood}}, \bibinfo
  {author} {\bibfnamefont {A.}~\bibnamefont {Das}}, \ and\ \bibinfo {author}
  {\bibfnamefont {T.}~\bibnamefont {Smidt}},\ }in\ \href
  {https://openreview.net/forum?id=mCOBKZmrzD} {\emph {\bibinfo {booktitle}
  {International Conference on Learning Representations}}}\ (\bibinfo {year}
  {2024})\BibitemShut {NoStop}%
\bibitem [{\citenamefont {Barroso-Luque}\ \emph {et~al.}(2024)\citenamefont
  {Barroso-Luque}, \citenamefont {Muhammed}, \citenamefont {Fu}, \citenamefont
  {Wood}, \citenamefont {Dzamba}, \citenamefont {Gao}, \citenamefont {Rizvi},
  \citenamefont {Zitnick},\ and\ \citenamefont {Ulissi}}]{barroso_omat24}%
  \BibitemOpen
  \bibfield  {author} {\bibinfo {author} {\bibfnamefont {L.}~\bibnamefont
  {Barroso-Luque}}, \bibinfo {author} {\bibfnamefont {S.}~\bibnamefont
  {Muhammed}}, \bibinfo {author} {\bibfnamefont {X.}~\bibnamefont {Fu}},
  \bibinfo {author} {\bibfnamefont {B.}~\bibnamefont {Wood}}, \bibinfo {author}
  {\bibfnamefont {M.}~\bibnamefont {Dzamba}}, \bibinfo {author} {\bibfnamefont
  {M.}~\bibnamefont {Gao}}, \bibinfo {author} {\bibfnamefont {A.}~\bibnamefont
  {Rizvi}}, \bibinfo {author} {\bibfnamefont {C.~L.}\ \bibnamefont {Zitnick}},
  \ and\ \bibinfo {author} {\bibfnamefont {Z.~W.}\ \bibnamefont {Ulissi}},\
  }\href@noop {} {\enquote {\bibinfo {title} {Open materials 2024 (omat24)
  inorganic materials dataset and models},}\ } (\bibinfo {year} {2024}),\
  \Eprint {http://arxiv.org/abs/arXiv:2410.12771} {arXiv:2410.12771}
  \BibitemShut {NoStop}%
\bibitem [{\citenamefont {Yang}\ \emph {et~al.}(2024)\citenamefont {Yang},
  \citenamefont {Hu}, \citenamefont {Zhou}, \citenamefont {Liu}, \citenamefont
  {Shi}, \citenamefont {Li}, \citenamefont {Li}, \citenamefont {Chen},
  \citenamefont {Chen}, \citenamefont {Zeni}, \citenamefont {Horton},
  \citenamefont {Pinsler}, \citenamefont {Fowler}, \citenamefont {Z\"{u}gner},
  \citenamefont {Xie}, \citenamefont {Smith}, \citenamefont {Sun},
  \citenamefont {Wang}, \citenamefont {Kong}, \citenamefont {Liu},
  \citenamefont {Hao},\ and\ \citenamefont {Lu}}]{yang2024mattersim}%
  \BibitemOpen
  \bibfield  {author} {\bibinfo {author} {\bibfnamefont {H.}~\bibnamefont
  {Yang}}, \bibinfo {author} {\bibfnamefont {C.}~\bibnamefont {Hu}}, \bibinfo
  {author} {\bibfnamefont {Y.}~\bibnamefont {Zhou}}, \bibinfo {author}
  {\bibfnamefont {X.}~\bibnamefont {Liu}}, \bibinfo {author} {\bibfnamefont
  {Y.}~\bibnamefont {Shi}}, \bibinfo {author} {\bibfnamefont {J.}~\bibnamefont
  {Li}}, \bibinfo {author} {\bibfnamefont {G.}~\bibnamefont {Li}}, \bibinfo
  {author} {\bibfnamefont {Z.}~\bibnamefont {Chen}}, \bibinfo {author}
  {\bibfnamefont {S.}~\bibnamefont {Chen}}, \bibinfo {author} {\bibfnamefont
  {C.}~\bibnamefont {Zeni}}, \bibinfo {author} {\bibfnamefont {M.}~\bibnamefont
  {Horton}}, \bibinfo {author} {\bibfnamefont {R.}~\bibnamefont {Pinsler}},
  \bibinfo {author} {\bibfnamefont {A.}~\bibnamefont {Fowler}}, \bibinfo
  {author} {\bibfnamefont {D.}~\bibnamefont {Z\"{u}gner}}, \bibinfo {author}
  {\bibfnamefont {T.}~\bibnamefont {Xie}}, \bibinfo {author} {\bibfnamefont
  {J.}~\bibnamefont {Smith}}, \bibinfo {author} {\bibfnamefont
  {L.}~\bibnamefont {Sun}}, \bibinfo {author} {\bibfnamefont {Q.}~\bibnamefont
  {Wang}}, \bibinfo {author} {\bibfnamefont {L.}~\bibnamefont {Kong}}, \bibinfo
  {author} {\bibfnamefont {C.}~\bibnamefont {Liu}}, \bibinfo {author}
  {\bibfnamefont {H.}~\bibnamefont {Hao}}, \ and\ \bibinfo {author}
  {\bibfnamefont {Z.}~\bibnamefont {Lu}},\ }\href@noop {} {\enquote {\bibinfo
  {title} {Mattersim: A deep learning atomistic model across elements,
  temperatures and pressures},}\ } (\bibinfo {year} {2024}),\ \Eprint
  {http://arxiv.org/abs/arXiv:2405.04967} {arXiv:2405.04967} \BibitemShut
  {NoStop}%
\bibitem [{\citenamefont {Takamoto}\ \emph
  {et~al.}(2022{\natexlab{a}})\citenamefont {Takamoto}, \citenamefont {Izumi},\
  and\ \citenamefont {Li}}]{TAKAMOTO2022111280}%
  \BibitemOpen
  \bibfield  {author} {\bibinfo {author} {\bibfnamefont {S.}~\bibnamefont
  {Takamoto}}, \bibinfo {author} {\bibfnamefont {S.}~\bibnamefont {Izumi}}, \
  and\ \bibinfo {author} {\bibfnamefont {J.}~\bibnamefont {Li}},\ }\href
  {\doibase https://doi.org/10.1016/j.commatsci.2022.111280} {\bibfield
  {journal} {\bibinfo  {journal} {Comput. Mater. Sci.}\ }\textbf {\bibinfo
  {volume} {207}},\ \bibinfo {pages} {111280} (\bibinfo {year}
  {2022}{\natexlab{a}})}\BibitemShut {NoStop}%
\bibitem [{\citenamefont {Takamoto}\ \emph
  {et~al.}(2022{\natexlab{b}})\citenamefont {Takamoto}, \citenamefont
  {Shinagawa}, \citenamefont {Motoki}, \citenamefont {Nakago}, \citenamefont
  {Li}, \citenamefont {Kurata}, \citenamefont {Watanabe}, \citenamefont
  {Yayama}, \citenamefont {Iriguchi}, \citenamefont {Asano}, \citenamefont
  {Onodera}, \citenamefont {Ishii}, \citenamefont {Kudo}, \citenamefont {Ono},
  \citenamefont {Sawada}, \citenamefont {Ishitani}, \citenamefont {Ong},
  \citenamefont {Yamaguchi}, \citenamefont {Kataoka}, \citenamefont {Hayashi},
  \citenamefont {Charoenphakdee},\ and\ \citenamefont
  {Ibuka}}]{10.1038/s41467-022-30687-9}%
  \BibitemOpen
  \bibfield  {author} {\bibinfo {author} {\bibfnamefont {S.}~\bibnamefont
  {Takamoto}}, \bibinfo {author} {\bibfnamefont {C.}~\bibnamefont {Shinagawa}},
  \bibinfo {author} {\bibfnamefont {D.}~\bibnamefont {Motoki}}, \bibinfo
  {author} {\bibfnamefont {K.}~\bibnamefont {Nakago}}, \bibinfo {author}
  {\bibfnamefont {W.}~\bibnamefont {Li}}, \bibinfo {author} {\bibfnamefont
  {I.}~\bibnamefont {Kurata}}, \bibinfo {author} {\bibfnamefont
  {T.}~\bibnamefont {Watanabe}}, \bibinfo {author} {\bibfnamefont
  {Y.}~\bibnamefont {Yayama}}, \bibinfo {author} {\bibfnamefont
  {H.}~\bibnamefont {Iriguchi}}, \bibinfo {author} {\bibfnamefont
  {Y.}~\bibnamefont {Asano}}, \bibinfo {author} {\bibfnamefont
  {T.}~\bibnamefont {Onodera}}, \bibinfo {author} {\bibfnamefont
  {T.}~\bibnamefont {Ishii}}, \bibinfo {author} {\bibfnamefont
  {T.}~\bibnamefont {Kudo}}, \bibinfo {author} {\bibfnamefont {H.}~\bibnamefont
  {Ono}}, \bibinfo {author} {\bibfnamefont {R.}~\bibnamefont {Sawada}},
  \bibinfo {author} {\bibfnamefont {R.}~\bibnamefont {Ishitani}}, \bibinfo
  {author} {\bibfnamefont {M.}~\bibnamefont {Ong}}, \bibinfo {author}
  {\bibfnamefont {T.}~\bibnamefont {Yamaguchi}}, \bibinfo {author}
  {\bibfnamefont {T.}~\bibnamefont {Kataoka}}, \bibinfo {author} {\bibfnamefont
  {A.}~\bibnamefont {Hayashi}}, \bibinfo {author} {\bibfnamefont
  {N.}~\bibnamefont {Charoenphakdee}}, \ and\ \bibinfo {author} {\bibfnamefont
  {T.}~\bibnamefont {Ibuka}},\ }\href {\doibase 10.1038/s41467-022-30687-9}
  {\bibfield  {journal} {\bibinfo  {journal} {Nat. Commun.}\ }\textbf {\bibinfo
  {volume} {13}},\ \bibinfo {pages} {2991} (\bibinfo {year}
  {2022}{\natexlab{b}})}\BibitemShut {NoStop}%
\bibitem [{\citenamefont {Jacobs}\ \emph {et~al.}(2025)\citenamefont {Jacobs},
  \citenamefont {Morgan}, \citenamefont {Attarian}, \citenamefont {Meng},
  \citenamefont {Shen}, \citenamefont {Wu}, \citenamefont {Xie}, \citenamefont
  {Yang}, \citenamefont {Artrith}, \citenamefont {Blaiszik}, \citenamefont
  {Ceder}, \citenamefont {Choudhary}, \citenamefont {Csanyi}, \citenamefont
  {Cubuk}, \citenamefont {Deng}, \citenamefont {Drautz}, \citenamefont {Fu},
  \citenamefont {Godwin}, \citenamefont {Honavar}, \citenamefont {Isayev},
  \citenamefont {Johansson}, \citenamefont {Kozinsky}, \citenamefont
  {Martiniani}, \citenamefont {Ong}, \citenamefont {Poltavsky}, \citenamefont
  {Schmidt}, \citenamefont {Takamoto}, \citenamefont {Thompson}, \citenamefont
  {Westermayr},\ and\ \citenamefont {Wood}}]{JACOBS2025101214}%
  \BibitemOpen
  \bibfield  {author} {\bibinfo {author} {\bibfnamefont {R.}~\bibnamefont
  {Jacobs}}, \bibinfo {author} {\bibfnamefont {D.}~\bibnamefont {Morgan}},
  \bibinfo {author} {\bibfnamefont {S.}~\bibnamefont {Attarian}}, \bibinfo
  {author} {\bibfnamefont {J.}~\bibnamefont {Meng}}, \bibinfo {author}
  {\bibfnamefont {C.}~\bibnamefont {Shen}}, \bibinfo {author} {\bibfnamefont
  {Z.}~\bibnamefont {Wu}}, \bibinfo {author} {\bibfnamefont {C.~Y.}\
  \bibnamefont {Xie}}, \bibinfo {author} {\bibfnamefont {J.~H.}\ \bibnamefont
  {Yang}}, \bibinfo {author} {\bibfnamefont {N.}~\bibnamefont {Artrith}},
  \bibinfo {author} {\bibfnamefont {B.}~\bibnamefont {Blaiszik}}, \bibinfo
  {author} {\bibfnamefont {G.}~\bibnamefont {Ceder}}, \bibinfo {author}
  {\bibfnamefont {K.}~\bibnamefont {Choudhary}}, \bibinfo {author}
  {\bibfnamefont {G.}~\bibnamefont {Csanyi}}, \bibinfo {author} {\bibfnamefont
  {E.~D.}\ \bibnamefont {Cubuk}}, \bibinfo {author} {\bibfnamefont
  {B.}~\bibnamefont {Deng}}, \bibinfo {author} {\bibfnamefont {R.}~\bibnamefont
  {Drautz}}, \bibinfo {author} {\bibfnamefont {X.}~\bibnamefont {Fu}}, \bibinfo
  {author} {\bibfnamefont {J.}~\bibnamefont {Godwin}}, \bibinfo {author}
  {\bibfnamefont {V.}~\bibnamefont {Honavar}}, \bibinfo {author} {\bibfnamefont
  {O.}~\bibnamefont {Isayev}}, \bibinfo {author} {\bibfnamefont
  {A.}~\bibnamefont {Johansson}}, \bibinfo {author} {\bibfnamefont
  {B.}~\bibnamefont {Kozinsky}}, \bibinfo {author} {\bibfnamefont
  {S.}~\bibnamefont {Martiniani}}, \bibinfo {author} {\bibfnamefont {S.~P.}\
  \bibnamefont {Ong}}, \bibinfo {author} {\bibfnamefont {I.}~\bibnamefont
  {Poltavsky}}, \bibinfo {author} {\bibfnamefont {K.}~\bibnamefont {Schmidt}},
  \bibinfo {author} {\bibfnamefont {S.}~\bibnamefont {Takamoto}}, \bibinfo
  {author} {\bibfnamefont {A.~P.}\ \bibnamefont {Thompson}}, \bibinfo {author}
  {\bibfnamefont {J.}~\bibnamefont {Westermayr}}, \ and\ \bibinfo {author}
  {\bibfnamefont {B.~M.}\ \bibnamefont {Wood}},\ }\href {\doibase
  https://doi.org/10.1016/j.cossms.2025.101214} {\bibfield  {journal} {\bibinfo
   {journal} {Curr. Opin. Solid State Mater. Sci.}\ }\textbf {\bibinfo {volume}
  {35}},\ \bibinfo {pages} {101214} (\bibinfo {year} {2025})}\BibitemShut
  {NoStop}%
\bibitem [{\citenamefont {Jacobsen}\ \emph {et~al.}(2018)\citenamefont
  {Jacobsen}, \citenamefont {J\o{}rgensen},\ and\ \citenamefont
  {Hammer}}]{PhysRevLett.120.026102}%
  \BibitemOpen
  \bibfield  {author} {\bibinfo {author} {\bibfnamefont {T.~L.}\ \bibnamefont
  {Jacobsen}}, \bibinfo {author} {\bibfnamefont {M.~S.}\ \bibnamefont
  {J\o{}rgensen}}, \ and\ \bibinfo {author} {\bibfnamefont {B.}~\bibnamefont
  {Hammer}},\ }\href {\doibase 10.1103/PhysRevLett.120.026102} {\bibfield
  {journal} {\bibinfo  {journal} {Phys. Rev. Lett.}\ }\textbf {\bibinfo
  {volume} {120}},\ \bibinfo {pages} {026102} (\bibinfo {year}
  {2018})}\BibitemShut {NoStop}%
\bibitem [{\citenamefont {Tong}\ \emph {et~al.}(2018)\citenamefont {Tong},
  \citenamefont {Xue}, \citenamefont {Lv}, \citenamefont {Wang},\ and\
  \citenamefont {Ma}}]{C8FD00055G}%
  \BibitemOpen
  \bibfield  {author} {\bibinfo {author} {\bibfnamefont {Q.}~\bibnamefont
  {Tong}}, \bibinfo {author} {\bibfnamefont {L.}~\bibnamefont {Xue}}, \bibinfo
  {author} {\bibfnamefont {J.}~\bibnamefont {Lv}}, \bibinfo {author}
  {\bibfnamefont {Y.}~\bibnamefont {Wang}}, \ and\ \bibinfo {author}
  {\bibfnamefont {Y.}~\bibnamefont {Ma}},\ }\href {\doibase 10.1039/C8FD00055G}
  {\bibfield  {journal} {\bibinfo  {journal} {Faraday Discuss.}\ }\textbf
  {\bibinfo {volume} {211}},\ \bibinfo {pages} {31} (\bibinfo {year}
  {2018})}\BibitemShut {NoStop}%
\bibitem [{\citenamefont {Wang}\ \emph {et~al.}(2020)\citenamefont {Wang},
  \citenamefont {Zhang}, \citenamefont {Zhang},\ and\ \citenamefont
  {Wang}}]{10.3389/fchem.2020.589795}%
  \BibitemOpen
  \bibfield  {author} {\bibinfo {author} {\bibfnamefont {H.}~\bibnamefont
  {Wang}}, \bibinfo {author} {\bibfnamefont {Y.}~\bibnamefont {Zhang}},
  \bibinfo {author} {\bibfnamefont {L.}~\bibnamefont {Zhang}}, \ and\ \bibinfo
  {author} {\bibfnamefont {H.}~\bibnamefont {Wang}},\ }\href {\doibase
  10.3389/fchem.2020.589795} {\bibfield  {journal} {\bibinfo  {journal} {Front.
  Chem.}\ }\textbf {\bibinfo {volume} {8}} (\bibinfo {year} {2020}),\
  10.3389/fchem.2020.589795}\BibitemShut {NoStop}%
\bibitem [{\citenamefont {Park}\ and\ \citenamefont
  {Wolverton}(2020)}]{PhysRevMaterials.4.063801}%
  \BibitemOpen
  \bibfield  {author} {\bibinfo {author} {\bibfnamefont {C.~W.}\ \bibnamefont
  {Park}}\ and\ \bibinfo {author} {\bibfnamefont {C.}~\bibnamefont
  {Wolverton}},\ }\href {\doibase 10.1103/PhysRevMaterials.4.063801} {\bibfield
   {journal} {\bibinfo  {journal} {Phys. Rev. Mater.}\ }\textbf {\bibinfo
  {volume} {4}},\ \bibinfo {pages} {063801} (\bibinfo {year}
  {2020})}\BibitemShut {NoStop}%
\bibitem [{\citenamefont {Kang}\ \emph {et~al.}(2022)\citenamefont {Kang},
  \citenamefont {Jeong}, \citenamefont {Hong}, \citenamefont {Hwang},
  \citenamefont {Yoon},\ and\ \citenamefont {Han}}]{Kang2022}%
  \BibitemOpen
  \bibfield  {author} {\bibinfo {author} {\bibfnamefont {S.}~\bibnamefont
  {Kang}}, \bibinfo {author} {\bibfnamefont {W.}~\bibnamefont {Jeong}},
  \bibinfo {author} {\bibfnamefont {C.}~\bibnamefont {Hong}}, \bibinfo {author}
  {\bibfnamefont {S.}~\bibnamefont {Hwang}}, \bibinfo {author} {\bibfnamefont
  {Y.}~\bibnamefont {Yoon}}, \ and\ \bibinfo {author} {\bibfnamefont
  {S.}~\bibnamefont {Han}},\ }\href {\doibase 10.1038/s41524-022-00792-w}
  {\bibfield  {journal} {\bibinfo  {journal} {npj Comput. Mater.}\ }\textbf
  {\bibinfo {volume} {8}},\ \bibinfo {pages} {108} (\bibinfo {year}
  {2022})}\BibitemShut {NoStop}%
\bibitem [{\citenamefont {Cheng}\ \emph {et~al.}(2022)\citenamefont {Cheng},
  \citenamefont {Gong},\ and\ \citenamefont {Yin}}]{Cheng2022}%
  \BibitemOpen
  \bibfield  {author} {\bibinfo {author} {\bibfnamefont {G.}~\bibnamefont
  {Cheng}}, \bibinfo {author} {\bibfnamefont {X.-G.}\ \bibnamefont {Gong}}, \
  and\ \bibinfo {author} {\bibfnamefont {W.-J.}\ \bibnamefont {Yin}},\ }\href
  {\doibase 10.1038/s41467-022-29241-4} {\bibfield  {journal} {\bibinfo
  {journal} {Nat. Commun.}\ }\textbf {\bibinfo {volume} {13}},\ \bibinfo
  {pages} {1492} (\bibinfo {year} {2022})}\BibitemShut {NoStop}%
\bibitem [{\citenamefont {Schmidt}\ \emph {et~al.}(2023)\citenamefont
  {Schmidt}, \citenamefont {Hoffmann}, \citenamefont {Wang}, \citenamefont
  {Borlido}, \citenamefont {Carriço}, \citenamefont {Cerqueira}, \citenamefont
  {Botti},\ and\ \citenamefont
  {Marques}}]{https://doi.org/10.1002/adma.202210788}%
  \BibitemOpen
  \bibfield  {author} {\bibinfo {author} {\bibfnamefont {J.}~\bibnamefont
  {Schmidt}}, \bibinfo {author} {\bibfnamefont {N.}~\bibnamefont {Hoffmann}},
  \bibinfo {author} {\bibfnamefont {H.-C.}\ \bibnamefont {Wang}}, \bibinfo
  {author} {\bibfnamefont {P.}~\bibnamefont {Borlido}}, \bibinfo {author}
  {\bibfnamefont {P.~J. M.~A.}\ \bibnamefont {Carriço}}, \bibinfo {author}
  {\bibfnamefont {T.~F.~T.}\ \bibnamefont {Cerqueira}}, \bibinfo {author}
  {\bibfnamefont {S.}~\bibnamefont {Botti}}, \ and\ \bibinfo {author}
  {\bibfnamefont {M.~A.~L.}\ \bibnamefont {Marques}},\ }\href {\doibase
  https://doi.org/10.1002/adma.202210788} {\bibfield  {journal} {\bibinfo
  {journal} {Adv. Mater.}\ }\textbf {\bibinfo {volume} {35}},\ \bibinfo {pages}
  {2210788} (\bibinfo {year} {2023})}\BibitemShut {NoStop}%
\bibitem [{\citenamefont {Chang}\ \emph {et~al.}(2024)\citenamefont {Chang},
  \citenamefont {Tamaki}, \citenamefont {Yokoyama}, \citenamefont {Wakasugi},
  \citenamefont {Yotsuhashi}, \citenamefont {Kusaba}, \citenamefont {Oganov},\
  and\ \citenamefont {Yoshida}}]{Chang2024}%
  \BibitemOpen
  \bibfield  {author} {\bibinfo {author} {\bibfnamefont {L.}~\bibnamefont
  {Chang}}, \bibinfo {author} {\bibfnamefont {H.}~\bibnamefont {Tamaki}},
  \bibinfo {author} {\bibfnamefont {T.}~\bibnamefont {Yokoyama}}, \bibinfo
  {author} {\bibfnamefont {K.}~\bibnamefont {Wakasugi}}, \bibinfo {author}
  {\bibfnamefont {S.}~\bibnamefont {Yotsuhashi}}, \bibinfo {author}
  {\bibfnamefont {M.}~\bibnamefont {Kusaba}}, \bibinfo {author} {\bibfnamefont
  {A.~R.}\ \bibnamefont {Oganov}}, \ and\ \bibinfo {author} {\bibfnamefont
  {R.}~\bibnamefont {Yoshida}},\ }\href {\doibase 10.1038/s41524-024-01471-8}
  {\bibfield  {journal} {\bibinfo  {journal} {npj Comput. Mater.}\ }\textbf
  {\bibinfo {volume} {10}},\ \bibinfo {pages} {298} (\bibinfo {year}
  {2024})}\BibitemShut {NoStop}%
\bibitem [{\citenamefont {Donaldson}\ \emph {et~al.}(2024)\citenamefont
  {Donaldson}, \citenamefont {Lawrence},\ and\ \citenamefont
  {Probert}}]{Donaldson2024AGA}%
  \BibitemOpen
  \bibfield  {author} {\bibinfo {author} {\bibfnamefont {S.}~\bibnamefont
  {Donaldson}}, \bibinfo {author} {\bibfnamefont {R.~A.}\ \bibnamefont
  {Lawrence}}, \ and\ \bibinfo {author} {\bibfnamefont {M.~I.~J.}\ \bibnamefont
  {Probert}}\ }(\bibinfo {year} {2024})\ \Eprint
  {http://arxiv.org/abs/arXiv:2404.14354} {arXiv:2404.14354} \BibitemShut
  {NoStop}%
\bibitem [{\citenamefont {Bl\"ochl}(1994)}]{PhysRevB.50.17953}%
  \BibitemOpen
  \bibfield  {author} {\bibinfo {author} {\bibfnamefont {P.~E.}\ \bibnamefont
  {Bl\"ochl}},\ }\href {\doibase 10.1103/PhysRevB.50.17953} {\bibfield
  {journal} {\bibinfo  {journal} {Phys. Rev. B}\ }\textbf {\bibinfo {volume}
  {50}},\ \bibinfo {pages} {17953} (\bibinfo {year} {1994})}\BibitemShut
  {NoStop}%
\bibitem [{\citenamefont {Kresse}\ and\ \citenamefont
  {Joubert}(1999)}]{PhysRevB.59.1758}%
  \BibitemOpen
  \bibfield  {author} {\bibinfo {author} {\bibfnamefont {G.}~\bibnamefont
  {Kresse}}\ and\ \bibinfo {author} {\bibfnamefont {D.}~\bibnamefont
  {Joubert}},\ }\href {\doibase 10.1103/PhysRevB.59.1758} {\bibfield  {journal}
  {\bibinfo  {journal} {Phys. Rev. B}\ }\textbf {\bibinfo {volume} {59}},\
  \bibinfo {pages} {1758} (\bibinfo {year} {1999})}\BibitemShut {NoStop}%
\bibitem [{\citenamefont {Perdew}\ \emph {et~al.}(1996)\citenamefont {Perdew},
  \citenamefont {Burke},\ and\ \citenamefont
  {Ernzerhof}}]{PhysRevLett.77.3865}%
  \BibitemOpen
  \bibfield  {author} {\bibinfo {author} {\bibfnamefont {J.~P.}\ \bibnamefont
  {Perdew}}, \bibinfo {author} {\bibfnamefont {K.}~\bibnamefont {Burke}}, \
  and\ \bibinfo {author} {\bibfnamefont {M.}~\bibnamefont {Ernzerhof}},\ }\href
  {\doibase 10.1103/PhysRevLett.77.3865} {\bibfield  {journal} {\bibinfo
  {journal} {Phys. Rev. Lett.}\ }\textbf {\bibinfo {volume} {77}},\ \bibinfo
  {pages} {3865} (\bibinfo {year} {1996})}\BibitemShut {NoStop}%
\bibitem [{\citenamefont {Kresse}\ and\ \citenamefont
  {Hafner}(1993)}]{PhysRevB.47.558}%
  \BibitemOpen
  \bibfield  {author} {\bibinfo {author} {\bibfnamefont {G.}~\bibnamefont
  {Kresse}}\ and\ \bibinfo {author} {\bibfnamefont {J.}~\bibnamefont
  {Hafner}},\ }\href {\doibase 10.1103/PhysRevB.47.558} {\bibfield  {journal}
  {\bibinfo  {journal} {Phys. Rev. B}\ }\textbf {\bibinfo {volume} {47}},\
  \bibinfo {pages} {558} (\bibinfo {year} {1993})}\BibitemShut {NoStop}%
\bibitem [{\citenamefont {Kresse}\ and\ \citenamefont
  {Furthm\"uller}(1996)}]{PhysRevB.54.11169}%
  \BibitemOpen
  \bibfield  {author} {\bibinfo {author} {\bibfnamefont {G.}~\bibnamefont
  {Kresse}}\ and\ \bibinfo {author} {\bibfnamefont {J.}~\bibnamefont
  {Furthm\"uller}},\ }\href {\doibase 10.1103/PhysRevB.54.11169} {\bibfield
  {journal} {\bibinfo  {journal} {Phys. Rev. B}\ }\textbf {\bibinfo {volume}
  {54}},\ \bibinfo {pages} {11169} (\bibinfo {year} {1996})}\BibitemShut
  {NoStop}%
\bibitem [{\citenamefont {Kresse}\ and\ \citenamefont
  {Furthm\"{u}ller}(1996)}]{KRESSE199615}%
  \BibitemOpen
  \bibfield  {author} {\bibinfo {author} {\bibfnamefont {G.}~\bibnamefont
  {Kresse}}\ and\ \bibinfo {author} {\bibfnamefont {J.}~\bibnamefont
  {Furthm\"{u}ller}},\ }\href {\doibase
  https://doi.org/10.1016/0927-0256(96)00008-0} {\bibfield  {journal} {\bibinfo
   {journal} {Comput. Mater. Sci.}\ }\textbf {\bibinfo {volume} {6}},\ \bibinfo
  {pages} {15} (\bibinfo {year} {1996})}\BibitemShut {NoStop}%
\bibitem [{\citenamefont {Dudarev}\ \emph {et~al.}(1998)\citenamefont
  {Dudarev}, \citenamefont {Botton}, \citenamefont {Savrasov}, \citenamefont
  {Humphreys},\ and\ \citenamefont {Sutton}}]{PhysRevB.57.1505}%
  \BibitemOpen
  \bibfield  {author} {\bibinfo {author} {\bibfnamefont {S.~L.}\ \bibnamefont
  {Dudarev}}, \bibinfo {author} {\bibfnamefont {G.~A.}\ \bibnamefont {Botton}},
  \bibinfo {author} {\bibfnamefont {S.~Y.}\ \bibnamefont {Savrasov}}, \bibinfo
  {author} {\bibfnamefont {C.~J.}\ \bibnamefont {Humphreys}}, \ and\ \bibinfo
  {author} {\bibfnamefont {A.~P.}\ \bibnamefont {Sutton}},\ }\href {\doibase
  10.1103/PhysRevB.57.1505} {\bibfield  {journal} {\bibinfo  {journal} {Phys.
  Rev. B}\ }\textbf {\bibinfo {volume} {57}},\ \bibinfo {pages} {1505}
  (\bibinfo {year} {1998})}\BibitemShut {NoStop}%
\bibitem [{\citenamefont {Jain}\ \emph {et~al.}(2013)\citenamefont {Jain},
  \citenamefont {Ong}, \citenamefont {Hautier}, \citenamefont {Chen},
  \citenamefont {Richards}, \citenamefont {Dacek}, \citenamefont {Cholia},
  \citenamefont {Gunter}, \citenamefont {Skinner}, \citenamefont {Ceder},\ and\
  \citenamefont {Persson}}]{10.1063/1.4812323}%
  \BibitemOpen
  \bibfield  {author} {\bibinfo {author} {\bibfnamefont {A.}~\bibnamefont
  {Jain}}, \bibinfo {author} {\bibfnamefont {S.~P.}\ \bibnamefont {Ong}},
  \bibinfo {author} {\bibfnamefont {G.}~\bibnamefont {Hautier}}, \bibinfo
  {author} {\bibfnamefont {W.}~\bibnamefont {Chen}}, \bibinfo {author}
  {\bibfnamefont {W.~D.}\ \bibnamefont {Richards}}, \bibinfo {author}
  {\bibfnamefont {S.}~\bibnamefont {Dacek}}, \bibinfo {author} {\bibfnamefont
  {S.}~\bibnamefont {Cholia}}, \bibinfo {author} {\bibfnamefont
  {D.}~\bibnamefont {Gunter}}, \bibinfo {author} {\bibfnamefont
  {D.}~\bibnamefont {Skinner}}, \bibinfo {author} {\bibfnamefont
  {G.}~\bibnamefont {Ceder}}, \ and\ \bibinfo {author} {\bibfnamefont {K.~A.}\
  \bibnamefont {Persson}},\ }\href {\doibase 10.1063/1.4812323} {\bibfield
  {journal} {\bibinfo  {journal} {APL Mater.}\ }\textbf {\bibinfo {volume}
  {1}},\ \bibinfo {pages} {011002} (\bibinfo {year} {2013})}\BibitemShut
  {NoStop}%
\bibitem [{\citenamefont {Wang}\ \emph {et~al.}(2006)\citenamefont {Wang},
  \citenamefont {Maxisch},\ and\ \citenamefont {Ceder}}]{PhysRevB.73.195107}%
  \BibitemOpen
  \bibfield  {author} {\bibinfo {author} {\bibfnamefont {L.}~\bibnamefont
  {Wang}}, \bibinfo {author} {\bibfnamefont {T.}~\bibnamefont {Maxisch}}, \
  and\ \bibinfo {author} {\bibfnamefont {G.}~\bibnamefont {Ceder}},\ }\href
  {\doibase 10.1103/PhysRevB.73.195107} {\bibfield  {journal} {\bibinfo
  {journal} {Phys. Rev. B}\ }\textbf {\bibinfo {volume} {73}},\ \bibinfo
  {pages} {195107} (\bibinfo {year} {2006})}\BibitemShut {NoStop}%
\bibitem [{\citenamefont {Mehl}\ \emph {et~al.}(2017)\citenamefont {Mehl},
  \citenamefont {Hicks}, \citenamefont {Toher}, \citenamefont {Levy},
  \citenamefont {Hanson}, \citenamefont {Hart},\ and\ \citenamefont
  {Curtarolo}}]{MEHL2017S1}%
  \BibitemOpen
  \bibfield  {author} {\bibinfo {author} {\bibfnamefont {M.~J.}\ \bibnamefont
  {Mehl}}, \bibinfo {author} {\bibfnamefont {D.}~\bibnamefont {Hicks}},
  \bibinfo {author} {\bibfnamefont {C.}~\bibnamefont {Toher}}, \bibinfo
  {author} {\bibfnamefont {O.}~\bibnamefont {Levy}}, \bibinfo {author}
  {\bibfnamefont {R.~M.}\ \bibnamefont {Hanson}}, \bibinfo {author}
  {\bibfnamefont {G.}~\bibnamefont {Hart}}, \ and\ \bibinfo {author}
  {\bibfnamefont {S.}~\bibnamefont {Curtarolo}},\ }\href {\doibase
  https://doi.org/10.1016/j.commatsci.2017.01.017} {\bibfield  {journal}
  {\bibinfo  {journal} {Comput. Mater. Sci.}\ }\textbf {\bibinfo {volume}
  {136}},\ \bibinfo {pages} {S1} (\bibinfo {year} {2017})}\BibitemShut
  {NoStop}%
\bibitem [{\citenamefont {Ong}\ \emph {et~al.}(2013)\citenamefont {Ong},
  \citenamefont {Richards}, \citenamefont {Jain}, \citenamefont {Hautier},
  \citenamefont {Kocher}, \citenamefont {Cholia}, \citenamefont {Gunter},
  \citenamefont {Chevrier}, \citenamefont {Persson},\ and\ \citenamefont
  {Ceder}}]{ONG2013314}%
  \BibitemOpen
  \bibfield  {author} {\bibinfo {author} {\bibfnamefont {S.~P.}\ \bibnamefont
  {Ong}}, \bibinfo {author} {\bibfnamefont {W.~D.}\ \bibnamefont {Richards}},
  \bibinfo {author} {\bibfnamefont {A.}~\bibnamefont {Jain}}, \bibinfo {author}
  {\bibfnamefont {G.}~\bibnamefont {Hautier}}, \bibinfo {author} {\bibfnamefont
  {M.}~\bibnamefont {Kocher}}, \bibinfo {author} {\bibfnamefont
  {S.}~\bibnamefont {Cholia}}, \bibinfo {author} {\bibfnamefont
  {D.}~\bibnamefont {Gunter}}, \bibinfo {author} {\bibfnamefont {V.~L.}\
  \bibnamefont {Chevrier}}, \bibinfo {author} {\bibfnamefont {K.~A.}\
  \bibnamefont {Persson}}, \ and\ \bibinfo {author} {\bibfnamefont
  {G.}~\bibnamefont {Ceder}},\ }\href {\doibase
  https://doi.org/10.1016/j.commatsci.2012.10.028} {\bibfield  {journal}
  {\bibinfo  {journal} {Comput. Mater. Sci.}\ }\textbf {\bibinfo {volume}
  {68}},\ \bibinfo {pages} {314} (\bibinfo {year} {2013})}\BibitemShut
  {NoStop}%
\bibitem [{\citenamefont {Wang}\ \emph {et~al.}(2021)\citenamefont {Wang},
  \citenamefont {Kingsbury}, \citenamefont {McDermott}, \citenamefont {Horton},
  \citenamefont {Jain}, \citenamefont {Ong}, \citenamefont {Dwaraknath},\ and\
  \citenamefont {Persson}}]{Wang2021}%
  \BibitemOpen
  \bibfield  {author} {\bibinfo {author} {\bibfnamefont {A.}~\bibnamefont
  {Wang}}, \bibinfo {author} {\bibfnamefont {R.}~\bibnamefont {Kingsbury}},
  \bibinfo {author} {\bibfnamefont {M.}~\bibnamefont {McDermott}}, \bibinfo
  {author} {\bibfnamefont {M.}~\bibnamefont {Horton}}, \bibinfo {author}
  {\bibfnamefont {A.}~\bibnamefont {Jain}}, \bibinfo {author} {\bibfnamefont
  {S.~P.}\ \bibnamefont {Ong}}, \bibinfo {author} {\bibfnamefont
  {S.}~\bibnamefont {Dwaraknath}}, \ and\ \bibinfo {author} {\bibfnamefont
  {K.~A.}\ \bibnamefont {Persson}},\ }\href {\doibase
  10.1038/s41598-021-94550-5} {\bibfield  {journal} {\bibinfo  {journal} {Sci.
  Rep.}\ }\textbf {\bibinfo {volume} {11}},\ \bibinfo {pages} {15496} (\bibinfo
  {year} {2021})}\BibitemShut {NoStop}%
\bibitem [{\citenamefont {Jain}\ \emph {et~al.}(2011)\citenamefont {Jain},
  \citenamefont {Hautier}, \citenamefont {Ong}, \citenamefont {Moore},
  \citenamefont {Fischer}, \citenamefont {Persson},\ and\ \citenamefont
  {Ceder}}]{PhysRevB.84.045115}%
  \BibitemOpen
  \bibfield  {author} {\bibinfo {author} {\bibfnamefont {A.}~\bibnamefont
  {Jain}}, \bibinfo {author} {\bibfnamefont {G.}~\bibnamefont {Hautier}},
  \bibinfo {author} {\bibfnamefont {S.~P.}\ \bibnamefont {Ong}}, \bibinfo
  {author} {\bibfnamefont {C.~J.}\ \bibnamefont {Moore}}, \bibinfo {author}
  {\bibfnamefont {C.~C.}\ \bibnamefont {Fischer}}, \bibinfo {author}
  {\bibfnamefont {K.~A.}\ \bibnamefont {Persson}}, \ and\ \bibinfo {author}
  {\bibfnamefont {G.}~\bibnamefont {Ceder}},\ }\href {\doibase
  10.1103/PhysRevB.84.045115} {\bibfield  {journal} {\bibinfo  {journal} {Phys.
  Rev. B}\ }\textbf {\bibinfo {volume} {84}},\ \bibinfo {pages} {045115}
  (\bibinfo {year} {2011})}\BibitemShut {NoStop}%
\bibitem [{\citenamefont {Trimarchi}\ \emph {et~al.}(2009)\citenamefont
  {Trimarchi}, \citenamefont {Freeman},\ and\ \citenamefont
  {Zunger}}]{PhysRevB.80.092101}%
  \BibitemOpen
  \bibfield  {author} {\bibinfo {author} {\bibfnamefont {G.}~\bibnamefont
  {Trimarchi}}, \bibinfo {author} {\bibfnamefont {A.~J.}\ \bibnamefont
  {Freeman}}, \ and\ \bibinfo {author} {\bibfnamefont {A.}~\bibnamefont
  {Zunger}},\ }\href {\doibase 10.1103/PhysRevB.80.092101} {\bibfield
  {journal} {\bibinfo  {journal} {Phys. Rev. B}\ }\textbf {\bibinfo {volume}
  {80}},\ \bibinfo {pages} {092101} (\bibinfo {year} {2009})}\BibitemShut
  {NoStop}%
\bibitem [{\citenamefont {Bartel}(2022)}]{Bartel2022}%
  \BibitemOpen
  \bibfield  {author} {\bibinfo {author} {\bibfnamefont {C.~J.}\ \bibnamefont
  {Bartel}},\ }\href {\doibase 10.1007/s10853-022-06915-4} {\bibfield
  {journal} {\bibinfo  {journal} {J. Mater. Sci.}\ }\textbf {\bibinfo {volume}
  {57}},\ \bibinfo {pages} {10475} (\bibinfo {year} {2022})}\BibitemShut
  {NoStop}%
\bibitem [{\citenamefont {Omee}\ \emph {et~al.}(2023)\citenamefont {Omee},
  \citenamefont {Wei},\ and\ \citenamefont
  {Hu}}]{omee2023crystalstructurepredictionusing}%
  \BibitemOpen
  \bibfield  {author} {\bibinfo {author} {\bibfnamefont {S.~S.}\ \bibnamefont
  {Omee}}, \bibinfo {author} {\bibfnamefont {L.}~\bibnamefont {Wei}}, \ and\
  \bibinfo {author} {\bibfnamefont {J.}~\bibnamefont {Hu}},\ }\href
  {https://arxiv.org/abs/2309.06710} {\enquote {\bibinfo {title} {Crystal
  structure prediction using neural network potential and age-fitness pareto
  genetic algorithm},}\ } (\bibinfo {year} {2023}),\ \Eprint
  {http://arxiv.org/abs/arXiv:2309.06710} {arXiv:2309.06710} \BibitemShut
  {NoStop}%
\bibitem [{\citenamefont {Omee}\ \emph {et~al.}(2025)\citenamefont {Omee},
  \citenamefont {Wei}, \citenamefont {Dey},\ and\ \citenamefont
  {Hu}}]{omee2025polymorphismcrystalstructureprediction}%
  \BibitemOpen
  \bibfield  {author} {\bibinfo {author} {\bibfnamefont {S.~S.}\ \bibnamefont
  {Omee}}, \bibinfo {author} {\bibfnamefont {L.}~\bibnamefont {Wei}}, \bibinfo
  {author} {\bibfnamefont {S.}~\bibnamefont {Dey}}, \ and\ \bibinfo {author}
  {\bibfnamefont {J.}~\bibnamefont {Hu}},\ }\href
  {https://arxiv.org/abs/2506.11332} {\enquote {\bibinfo {title} {Polymorphism
  crystal structure prediction with adaptive space group diversity control},}\
  } (\bibinfo {year} {2025}),\ \Eprint {http://arxiv.org/abs/2506.11332}
  {arXiv:2506.11332 [cond-mat.mtrl-sci]} \BibitemShut {NoStop}%
\bibitem [{\citenamefont {Goldberg}\ \emph {et~al.}(1987)\citenamefont
  {Goldberg}, \citenamefont {Richardson} \emph {et~al.}}]{goldberg1987genetic}%
  \BibitemOpen
  \bibfield  {author} {\bibinfo {author} {\bibfnamefont {D.~E.}\ \bibnamefont
  {Goldberg}}, \bibinfo {author} {\bibfnamefont {J.}~\bibnamefont
  {Richardson}},  \emph {et~al.},\ }in\ \href@noop {} {\emph {\bibinfo
  {booktitle} {Genetic algorithms and their applications: Proceedings of the
  Second International Conference on Genetic Algorithms}}},\ Vol.\ \bibinfo
  {volume} {4149}\ (\bibinfo {organization} {Lawrence Erlbaum, Hillsdale, NJ},\
  \bibinfo {year} {1987})\ pp.\ \bibinfo {pages} {414--425}\BibitemShut
  {NoStop}%
\bibitem [{\citenamefont {Deb}\ \emph {et~al.}(2002)\citenamefont {Deb},
  \citenamefont {Pratap}, \citenamefont {Agarwal},\ and\ \citenamefont
  {Meyarivan}}]{996017}%
  \BibitemOpen
  \bibfield  {author} {\bibinfo {author} {\bibfnamefont {K.}~\bibnamefont
  {Deb}}, \bibinfo {author} {\bibfnamefont {A.}~\bibnamefont {Pratap}},
  \bibinfo {author} {\bibfnamefont {S.}~\bibnamefont {Agarwal}}, \ and\
  \bibinfo {author} {\bibfnamefont {T.}~\bibnamefont {Meyarivan}},\ }\href
  {\doibase 10.1109/4235.996017} {\bibfield  {journal} {\bibinfo  {journal}
  {IEEE Trans. Evol. Comput.}\ }\textbf {\bibinfo {volume} {6}},\ \bibinfo
  {pages} {182} (\bibinfo {year} {2002})}\BibitemShut {NoStop}%
\bibitem [{\citenamefont {Deb}\ and\ \citenamefont {Jain}(2014)}]{6600851}%
  \BibitemOpen
  \bibfield  {author} {\bibinfo {author} {\bibfnamefont {K.}~\bibnamefont
  {Deb}}\ and\ \bibinfo {author} {\bibfnamefont {H.}~\bibnamefont {Jain}},\
  }\href {\doibase 10.1109/TEVC.2013.2281535} {\bibfield  {journal} {\bibinfo
  {journal} {IEEE Trans. Evol. Comput.}\ }\textbf {\bibinfo {volume} {18}},\
  \bibinfo {pages} {577} (\bibinfo {year} {2014})}\BibitemShut {NoStop}%
\bibitem [{\citenamefont {Akiba}\ \emph {et~al.}(2019)\citenamefont {Akiba},
  \citenamefont {Sano}, \citenamefont {Yanase}, \citenamefont {Ohta},\ and\
  \citenamefont {Koyama}}]{akiba2019optuna}%
  \BibitemOpen
  \bibfield  {author} {\bibinfo {author} {\bibfnamefont {T.}~\bibnamefont
  {Akiba}}, \bibinfo {author} {\bibfnamefont {S.}~\bibnamefont {Sano}},
  \bibinfo {author} {\bibfnamefont {T.}~\bibnamefont {Yanase}}, \bibinfo
  {author} {\bibfnamefont {T.}~\bibnamefont {Ohta}}, \ and\ \bibinfo {author}
  {\bibfnamefont {M.}~\bibnamefont {Koyama}},\ }in\ \href {\doibase
  10.1145/3292500.3330701} {\emph {\bibinfo {booktitle} {The 25th ACM SIGKDD
  International Conference on Knowledge Discovery \& Data Mining}}}\ (\bibinfo
  {year} {2019})\ pp.\ \bibinfo {pages} {2623--2631}\BibitemShut {NoStop}%
\bibitem [{\citenamefont {Larsen}\ \emph {et~al.}(2017)\citenamefont {Larsen},
  \citenamefont {Mortensen}, \citenamefont {Blomqvist}, \citenamefont
  {Castelli}, \citenamefont {Christensen}, \citenamefont {Du\l{}ak},
  \citenamefont {Friis}, \citenamefont {Groves}, \citenamefont {Hammer},
  \citenamefont {Hargus}, \citenamefont {Hermes}, \citenamefont {Jennings},
  \citenamefont {Jensen}, \citenamefont {Kermode}, \citenamefont {Kitchin},
  \citenamefont {Kolsbjerg}, \citenamefont {Kubal}, \citenamefont {Kaasbjerg},
  \citenamefont {Lysgaard}, \citenamefont {Maronsson}, \citenamefont {Maxson},
  \citenamefont {Olsen}, \citenamefont {Pastewka}, \citenamefont {Peterson},
  \citenamefont {Rostgaard}, \citenamefont {Schi\o{}tz}, \citenamefont
  {Sch\"{u}tt}, \citenamefont {Strange}, \citenamefont {Thygesen},
  \citenamefont {Vegge}, \citenamefont {Vilhelmsen}, \citenamefont {Walter},
  \citenamefont {Zeng},\ and\ \citenamefont {Jacobsen}}]{ase2017}%
  \BibitemOpen
  \bibfield  {author} {\bibinfo {author} {\bibfnamefont {A.~H.}\ \bibnamefont
  {Larsen}}, \bibinfo {author} {\bibfnamefont {J.~J.}\ \bibnamefont
  {Mortensen}}, \bibinfo {author} {\bibfnamefont {J.}~\bibnamefont
  {Blomqvist}}, \bibinfo {author} {\bibfnamefont {I.~E.}\ \bibnamefont
  {Castelli}}, \bibinfo {author} {\bibfnamefont {R.}~\bibnamefont
  {Christensen}}, \bibinfo {author} {\bibfnamefont {M.}~\bibnamefont
  {Du\l{}ak}}, \bibinfo {author} {\bibfnamefont {J.}~\bibnamefont {Friis}},
  \bibinfo {author} {\bibfnamefont {M.~N.}\ \bibnamefont {Groves}}, \bibinfo
  {author} {\bibfnamefont {B.}~\bibnamefont {Hammer}}, \bibinfo {author}
  {\bibfnamefont {C.}~\bibnamefont {Hargus}}, \bibinfo {author} {\bibfnamefont
  {E.~D.}\ \bibnamefont {Hermes}}, \bibinfo {author} {\bibfnamefont {P.~C.}\
  \bibnamefont {Jennings}}, \bibinfo {author} {\bibfnamefont {P.~B.}\
  \bibnamefont {Jensen}}, \bibinfo {author} {\bibfnamefont {J.}~\bibnamefont
  {Kermode}}, \bibinfo {author} {\bibfnamefont {J.~R.}\ \bibnamefont
  {Kitchin}}, \bibinfo {author} {\bibfnamefont {E.~L.}\ \bibnamefont
  {Kolsbjerg}}, \bibinfo {author} {\bibfnamefont {J.}~\bibnamefont {Kubal}},
  \bibinfo {author} {\bibfnamefont {K.}~\bibnamefont {Kaasbjerg}}, \bibinfo
  {author} {\bibfnamefont {S.}~\bibnamefont {Lysgaard}}, \bibinfo {author}
  {\bibfnamefont {J.~B.}\ \bibnamefont {Maronsson}}, \bibinfo {author}
  {\bibfnamefont {T.}~\bibnamefont {Maxson}}, \bibinfo {author} {\bibfnamefont
  {T.}~\bibnamefont {Olsen}}, \bibinfo {author} {\bibfnamefont
  {L.}~\bibnamefont {Pastewka}}, \bibinfo {author} {\bibfnamefont
  {A.}~\bibnamefont {Peterson}}, \bibinfo {author} {\bibfnamefont
  {C.}~\bibnamefont {Rostgaard}}, \bibinfo {author} {\bibfnamefont
  {J.}~\bibnamefont {Schi\o{}tz}}, \bibinfo {author} {\bibfnamefont
  {O.}~\bibnamefont {Sch\"{u}tt}}, \bibinfo {author} {\bibfnamefont
  {M.}~\bibnamefont {Strange}}, \bibinfo {author} {\bibfnamefont {K.~S.}\
  \bibnamefont {Thygesen}}, \bibinfo {author} {\bibfnamefont {T.}~\bibnamefont
  {Vegge}}, \bibinfo {author} {\bibfnamefont {L.}~\bibnamefont {Vilhelmsen}},
  \bibinfo {author} {\bibfnamefont {M.}~\bibnamefont {Walter}}, \bibinfo
  {author} {\bibfnamefont {Z.}~\bibnamefont {Zeng}}, \ and\ \bibinfo {author}
  {\bibfnamefont {K.~W.}\ \bibnamefont {Jacobsen}},\ }\href {\doibase
  10.1088/1361-648X/aa680e} {\bibfield  {journal} {\bibinfo  {journal} {J.
  Phys. Condens. Matter.}\ }\textbf {\bibinfo {volume} {29}},\ \bibinfo {pages}
  {273002} (\bibinfo {year} {2017})}\BibitemShut {NoStop}%
\bibitem [{\citenamefont {Lyakhov}\ \emph {et~al.}(2010)\citenamefont
  {Lyakhov}, \citenamefont {Oganov},\ and\ \citenamefont
  {Valle}}]{LYAKHOV20101623}%
  \BibitemOpen
  \bibfield  {author} {\bibinfo {author} {\bibfnamefont {A.~O.}\ \bibnamefont
  {Lyakhov}}, \bibinfo {author} {\bibfnamefont {A.~R.}\ \bibnamefont {Oganov}},
  \ and\ \bibinfo {author} {\bibfnamefont {M.}~\bibnamefont {Valle}},\ }\href
  {\doibase https://doi.org/10.1016/j.cpc.2010.06.007} {\bibfield  {journal}
  {\bibinfo  {journal} {Comput. Phys. Commun.}\ }\textbf {\bibinfo {volume}
  {181}},\ \bibinfo {pages} {1623} (\bibinfo {year} {2010})}\BibitemShut
  {NoStop}%
\bibitem [{\citenamefont {Fredericks}\ \emph {et~al.}(2021)\citenamefont
  {Fredericks}, \citenamefont {Parrish}, \citenamefont {Sayre},\ and\
  \citenamefont {Zhu}}]{FREDERICKS2021107810}%
  \BibitemOpen
  \bibfield  {author} {\bibinfo {author} {\bibfnamefont {S.}~\bibnamefont
  {Fredericks}}, \bibinfo {author} {\bibfnamefont {K.}~\bibnamefont {Parrish}},
  \bibinfo {author} {\bibfnamefont {D.}~\bibnamefont {Sayre}}, \ and\ \bibinfo
  {author} {\bibfnamefont {Q.}~\bibnamefont {Zhu}},\ }\href {\doibase
  https://doi.org/10.1016/j.cpc.2020.107810} {\bibfield  {journal} {\bibinfo
  {journal} {Comput. Phys. Commun.}\ }\textbf {\bibinfo {volume} {261}},\
  \bibinfo {pages} {107810} (\bibinfo {year} {2021})}\BibitemShut {NoStop}%
\bibitem [{\citenamefont {Valle}\ and\ \citenamefont
  {Oganov}(2008)}]{valle2008crystal}%
  \BibitemOpen
  \bibfield  {author} {\bibinfo {author} {\bibfnamefont {M.}~\bibnamefont
  {Valle}}\ and\ \bibinfo {author} {\bibfnamefont {A.~R.}\ \bibnamefont
  {Oganov}},\ }in\ \href@noop {} {\emph {\bibinfo {booktitle} {2008 IEEE
  Symposium on Visual Analytics Science and Technology}}}\ (\bibinfo
  {organization} {IEEE},\ \bibinfo {year} {2008})\ pp.\ \bibinfo {pages}
  {11--18}\BibitemShut {NoStop}%
\bibitem [{\citenamefont {Oganov}\ and\ \citenamefont
  {Valle}(2009)}]{oganov2009quantify}%
  \BibitemOpen
  \bibfield  {author} {\bibinfo {author} {\bibfnamefont {A.~R.}\ \bibnamefont
  {Oganov}}\ and\ \bibinfo {author} {\bibfnamefont {M.}~\bibnamefont {Valle}},\
  }\href@noop {} {\bibfield  {journal} {\bibinfo  {journal} {The Journal of
  chemical physics}\ }\textbf {\bibinfo {volume} {130}} (\bibinfo {year}
  {2009})}\BibitemShut {NoStop}%
\bibitem [{\citenamefont {Oganov}\ \emph {et~al.}(2010)\citenamefont {Oganov},
  \citenamefont {Ma}, \citenamefont {Lyakhov}, \citenamefont {Valle},\ and\
  \citenamefont {Gatti}}]{10.2138/rmg.2010.71.13}%
  \BibitemOpen
  \bibfield  {author} {\bibinfo {author} {\bibfnamefont {A.~R.}\ \bibnamefont
  {Oganov}}, \bibinfo {author} {\bibfnamefont {Y.}~\bibnamefont {Ma}}, \bibinfo
  {author} {\bibfnamefont {A.~O.}\ \bibnamefont {Lyakhov}}, \bibinfo {author}
  {\bibfnamefont {M.}~\bibnamefont {Valle}}, \ and\ \bibinfo {author}
  {\bibfnamefont {C.}~\bibnamefont {Gatti}},\ }\href {\doibase
  10.2138/rmg.2010.71.13} {\bibfield  {journal} {\bibinfo  {journal} {Rev.
  Mineral. Geochem.}\ }\textbf {\bibinfo {volume} {71}},\ \bibinfo {pages}
  {271} (\bibinfo {year} {2010})}\BibitemShut {NoStop}%
\bibitem [{\citenamefont {Zhu}\ \emph {et~al.}(2012)\citenamefont {Zhu},
  \citenamefont {Oganov},\ and\ \citenamefont {Lyakhov}}]{C2CE06642D}%
  \BibitemOpen
  \bibfield  {author} {\bibinfo {author} {\bibfnamefont {Q.}~\bibnamefont
  {Zhu}}, \bibinfo {author} {\bibfnamefont {A.~R.}\ \bibnamefont {Oganov}}, \
  and\ \bibinfo {author} {\bibfnamefont {A.~O.}\ \bibnamefont {Lyakhov}},\
  }\href {\doibase 10.1039/C2CE06642D} {\bibfield  {journal} {\bibinfo
  {journal} {CrystEngComm}\ }\textbf {\bibinfo {volume} {14}},\ \bibinfo
  {pages} {3596} (\bibinfo {year} {2012})}\BibitemShut {NoStop}%
\bibitem [{\citenamefont {Real}\ \emph {et~al.}(2019)\citenamefont {Real},
  \citenamefont {Aggarwal}, \citenamefont {Huang},\ and\ \citenamefont
  {Le}}]{real2019regularized}%
  \BibitemOpen
  \bibfield  {author} {\bibinfo {author} {\bibfnamefont {E.}~\bibnamefont
  {Real}}, \bibinfo {author} {\bibfnamefont {A.}~\bibnamefont {Aggarwal}},
  \bibinfo {author} {\bibfnamefont {Y.}~\bibnamefont {Huang}}, \ and\ \bibinfo
  {author} {\bibfnamefont {Q.~V.}\ \bibnamefont {Le}},\ }in\ \href
  {https://dl.acm.org/doi/abs/10.1609/aaai.v33i01.33014780} {\emph {\bibinfo
  {booktitle} {Proceedings of the AAAI Conference on artificial
  intelligence}}},\ Vol.~\bibinfo {volume} {33}\ (\bibinfo {year} {2019})\ pp.\
  \bibinfo {pages} {4780--4789}\BibitemShut {NoStop}%
\bibitem [{Note1()}]{Note1}%
  \BibitemOpen
  \bibinfo {note} {The modified crossover and mutation methods are released and
  available at \protect \url
  {https://pypi.org/project/pfn-ase-extras/}.}\BibitemShut {Stop}%
\bibitem [{\citenamefont {Vilhelmsen}\ and\ \citenamefont
  {Hammer}(2012)}]{PhysRevLett.108.126101}%
  \BibitemOpen
  \bibfield  {author} {\bibinfo {author} {\bibfnamefont {L.~B.}\ \bibnamefont
  {Vilhelmsen}}\ and\ \bibinfo {author} {\bibfnamefont {B.}~\bibnamefont
  {Hammer}},\ }\href {\doibase 10.1103/PhysRevLett.108.126101} {\bibfield
  {journal} {\bibinfo  {journal} {Phys. Rev. Lett.}\ }\textbf {\bibinfo
  {volume} {108}},\ \bibinfo {pages} {126101} (\bibinfo {year}
  {2012})}\BibitemShut {NoStop}%
\bibitem [{\citenamefont {Clark}\ and\ \citenamefont
  {Evans}(1954)}]{clark1954distance}%
  \BibitemOpen
  \bibfield  {author} {\bibinfo {author} {\bibfnamefont {P.~J.}\ \bibnamefont
  {Clark}}\ and\ \bibinfo {author} {\bibfnamefont {F.~C.}\ \bibnamefont
  {Evans}},\ }\href {\doibase 10.2307/1931034} {\bibfield  {journal} {\bibinfo
  {journal} {Ecology}\ }\textbf {\bibinfo {volume} {35}},\ \bibinfo {pages}
  {445} (\bibinfo {year} {1954})}\BibitemShut {NoStop}%
\bibitem [{\citenamefont {Zeni}\ \emph {et~al.}(2025)\citenamefont {Zeni},
  \citenamefont {Pinsler}, \citenamefont {Z{\"u}gner}, \citenamefont {Fowler},
  \citenamefont {Horton}, \citenamefont {Fu}, \citenamefont {Wang},
  \citenamefont {Shysheya}, \citenamefont {Crabb{\'e}}, \citenamefont {Ueda},
  \citenamefont {Sordillo}, \citenamefont {Sun}, \citenamefont {Smith},
  \citenamefont {Nguyen}, \citenamefont {Schulz}, \citenamefont {Lewis},
  \citenamefont {Huang}, \citenamefont {Lu}, \citenamefont {Zhou},
  \citenamefont {Yang}, \citenamefont {Hao}, \citenamefont {Li}, \citenamefont
  {Yang}, \citenamefont {Li}, \citenamefont {Tomioka},\ and\ \citenamefont
  {Xie}}]{MatterGen2025}%
  \BibitemOpen
  \bibfield  {author} {\bibinfo {author} {\bibfnamefont {C.}~\bibnamefont
  {Zeni}}, \bibinfo {author} {\bibfnamefont {R.}~\bibnamefont {Pinsler}},
  \bibinfo {author} {\bibfnamefont {D.}~\bibnamefont {Z{\"u}gner}}, \bibinfo
  {author} {\bibfnamefont {A.}~\bibnamefont {Fowler}}, \bibinfo {author}
  {\bibfnamefont {M.}~\bibnamefont {Horton}}, \bibinfo {author} {\bibfnamefont
  {X.}~\bibnamefont {Fu}}, \bibinfo {author} {\bibfnamefont {Z.}~\bibnamefont
  {Wang}}, \bibinfo {author} {\bibfnamefont {A.}~\bibnamefont {Shysheya}},
  \bibinfo {author} {\bibfnamefont {J.}~\bibnamefont {Crabb{\'e}}}, \bibinfo
  {author} {\bibfnamefont {S.}~\bibnamefont {Ueda}}, \bibinfo {author}
  {\bibfnamefont {R.}~\bibnamefont {Sordillo}}, \bibinfo {author}
  {\bibfnamefont {L.}~\bibnamefont {Sun}}, \bibinfo {author} {\bibfnamefont
  {J.}~\bibnamefont {Smith}}, \bibinfo {author} {\bibfnamefont
  {B.}~\bibnamefont {Nguyen}}, \bibinfo {author} {\bibfnamefont
  {H.}~\bibnamefont {Schulz}}, \bibinfo {author} {\bibfnamefont
  {S.}~\bibnamefont {Lewis}}, \bibinfo {author} {\bibfnamefont {C.-W.}\
  \bibnamefont {Huang}}, \bibinfo {author} {\bibfnamefont {Z.}~\bibnamefont
  {Lu}}, \bibinfo {author} {\bibfnamefont {Y.}~\bibnamefont {Zhou}}, \bibinfo
  {author} {\bibfnamefont {H.}~\bibnamefont {Yang}}, \bibinfo {author}
  {\bibfnamefont {H.}~\bibnamefont {Hao}}, \bibinfo {author} {\bibfnamefont
  {J.}~\bibnamefont {Li}}, \bibinfo {author} {\bibfnamefont {C.}~\bibnamefont
  {Yang}}, \bibinfo {author} {\bibfnamefont {W.}~\bibnamefont {Li}}, \bibinfo
  {author} {\bibfnamefont {R.}~\bibnamefont {Tomioka}}, \ and\ \bibinfo
  {author} {\bibfnamefont {T.}~\bibnamefont {Xie}},\ }\href {\doibase
  10.1038/s41586-025-08628-5} {\bibfield  {journal} {\bibinfo  {journal}
  {Nature}\ } (\bibinfo {year} {2025}),\
  10.1038/s41586-025-08628-5}\BibitemShut {NoStop}%
\bibitem [{Note2()}]{Note2}%
  \BibitemOpen
  \bibinfo {note} {These crystal structures in xyz formats are available at
  Supplemental Materials.}\BibitemShut {Stop}%
\bibitem [{\citenamefont {Togo}\ \emph {et~al.}(2024)\citenamefont {Togo},
  \citenamefont {Shinohara},\ and\ \citenamefont {Isao}}]{spglib}%
  \BibitemOpen
  \bibfield  {author} {\bibinfo {author} {\bibfnamefont {A.}~\bibnamefont
  {Togo}}, \bibinfo {author} {\bibfnamefont {K.}~\bibnamefont {Shinohara}}, \
  and\ \bibinfo {author} {\bibfnamefont {T.}~\bibnamefont {Isao}},\ }\href
  {\doibase 10.1080/27660400.2024.2384822} {\bibfield  {journal} {\bibinfo
  {journal} {Sci. Technol. Adv. Mater., Meth.}\ }\textbf {\bibinfo {volume}
  {4}},\ \bibinfo {pages} {2384822} (\bibinfo {year} {2024})}\BibitemShut
  {NoStop}%
\bibitem [{\citenamefont {Hicks}\ \emph {et~al.}(2019)\citenamefont {Hicks},
  \citenamefont {Mehl}, \citenamefont {Gossett}, \citenamefont {Toher},
  \citenamefont {Levy}, \citenamefont {Hanson}, \citenamefont {Hart},\ and\
  \citenamefont {Curtarolo}}]{HICKS2019S1}%
  \BibitemOpen
  \bibfield  {author} {\bibinfo {author} {\bibfnamefont {D.}~\bibnamefont
  {Hicks}}, \bibinfo {author} {\bibfnamefont {M.~J.}\ \bibnamefont {Mehl}},
  \bibinfo {author} {\bibfnamefont {E.}~\bibnamefont {Gossett}}, \bibinfo
  {author} {\bibfnamefont {C.}~\bibnamefont {Toher}}, \bibinfo {author}
  {\bibfnamefont {O.}~\bibnamefont {Levy}}, \bibinfo {author} {\bibfnamefont
  {R.~M.}\ \bibnamefont {Hanson}}, \bibinfo {author} {\bibfnamefont
  {G.}~\bibnamefont {Hart}}, \ and\ \bibinfo {author} {\bibfnamefont
  {S.}~\bibnamefont {Curtarolo}},\ }\href {\doibase
  https://doi.org/10.1016/j.commatsci.2018.10.043} {\bibfield  {journal}
  {\bibinfo  {journal} {Comput. Mater. Sc.}\ }\textbf {\bibinfo {volume}
  {161}},\ \bibinfo {pages} {S1} (\bibinfo {year} {2019})}\BibitemShut
  {NoStop}%
\bibitem [{\citenamefont {Hicks}\ \emph {et~al.}(2021)\citenamefont {Hicks},
  \citenamefont {Mehl}, \citenamefont {Esters}, \citenamefont {Oses},
  \citenamefont {Levy}, \citenamefont {Hart}, \citenamefont {Toher},\ and\
  \citenamefont {Curtarolo}}]{HICKS2021110450}%
  \BibitemOpen
  \bibfield  {author} {\bibinfo {author} {\bibfnamefont {D.}~\bibnamefont
  {Hicks}}, \bibinfo {author} {\bibfnamefont {M.~J.}\ \bibnamefont {Mehl}},
  \bibinfo {author} {\bibfnamefont {M.}~\bibnamefont {Esters}}, \bibinfo
  {author} {\bibfnamefont {C.}~\bibnamefont {Oses}}, \bibinfo {author}
  {\bibfnamefont {O.}~\bibnamefont {Levy}}, \bibinfo {author} {\bibfnamefont
  {G.~L.}\ \bibnamefont {Hart}}, \bibinfo {author} {\bibfnamefont
  {C.}~\bibnamefont {Toher}}, \ and\ \bibinfo {author} {\bibfnamefont
  {S.}~\bibnamefont {Curtarolo}},\ }\href {\doibase
  https://doi.org/10.1016/j.commatsci.2021.110450} {\bibfield  {journal}
  {\bibinfo  {journal} {Comput. Mater. Sci.}\ }\textbf {\bibinfo {volume}
  {199}},\ \bibinfo {pages} {110450} (\bibinfo {year} {2021})}\BibitemShut
  {NoStop}%
\bibitem [{\citenamefont {Eckert}\ \emph {et~al.}(2024)\citenamefont {Eckert},
  \citenamefont {Divilov}, \citenamefont {Mehl}, \citenamefont {Hicks},
  \citenamefont {Zettel}, \citenamefont {Esters}, \citenamefont {Campilongo},\
  and\ \citenamefont {Curtarolo}}]{ECKERT2024112988}%
  \BibitemOpen
  \bibfield  {author} {\bibinfo {author} {\bibfnamefont {H.}~\bibnamefont
  {Eckert}}, \bibinfo {author} {\bibfnamefont {S.}~\bibnamefont {Divilov}},
  \bibinfo {author} {\bibfnamefont {M.~J.}\ \bibnamefont {Mehl}}, \bibinfo
  {author} {\bibfnamefont {D.}~\bibnamefont {Hicks}}, \bibinfo {author}
  {\bibfnamefont {A.~C.}\ \bibnamefont {Zettel}}, \bibinfo {author}
  {\bibfnamefont {M.}~\bibnamefont {Esters}}, \bibinfo {author} {\bibfnamefont
  {X.}~\bibnamefont {Campilongo}}, \ and\ \bibinfo {author} {\bibfnamefont
  {S.}~\bibnamefont {Curtarolo}},\ }\href {\doibase
  https://doi.org/10.1016/j.commatsci.2024.112988} {\bibfield  {journal}
  {\bibinfo  {journal} {Comput. Mater. Sci.}\ }\textbf {\bibinfo {volume}
  {240}},\ \bibinfo {pages} {112988} (\bibinfo {year} {2024})}\BibitemShut
  {NoStop}%
\bibitem [{\citenamefont {Goodall}\ \emph {et~al.}(2022)\citenamefont
  {Goodall}, \citenamefont {Parackal}, \citenamefont {Faber}, \citenamefont
  {Armiento},\ and\ \citenamefont {Lee}}]{goodall_2022_rapid}%
  \BibitemOpen
  \bibfield  {author} {\bibinfo {author} {\bibfnamefont {R.~E.}\ \bibnamefont
  {Goodall}}, \bibinfo {author} {\bibfnamefont {A.~S.}\ \bibnamefont
  {Parackal}}, \bibinfo {author} {\bibfnamefont {F.~A.}\ \bibnamefont {Faber}},
  \bibinfo {author} {\bibfnamefont {R.}~\bibnamefont {Armiento}}, \ and\
  \bibinfo {author} {\bibfnamefont {A.~A.}\ \bibnamefont {Lee}},\ }\href
  {https://www.science.org/doi/10.1126/sciadv.abn4117} {\bibfield  {journal}
  {\bibinfo  {journal} {Sci. Adv.}\ }\textbf {\bibinfo {volume} {8}},\ \bibinfo
  {pages} {eabn4117} (\bibinfo {year} {2022})}\BibitemShut {NoStop}%
\end{thebibliography}%

\end{document}